\documentclass[aps,prl,reprint,superscriptaddress]{revtex4-2}%

\usepackage{amsfonts}
\usepackage{amssymb}
\usepackage{amsmath}
\usepackage{siunitx}
\usepackage{epsfig}
\usepackage{color}
\usepackage{graphics, graphicx}
\usepackage{bbold}
\usepackage{psfrag}
\usepackage{mathcomp}
\usepackage{verbatim}
\usepackage{float}

\usepackage[colorlinks,citecolor=blue]{hyperref}
\usepackage{subcaption}

\makeatletter

\newcommand{\Rmnum}[1]{\expandafter\@slowromancap\romannumeral #1@}
\makeatother

\usepackage{xr-hyper}

\usepackage[utf8]{inputenc}
\usepackage{dcolumn}
\usepackage{bm}
\usepackage{adjustbox}
\usepackage{booktabs}

\graphicspath{{Figure/}}

\usepackage{cleveref}
\usepackage{braket}

\crefname{subequation}{Eq.}{Eqs.}

\usepackage{caption}

\renewcommand{\figurename}{FIG.}
\usepackage[english]{babel}
\usepackage{ragged2e}

\makeatletter
\renewcommand{\fnum@figure}{\figurename~\thefigure}
\long\def\@makecaption#1#2{
  \vskip\abovecaptionskip
  \justifying\normalfont\small
  \sbox\@tempboxa{#1: #2}
  \ifdim \wd\@tempboxa >\hsize
    #1: #2\par
  \else
    \global \@minipagefalse
    \hb@xt@\hsize{\hfil\box\@tempboxa\hfil}
  \fi
  \vskip\belowcaptionskip}
\makeatother

\makeatletter
\renewcommand\section{\@startsection {section}{1}{\z@}
    {-1.5ex \@plus -1ex \@minus -.2ex}
    {1.0ex}
    {\normalfont\small\bfseries\centering}}
\makeatother

\begin{document}



\title{Noise-Robust Quantum State Characterization for Remote State Preparation with Deep Learning}


\author{Bo Tang}
\altaffiliation{These authors contributed equally to this work.}
\affiliation{State Key Laboratory of Photonics and Communications, School of Physics and Astronomy, Shanghai Jiao Tong University, Shanghai 200240, China}

\author{Zixuan Liao}
\altaffiliation{These authors contributed equally to this work.}
\affiliation{State Key Laboratory of Photonics and Communications, School of Physics and Astronomy, Shanghai Jiao Tong University, Shanghai 200240, China}

\author{Hao Li}
\altaffiliation{These authors contributed equally to this work.}
\affiliation{State Key Laboratory of Photonics and Communications, School of Physics and Astronomy, Shanghai Jiao Tong University, Shanghai 200240, China}

\author{Yilin Yang}%
\affiliation{State Key Laboratory of Photonics and Communications, School of Physics and Astronomy, Shanghai Jiao Tong University, Shanghai 200240, China}
\author{Jiani Lei}%
\affiliation{State Key Laboratory of Photonics and Communications, School of Physics and Astronomy, Shanghai Jiao Tong University, Shanghai 200240, China}
\author{Zengya Li}
\affiliation{State Key Laboratory of Photonics and Communications, School of Physics and Astronomy, Shanghai Jiao Tong University, Shanghai 200240, China}
\author{Jing Qiu}
\affiliation{State Key Laboratory of Photonics and Communications, School of Physics and Astronomy, Shanghai Jiao Tong University, Shanghai 200240, China}
\author{Zhaohui Dong}
\affiliation{State Key Laboratory of Photonics and Communications, School of Physics and Astronomy, Shanghai Jiao Tong University, Shanghai 200240, China}
\author{Zhengyang Mao}
\affiliation{State Key Laboratory of Photonics and Communications, School of Physics and Astronomy, Shanghai Jiao Tong University, Shanghai 200240, China}

\author{Yuanhua Li}
\email{lyhua1984@shiep.edu.cn}
\affiliation{Department of Physics, Shanghai Key Laboratory of Materials Protection and Advanced Materials in Electric Power, Shanghai University of Electric Power, Shanghai 200090, China}

\author{Yuanlin Zheng}
\email{ylzheng@sjtu.edu.cn}
\affiliation{State Key Laboratory of Photonics and Communications, School of Physics and Astronomy, Shanghai Jiao Tong University, Shanghai 200240, China}
\affiliation{Hefei National Laboratory, Hefei 230088, China}
\affiliation{Shanghai Research Center for Quantum Sciences, Shanghai 201315, China}

\author{Xianfeng Chen}
\email{xfchen@sjtu.edu.cn}
\affiliation{State Key Laboratory of Photonics and Communications, School of Physics and Astronomy, Shanghai Jiao Tong University, Shanghai 200240, China}
\affiliation{Hefei National Laboratory, Hefei 230088, China}
\affiliation{Shanghai Research Center for Quantum Sciences, Shanghai 201315, China}
\affiliation{Collaborative Innovation Center of Light Manipulations and Applications, Shandong Normal University, Jinan 250358, China}

\noaffiliation

\date{\today}

\begin{abstract}

Quantum communication underpins secure information processing and scalable quantum networks. In particular, remote state preparation (RSP) enables efficient quantum state transfer, but accurately estimating target states under complex noise remains challenging. 
Here, we propose a Transformer-based Quantum State Characterizer (TQSC) model for noisy RSP experiments. 
Our model reconstructs experimentally prepared pure and mixed photonic polarization states from noisy measurements in complex scattering environments, while its attention patterns provide physically grounded insights into correlations among the measured observables.
The method achieves a mean estimator-target fidelity exceeding 99.999\% under complex scattering and dynamic Gaussian noise, while its robustness and generalization are further examined using Qiskit-simulated Bloch-ball states.
Furthermore, in a practical MNIST image transmission task with held-out states, the decoded bit error rate is reduced from 50.34\% to zero after TQSC post-processing.
The TQSC model enables accurate tomographic characterization under dynamic noise and provides physically grounded post-hoc insights, holding promise for intelligent quantum information processing applications.

\end{abstract}


\maketitle


\section{Introduction}
Quantum information science drives advancements in communication, computing, and sensing by harnessing quantum mechanics \cite{main2025Distributed,pittaluga2025Longdistancea,bhattacharyya2024imaging}. Quantum communication is the cornerstone of the future quantum internet, enabling secure information transfer \cite{kimble2008quantum,yang2025300}. Reliable distribution of quantum states between distant nodes is a fundamental task for practical quantum networks \cite{wehner2018quantum}. Among protocols, remote state preparation (RSP) is a promising approach \cite{pati2000minimum,lo2000classical}. RSP enables Alice to remotely prepare a known quantum state at Bob's site using shared entanglement and classical communication, offering reduced classical resources than teleportation \cite{bennett1993teleporting} and improved security over direct transmission \cite{pogorzalek2019secure}. Firstly demonstrated in liquid-state nuclear magnetic resonance \cite{peng2003experimental}, RSP is extended to single- and multi-qubit states 
\cite{peters2005remote,xiang2005RemotePreparationMixed,raadmark2013experimental}.
Recent advances include photonic states with non-classical features \cite{liu2022experimental}, high-dimensional states \cite{erhard2020advances} using hybrid entanglement \cite{erhard2015real}, orbital-angular-momentum lattices \cite{cameron2021remote}, and metasurfaces \cite{ning2025high}, and implementations in hybrid platforms \cite{sun2021remote}.

Reliable quantum state preparation and characterization is central to practical quantum information processing. In RSP protocols, pure states are essential for computation, communication, and metrology, 
while mixed states offer noise resilience and resource efficiency, enabling advantages in realistic quantum communication and computation \cite{chitambar2019quantum,modi2012classical}.
Accurate reconstruction through quantum state tomography (QST) and fidelity estimation is therefore critical \cite{zhang2021direct,qin2024experimental}. Nevertheless, hardware imperfections---including decoherence \cite{zurek2003decoherence}, dynamic scattering \cite{defienne2018adaptive}, control errors \cite{magesan2011scalable}, and state preparation and measurement (SPAM) errors \cite{blume2017demonstration}---invariably constrain the achievable fidelity.

To combat complex noise environments, machine learning has emerged as a powerful tool for quantum estimation and control \cite{ma2025machine,torlai2020machine}. Specifically, learning-based approaches have been applied to quantum error mitigation \cite{strikis2021learning,cai2023quantum,liao2024machine,liao2025noise}, dynamic state preparation \cite{wang2024adaptive,li2023enhanced}, and QST, where neural networks map noisy measurement statistics to density matrices \cite{torlai2018neural,carrasquilla2019reconstructing,palmieri2020experimental,ahmed2021quantum,hu2026error}. 
The Transformer \cite{vaswani2017attention}, featuring self-attention to model sequential dependencies and multi-head attention to capture diverse contextual relationships, has demonstrated strong multimodal performance \cite{dosovitskiy2021an,dong2018speech}. 
In quantum physics, this architecture has been employed across a wide range of problems \cite{zhang2025survey}, including 
electronic structures \cite{sobral2025physics}, quantum correlations \cite{zhang2023transformer}, noise-robust quantum communication \cite{Li2025Language}, automated circuit generation \cite{daimon2024quantum}, scalable quantum error correction \cite{Bausch2024Learning}.

Previous attention-based QST methods have learned measurement distributions \cite{cha2022attention}, denoised LI/MLE estimates through Cholesky representations under simulated noise \cite{palmieri2024enhancing}, and reconstructed states from structured measurements with IBM-device validation \cite{ma2025tomography}. However, these studies do not address quantum-state reconstruction in real physical environments involving complex scattering and dynamically varying noise. Moreover, conventional `black-box' neural networks provide limited physical insight, making it difficult to fully trust the reconstructed outputs. High model performance does not necessarily indicate that the model has learned meaningful and task-relevant features \cite{ribeiro2016should,lapuschkin2019unmasking,degrave2021ai}.

Here, we propose a Transformer-based Quantum State Characterizer (TQSC) model with built-in physical constraints for quantum state reconstruction, designed to improve reconstruction fidelity and robustness in noisy RSP while enabling physically grounded post-hoc interpretation. Using photonic polarization as the experimental platform, TQSC reconstructs density matrices from noisy measurements under both MMF transmission alone and MMF transmission with additional dynamic perturbations. Beyond reconstruction fidelity, we further analyze the learned internal representation using statistical, intervention-based, and end-to-end attribution methods rather than relying on attention maps alone. Experimentally, TQSC substantially improves reconstruction performance, achieving an average fidelity above 99.999\%.

These results demonstrate a measurement-level and experimentally validated framework for robust quantum-state reconstruction in dynamically perturbed optical RSP systems. More broadly, reliable state reconstruction under realistic experimental noise and perturbations may support state monitoring, validation, and diagnostics in practical quantum networks and distributed quantum information processing.

\begin{figure*}[htbp]
    \centering
    \captionsetup[figure]{justification=justified}
    \includegraphics[width=0.95\textwidth]{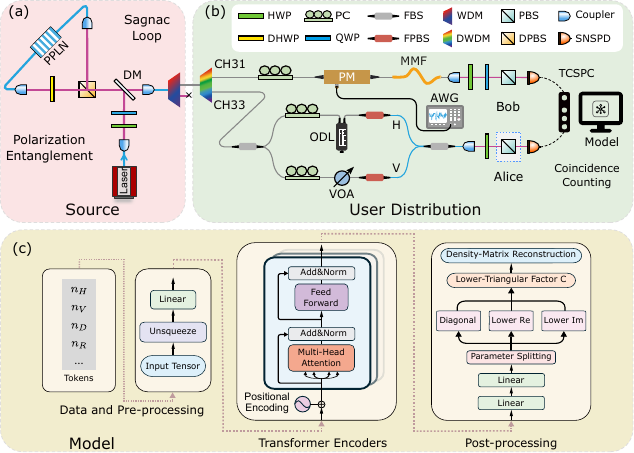}
    \caption{Schematic diagram of the experimental setup for RSP. (a) Polarization entanglement source based on Sagnac loop. (b) RSP of user distribution device. PPLN, periodically poled lithium niobate waveguide; DM, dichroic mirror; HWP, half-wave plate; DHWP, dual-wavelength half-wave plate; QWP, quarter-wave plate; PC, polarizaton controller; FBS, fiber beam splitter; FPBS, fiber polarization beam splitter; WDM, wavelength division multiplexing; DWDM, dense wavelength division multiplexing; PM, phase modulator; AWG, arbitrary waveform generator; PBS, polarizing beam splitter; DPBS, dual-polarizing beam splitter; MMF, multimode fiber; ODL, optical delay line; VOA, variable optical attenuator; SNSPD, Superconducting nanowire single-photon detector; TCSPC, timecorrelated single-photon counting. 
    (c) Multidimensional measurement data are processed through a preprocessing module, a three-layer Transformer Encoders module for feature extraction, and a post-processing module that reconstructs the predicted density matrix. The loss function quantifies the difference between predicted and actual labels, minimized via the Adam optimizer.
    }
    \label{setup}
\end{figure*}

\section{The Scheme}
The positive operator-valued measure (POVM) provides a fundamental framework for quantum measurement theory~ \cite{kraus1983states}. In our RSP protocol, the probability of obtaining measurement outcome $m$ is given by
\begin{equation}
    p_m = \mathrm{Tr}(M_m^\dagger M_m \rho),
\end{equation}
where the measurement operators $\{M_m\}$ satisfy the completeness condition
$\sum_{m} M_m^\dagger M_m = I$.
Following the theoretical framework in Ref.~\cite{wu201016DeterministicRemotePreparation}, but with a distinct entanglement source, we implement the RSP protocol for mixed states. Taking a mixed state as an example, Bob's target mixed state to be transmitted is expressed as
\begin{equation}
    \rho_{B} = \alpha^2 |H\rangle\langle H| + \beta^2 |V\rangle\langle V|,
    \label{eq:Bob's_target_mixed_state}
\end{equation}
     (where $\alpha^2 + \beta^2 = 1$). The maximally entangled Bell state generated by our entanglement source takes the form
\begin{equation}
    |\Phi^+\rangle_{AB} = \frac{1}{\sqrt{2}} \left( |H\rangle_A |H\rangle_B + |V\rangle_A |V\rangle_B \right).
    \label{Bell}
\end{equation}
After passing through the POVM-based pre-processing module, projective measurements on specific polarization states are implemented using analyzers. 
When Alice's photon is projected onto the $|H\rangle$ state, the remote preparation of Bob's photon collapses to the desired state $\hat{\rho}_B^I$ (see \cref{eq:Bob's_target_mixed_state}). For other measurement outcomes, Bob applies appropriate local unitary operations $\{ \hat{\sigma}_x, \hat{\sigma}_z,\hat{\sigma}_y\}$ to recover the target state. 

The experimental implementation of RSP is inevitably affected by various physical imperfections. Quantum noise originates from uncontrollable interactions between a quantum system and its environment, leading to deviations from ideal unitary evolution. We describe this noise using the Kraus operator formalism of quantum channels. 
In our analysis of RSP, we consider four primary sources of noise: photon loss, decoherence, fiber-induced noise, and detector imperfections. The final state of Bob, incorporating these noise effects, is given by
\begin{equation}
\rho_B^{\text{final}} =
\sum_{k'} P(k') \mathcal{E}_{k'}
\left(
U_{k'} \rho_{B,\text{ideal}}^{(k')} U_{k'}^\dagger
\right).
\end{equation}
Here, $P(k')$ denotes the experimentally observed probability of Alice obtaining measurement outcome $k'$,
$\rho_{B,\mathrm{ideal}}^{(k')}$ is Bob's ideal conditional state associated with
outcome $k'$ before the correction operation, $U_{k'}$ is the corresponding correction
unitary applied by Bob, 
and $\mathcal{E}_{k'}$ represents an outcome-dependent effective trace-preserving noise map for the entire noisy RSP branch, not only for post-correction noise. The final state is obtained by
averaging the noise-affected corrected states over all possible measurement outcomes.

Our goal is to suppress this noise. In the context of neural networks, the entire process can be viewed as a complex quantum-to-classical channel, denoted by $\mathcal{C}$, which maps an ideal target quantum state $\rho_{\rm target}$ onto a set of classical measurement outcomes.
The data vector thus constitutes a noise-corrupted quantum state:
\begin{equation}
    \mathbf{M} = \mathcal{C}(\rho_{\text{target}}).
\end{equation}

The map $\mathcal{C}$ formally integrates the four previously discussed noise sources, representing them as a unified framework of coherent and incoherent dynamic disturbances.

To counteract these effects, we propose a TQSC framework that employs a neural network to parameterize an inverse channel, denoted as $\Phi_{\text{TQSC}}(\cdot; \theta)$. The predicted quantum state $\rho_{\text{pred}}$ is obtained from the measurement data $\mathbf{M}$ via the inverse map:
\begin{equation}
    \rho_{\text{pred}} = \Phi_{\text{TQSC}}(\mathbf{M}; \theta).
    \label{eq:inverse_map}
\end{equation}
Further details on the POVM implementation, RSP communication noise, and the model are provided in the Supplemental Material (SM).

\section{Experimental Setup}
The experimental setup (Fig.~\ref{setup}) utilizes a polarization Sagnac interferometric loop with an integrated spontaneous parametric down-conversion (SPDC) source to generate phase-stable polarization-entangled photon pairs. A femtosecond laser pumps a periodically poled lithium niobate (PPLN) waveguide within the loop, producing degenerate photon pairs at the telecom wavelength. Waveplates optimize the source for maximal Bell state generation (Eq. \ref{Bell}). Subsequent wavelength filtering routes photons to separate receivers for Alice and Bob.

Alice's detection module employs a beam splitter to divide the incoming photon. One path incorporates a tunable delay for synchronization, while the other includes a variable attenuator to adjust the intensity ratio $\eta=\alpha^{2}/\beta^{2}$ between measurement bases. Polarization controllers (PC) compensate fiber effects before polarization projection. A hybrid detection scheme combines specific polarization outputs at a second beam splitter for state analysis, followed by final projective measurement. 

Bob's receiver includes a PC for fiber-effect compensation, followed by a segment of MMF mimicking spatially scrambling in scattering media \cite{yu2024high,matthes2021learning}. Dynamic noise is generated by driving the phase modulator (PM) with signals from the arbitrary waveform generator (AWG). QST is performed with a polarization analyzer composed of a quarter-wave plate (QWP), a half-wave plate (HWP), and a polarizing beam splitter (PBS) arranged in sequence. 
For preparing pure states, the PBS in Alice's polarization analyzer is retained and the QWP at the entanglement source is adjusted to control the phase. For mixed states, the PBS in Alice's analyzer is removed.
Both users employ superconducting nanowire single-photon detectors (SNSPDs). Detection events are recorded with time-correlated single-photon counting (TCSPC). 

The structure of the TQSC is illustrated in Fig.~\ref{setup}(c). The model takes a 13-dimensional input vector representing measurement data for quantum states and outputs the corresponding density matrix. 

Input data points are treated as tokens. After preprocessing with linear transformation and dimension adjustment, the data passes through a stack of Transformer encoders. These incorporate positional encoding and multi-head self-attention to capture dependencies between sequence elements, with residual connections and layer normalization stabilizing the learning process. A feedforward network then extracts deeper nonlinear features.

In the post-processing stage, the extracted features are linearly transformed into real factor parameters and split into diagonal, lower-real, and lower-imaginary components. These components form a lower-triangular complex factor C, and the final density matrix is reconstructed as $\rho=CC^\dagger/\mathrm{Tr}(CC^\dagger)$.

Model training minimizes the Mean Squared Error (MSE)  between predicted and true density matrices using the Adam optimizer. 
Details of the experimental setup, model architecture, and training parameters are provided in the SM.
\begin{figure}[tbp]
    \centering
    \includegraphics{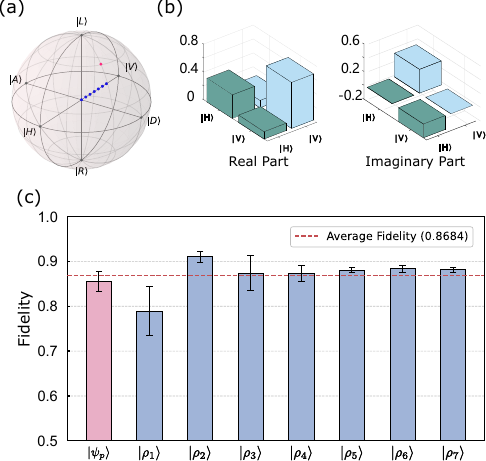}
    \caption{Characterization and verification of RSP.
    (a) Remote quantum states are prepared on the Bloch ball, with the pure state shown as a red dot and mixed states shown as blue dots \cite{LAMBERT20261}.
    (b) The real and imaginary parts of the density matrix with $\eta=1$.
    (c) Reconstruction fidelity of the pure state and mixed states prepared via RSP through an MMF.}
    \label{Fidelity}
\end{figure}
\section{Result}

Initially, we prepare a set of mixed states with $\eta$ ranging from 1 to 7, given by
\begin{equation}
    \rho_\eta = \frac{1}{\eta+1} (|H\rangle\langle H| + \eta|V\rangle\langle V|).
\end{equation}
 Furthermore, we prepare a pure state $|\psi_p\rangle$ as
\begin{equation}
    |\psi_p\rangle = \frac{1}{\sqrt{10}} \left( 3|H\rangle + e^{i\pi/3}|V\rangle \right).
\end{equation}
The corresponding data points are shown in Fig.~\ref{Fidelity}(a).

QST uses the projection bases $H$, $V$, $D$, and $R$, corresponding to horizontal, vertical, diagonal, and right-circular polarizations, respectively.
We prepare eight quantum states, averaging five measurements per basis without the MMF to benchmark state-preparation fidelity, and utilize single-shot measurements for all subsequent MMF data.

To verify the prepared states, we reconstruct the density matrix $\rho_B$ using the maximum likelihood method \cite{james2001measurement} and calculate the fidelity $\langle \overline{F} \rangle$ with the target prepared state $\rho_p$ \cite{altepeter2005photonic}.
The fidelity is computed using the following formula:
\begin{equation}
F(\rho_p, \rho_B) = \left( \operatorname{Tr} \left( \sqrt{ \sqrt{\rho_p} \, \rho_B \, \sqrt{\rho_p} } \right) \right)^2.
\end{equation}

Without the MMF, the average fidelity of these states reaches $(99.25\pm0.20)\%$, demonstrating the high state-preparation capability of our experimental setup. Upon introducing the MMF, however, the fidelities of the states decrease to varying degrees, as illustrated in Fig.~\ref{Fidelity}(c).
As an example, the real and imaginary parts of the density matrix for the case $\eta=1$ are reconstructed and shown in Fig.~\ref{Fidelity}(b).

The impact of dynamic disturbances on the system coherence is then evaluated.
In the absence of noise, the prepared quantum states exhibit a purity of $1$ and a
fidelity of $(99.23 \pm 0.20)\%$. When zero-mean Gaussian dynamic noise ($\sigma = \SI{1}{\volt}$) generated by the AWG is introduced by driving the PM ($V_\pi = \SI{5}{\volt}$) in the MMF setup, the purity decreases to $0.6724 \pm 0.0698$, while the fidelity drops to $(38.31 \pm 12.30)\%$, indicating significant decoherence and fidelity degradation under the applied noise conditions.

\begin{figure}[tbp]
    \centering
    \includegraphics{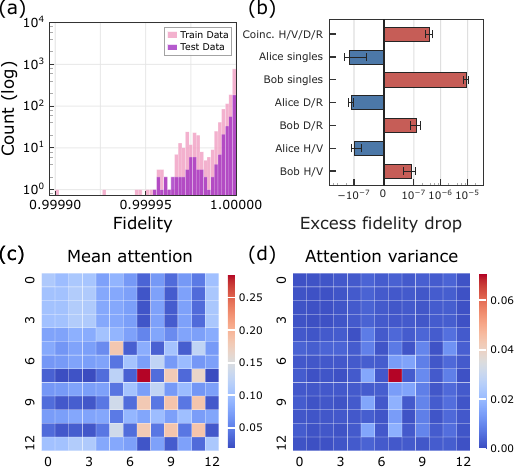}
    \caption{
Performance and attention analysis of the TQSC model.
(a) Fidelity distributions for the training and test datasets.
(b) Excess fidelity drop from targeted token-group interventions relative to size-matched random ablations.
(c) Mean attention map of Layer 3 Head 4 (L3H4). 
(d) Attention variance map of L3H4. In (c) and (d), the query (y) and key (x) axes index the 13 input tokens (0--12), comprising H/V/D/R coincidence counts, the corresponding Alice and Bob single-photon counts, and measurement time.
}
    \label{NN_result}
\end{figure}

To overcome these limitations, our model is comprehensively trained on pure states, mixed states, and pure states under dynamic noise, achieving an overall average estimation fidelity of $0.99999497 \pm 8.78 \times 10^{-6}$ when evaluated on both training and test sets.
As illustrated in Fig.~\ref{NN_result}(a),
the estimation fidelity distributions for both the training and test sets are tightly clustered around values approaching 1, indicating precise agreement between predicted results and theoretical values. 
To benchmark the reconstruction performance, we compare the TQSC model with representative baselines; it achieves the highest average test fidelity, with a slight advantage over the parameter-matched MLP baseline. 

Furthermore, the model demonstrates robust generalization to new parameter values, maintaining estimation fidelities above 99\% in tests involving both interpolation ($\eta=4.5$)  and leave-one-out validation ($\eta=6$).
In addition, we test the model on Qiskit-simulated Bloch-ball states beyond the experimentally measured state family, where it also achieves estimation fidelities exceeding 99\%.
This result demonstrates that the model maintains high estimation fidelity on unseen test data with a slight performance degradation, indicating strong generalization ability and robustness without evident signs of overfitting. Using only single-shot measurement data as input, the TQSC model reliably reconstructs quantum states with high estimation fidelity.

Recent MPO-based methods target large-scale mixed-state learning under noise \cite{votto2026learning}, rather than direct pure-state optimization under dynamic noise.

Admittedly, the train set scattered high-fidelity outliers suggest potential sensitivity of the model to edge-case data. Additionally, the limited sample size of the test set ($n < 400$) may necessitate expansion of the validation set to enhance statistical inference reliability.

\begin{figure}[tbp]
    \centering
    \includegraphics{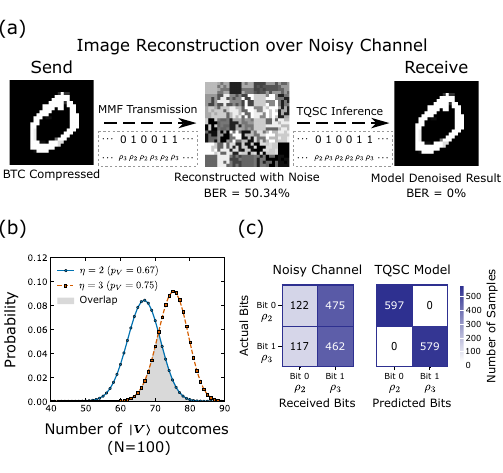}
    \caption{TQSC-based image transmission and reconstruction. 
(a) Degradation and recovery of BTC-compressed MNIST images through the noisy channel. 
(b) Probability distributions of $|V\rangle$ measurement outcomes ($N=100$) for mixed states $\rho_2$ and $\rho_3$. 
(c) Confusion matrices for the noisy channel (left) and TQSC predictions (right).}
    \label{Fig:Image_Transmission}
\end{figure}

L3H4-targeted token-group ablations in Fig.~\ref{NN_result}(b) reveal a nonuniform local feature dependence, dominated by Bob-side single counts, with smaller positive contributions from coincidence and Bob polarization features.

To gain insight into how the TQSC model integrates experimental information when optimizing quantum-state fidelity, we analyze the attention maps of Layer 3 Head 4 (L3H4), a deeper Transformer layer that may capture higher-order, task-relevant dependencies among input features \cite{jawahar2019does}. As shown in Fig.~\ref{NN_result}(c), the mean attention map exhibits localized high-attention regions, indicating preferential interactions among selected query--key pairs. The attention variance in Fig.~\ref{NN_result}(d) is concentrated on a smaller subset of pairs, showing that only selected interactions vary appreciably across samples. Together, these features indicate selective rather than global information integration.

Quantitative analysis of these localized attention patterns further reveals enhanced Bob H/V--Bob D/R and Coinc.\ H/V--Coinc.\ D/R interactions, consistent with the model integrating complementary population- and coherence/phase-sensitive information. In addition, interactions between the coincidence features and acquisition time suggest that the model interprets the coincidence statistics in conjunction with the measurement duration, which determines the degree of statistical averaging and hence the reliability of the measured counts.
Together, these observations suggest that the deeper-layer attention does not merely emphasize individual observables, but captures structured correlations among multiple experimentally relevant features. Quantitative attention and intervention analyses are provided in the SM. 

Ultimately, to further validate the practicality of the proposed scheme, we conduct an image transmission experiment under the same system configuration as described above. In the complex scattering environment of MMF, mixed states $\rho_2$ and $\rho_3$ are respectively employed to encode bits 0 and 1, and fidelity-based criteria are used for bit decision. 
In Fig.~\ref{Fig:Image_Transmission}(b), with the state fidelity of $F(\rho_2, \rho_3) \approx 99.16\%$, measurement-induced Gaussian noise causes the distributions to overlap, making the states difficult to resolve even without the MMF.
The $N=100$ measurements are used only to characterize the projection-outcome distributions and are not used for decoding or BER evaluation. 

A grayscale image of the handwritten digit ``0"  from the MNIST dataset is selected. After BTC compression \cite{delp2003image}, the $28\times28$ pixel image is encoded into 1176 bits, followed by quantum state preparation and transmission using the experimental setup. 
The transmitted quantum-state data are all held-out states unseen by the model, with each state measured only once.
As shown in Fig.~\ref{Fig:Image_Transmission}(a), the original bit error rate (BER) is 50.34\%; after applying the TQSC model, the BER is reduced to 0\%. As depicted in Fig.~\ref{Fig:Image_Transmission}(c), the confusion matrices show bit errors in the noisy channel and accurate recovery by the TQSC model.
These results demonstrate the strong practicality and generalization capability of the model. 

Additional results on baseline comparisons, generalization validation, the explanation of “key” and “query”, and further details of the image transmission experiments are provided in the SM.

\section{Conclusion}
In summary, we have proposed and implemented a TQSC-assisted tomographic characterization scheme for mixed- and pure-state reconstruction under complex scattering, with additional dynamic Gaussian noise for pure states. Through this framework, our approach achieves an average estimator-target fidelity exceeding 99.999\% between the reconstructed and target states in the RSP experiments, while simultaneously providing interpretable data-driven insights into measurement correlations.
Although the training phase requires a dataset of paired $(M, \rho_{\text{target}})$ examples, during the inference stage, the scheme does not require additional quantum operations and performs classical tomographic denoising based on a single acquired measurement vector. 

Moreover, it demonstrates high single-acquisition efficiency, generalization across the tested conditions, and interpretability, providing a data-driven approach to quantum state estimation in complex noisy environments.
While our current implementation demonstrates remarkable performance in a controlled environment, we acknowledge that its scalability and resilience against more complex noise models warrant further investigation. Future work will focus on integrating this scheme as a post-processing and diagnostic tool in larger, multi-node quantum networks and exploring its compatibility with different quantum information protocols. These continued efforts are expected to further unlock the potential of our approach, paving the way for its broader application in intelligent quantum communication and quantum state characterization.
\section{ACKNOWLEDGMENTS}
This work is supported in part by the National Natural Science Foundation of China (Grants No. 62375164, No. 12574360 and No. 12192252), the Foundation for Shanghai Municipal Science Technology Major Project (Grant No. 2019SHZDZX01-ZX06), Quantum Science and Technology-National Science and Technology Major Project (Grant No. 2021ZD0300802), the Shuguang Program of Shanghai Education Development Foundation and Shanghai Municipal Education Commission (Grant No. 24SG53), and Guangdong Provincial Quantum Science Strategic Initiative (Grant No. GDZX2403003) and Shanghai Oriental Talent Plan Youth Project (Grant no. QNKJ2025013).

\bibliography{Reference.bib}

\end{document}




\title{Supplementary Material for: Noise-Robust Quantum State Characterization for Remote State Preparation with Deep Learning}


\author{Bo Tang}
\altaffiliation{These authors contributed equally to this work.}
\affiliation{State Key Laboratory of Photonics and Communications, School of Physics and Astronomy, Shanghai Jiao Tong University, Shanghai 200240, China}

\author{Zixuan Liao}
\altaffiliation{These authors contributed equally to this work.}
\affiliation{State Key Laboratory of Photonics and Communications, School of Physics and Astronomy, Shanghai Jiao Tong University, Shanghai 200240, China}

\author{Hao Li}
\altaffiliation{These authors contributed equally to this work.}
\affiliation{State Key Laboratory of Photonics and Communications, School of Physics and Astronomy, Shanghai Jiao Tong University, Shanghai 200240, China}

\author{Yilin Yang}
\affiliation{State Key Laboratory of Photonics and Communications, School of Physics and Astronomy, Shanghai Jiao Tong University, Shanghai 200240, China}
\author{Jiani Lei}
\affiliation{State Key Laboratory of Photonics and Communications, School of Physics and Astronomy, Shanghai Jiao Tong University, Shanghai 200240, China}
\author{Zengya Li}
\affiliation{State Key Laboratory of Photonics and Communications, School of Physics and Astronomy, Shanghai Jiao Tong University, Shanghai 200240, China}
\author{Jing Qiu}
\affiliation{State Key Laboratory of Photonics and Communications, School of Physics and Astronomy, Shanghai Jiao Tong University, Shanghai 200240, China}
\author{Zhaohui Dong}
\affiliation{State Key Laboratory of Photonics and Communications, School of Physics and Astronomy, Shanghai Jiao Tong University, Shanghai 200240, China}
\author{Zhengyang Mao}
\affiliation{State Key Laboratory of Photonics and Communications, School of Physics and Astronomy, Shanghai Jiao Tong University, Shanghai 200240, China}

\author{Yuanhua Li}
\email{lyhua1984@shiep.edu.cn}
\affiliation{Department of Physics, Shanghai Key Laboratory of Materials Protection and Advanced Materials in Electric Power, Shanghai University of Electric Power, Shanghai 200090, China}

\author{Yuanlin Zheng}
\email{ylzheng@sjtu.edu.cn}
\affiliation{State Key Laboratory of Photonics and Communications, School of Physics and Astronomy, Shanghai Jiao Tong University, Shanghai 200240, China}
\affiliation{Hefei National Laboratory, Hefei 230088, China}
\affiliation{Shanghai Research Center for Quantum Sciences, Shanghai 201315, China}

\author{Xianfeng Chen}
\email{xfchen@sjtu.edu.cn}
\affiliation{State Key Laboratory of Photonics and Communications, School of Physics and Astronomy, Shanghai Jiao Tong University, Shanghai 200240, China}
\affiliation{Hefei National Laboratory, Hefei 230088, China}
\affiliation{Shanghai Research Center for Quantum Sciences, Shanghai 201315, China}
\affiliation{Collaborative Innovation Center of Light Manipulations and Applications, Shandong Normal University, Jinan 250358, China}

\noaffiliation

\date{\today}

\maketitle

\setcounter{equation}{0}
\setcounter{figure}{0}
\setcounter{table}{0}
\setcounter{page}{1}
\renewcommand{\theequation}{S\arabic{equation}}
\renewcommand{\thefigure}{S\arabic{figure}}
\renewcommand{\thetable}{S\arabic{table}}
\clearpage
\section{Preparation of   Mixed State Theory Scheme}
\label{app:rsp}
Consider that the desired mixed state is \cite{wu201016DeterministicRemotePreparation}
\begin{equation}
\rho_B = p^2 |\varphi_B\rangle\langle\varphi_B| + q^2 |\varphi_B^\perp\rangle\langle\varphi_B^\perp|
\end{equation}
with
\begin{align}
|\varphi_B\rangle &= \alpha |H_B\rangle + \beta e^{i\phi} |V_B\rangle, \\
|\varphi_B^\perp\rangle &= \beta e^{-i\phi} |H_B\rangle - \alpha |V_B\rangle,
\end{align}
where $\alpha,\beta \in \mathbb{C}$ with $|\alpha|^2 + |\beta|^2 = 1$, and $p,q$ are probability amplitudes satisfying $p^2 + q^2 = 1$.

Without loss of generality, we assume that $p,q$ are real numbers, $p^2 + q^2 = 1$, and $\alpha$, $\beta$, $\phi$ are the same as before. To prepare arbitrary mixed states we need to achieve complete control over all five parameters. In Alice's setup, two polarization controllers, PC1 and PC2, are inserted in the two distinct optical paths to rotate the polarization states as
\begin{align}
|H\rangle &\to p|H\rangle + q|V\rangle, \\
|V\rangle &\to p|H\rangle + q|V\rangle.
\end{align}

To achieve independent polarization control in both paths, two additional polarization controllers, $\mathrm{PC1}'$ and $\mathrm{PC2}'$ , are employed. These are adjusted to implement the transformation
\begin{align}
|H\rangle &\to \alpha |H\rangle + \beta e^{i\phi}|V\rangle, \\
|V\rangle &\to \beta e^{-i\phi}|H\rangle - \alpha |V\rangle.
\end{align}

Then POVM measurement described by $M_1$ and $M_2$ are performed on the H-polarized and V-polarized components. 

Then the Alice's side HWP ($\sim$22.5$^\circ$) and the detectors perform the projection measurement , which projects Bob's photon onto one of the four mixed states:
\begin{subequations}
\begin{align}
\hat{\rho}_B^I &= p^2 |\varphi_B\rangle\langle\varphi_B| + q^2 |\varphi_B^\perp\rangle\langle\varphi_B^\perp|,  \\
\hat{\rho}_B^{X} &= p^2 (\hat{\sigma}_x |\varphi_B\rangle)(\langle\varphi_B|\hat{\sigma}_x) + q^2 (\hat{\sigma}_x |\varphi_B^\perp\rangle)(\langle\varphi_B^\perp|\hat{\sigma}_x), \\
\hat{\rho}_B^Y &=  p^2 (\hat{\sigma}_y |\varphi_B\rangle)(\langle\varphi_B|\hat{\sigma}_y) + q^2 (\hat{\sigma}_y |\varphi_B^\perp\rangle)(\langle\varphi_B^\perp|\hat{\sigma}_y),  \\
\hat{\rho}_B^Z &= p^2 (\hat{\sigma}_z |\varphi_B\rangle)(\langle\varphi_B|\hat{\sigma}_z) + q^2 (\hat{\sigma}_z |\varphi_B^\perp\rangle)(\langle\varphi_B^\perp|\hat{\sigma}_z). 
\end{align}
\end{subequations}
Bob can obtain the desired mixed state by applying local unitary operations $\hat{I}$, $\hat{\sigma}_x$, $\hat{\sigma}_y$  or $\hat{\sigma}_z$ according to Alice's measurement results. The required classical communication is two bits.

For simplicity, we do not use $\mathrm{PC1}'$ and $\mathrm{PC2}'$ to adjust the transformation. The final mixed states are as follows:
\begin{subequations}
\begin{align}
\hat{\rho}_B^I &= \alpha^2 |H\rangle\langle H| + \beta^2 |V\rangle\langle V|,  \\
\hat{\rho}_B^{X} &= \alpha^2 (\hat{\sigma}_x |H\rangle)(\langle H|\hat{\sigma}_x) + \beta^2 (\hat{\sigma}_x |V\rangle)(\langle V|\hat{\sigma}_x),  \\
\hat{\rho}_B^Y &=  \alpha^2 (\hat{\sigma}_y |H\rangle)(\langle H|\hat{\sigma}_y) + \beta^2 (\hat{\sigma}_y |V\rangle)(\langle V|\hat{\sigma}_y), \\
\hat{\rho}_B^Z &= \alpha^2 (\hat{\sigma}_z |H\rangle)(\langle H|\hat{\sigma}_z) + \beta^2 (\hat{\sigma}_z |V\rangle)(\langle V|\hat{\sigma}_z).
\end{align}
\end{subequations}

For pure states, the derivation follows analogously to the mixed-state case (or refer to Ref.\cite{liu2007experimental}), similarly leading to:
\begin{equation}
\begin{aligned}
|\phi\rangle &= \frac{1}{2} \{ |H_{2'}^A\rangle (\alpha|H^B\rangle + \beta e^{i\varphi}|V^B\rangle) \\
&\qquad - |V_{2'}^A\rangle [\hat{\sigma}_z(\alpha|H^B\rangle + \beta e^{i\varphi}|V^B\rangle)] 
\\
&\qquad+ |H_{1'}^A\rangle [\hat{\sigma}_x(\alpha|H^B\rangle + \beta e^{i\varphi}|V^B\rangle)]  \\
&\qquad - |V_{1'}^A\rangle [i\hat{\sigma}_y(\alpha|H^B\rangle + \beta e^{i\varphi}|V^B\rangle)] \},
\end{aligned}
\end{equation}

In a similar manner, Bob can obtain the desired pure state by applying local unitary operations $\hat{I}$, $\hat{\sigma}_x$, $\hat{\sigma}_y$, or $\hat{\sigma}_z$ according to Alice's measurement results. The required classical communication is two bits.
\begin{subequations}
\begin{align}
{|\psi\rangle}_B^I &=\frac{1}{2} (\alpha|H^B\rangle + \beta e^{i\varphi}|V^B\rangle),  
\\
{|\psi\rangle}_B^{X} &=\frac{1}{2} \hat{\sigma}_x(\alpha|H^B\rangle + \beta e^{i\varphi}|V^B\rangle),  \\
{|\psi\rangle}_B^Y &=\frac{1}{2}
i\hat{\sigma}_y(\alpha|H^B\rangle + \beta e^{i\varphi}|V^B\rangle), \\
{|\psi\rangle}_B^Z &=\frac{1}{2}
\hat{\sigma}_z(\alpha|H^B\rangle + \beta e^{i\varphi}|V^B\rangle).
\end{align}
\end{subequations}

\section{Quantum Noise Models}
\label{app:noise_model}

Quantum noise arises from the uncontrollable interaction between a quantum system and its environment, causing the system's evolution to deviate from ideal unitary dynamics. We employ the framework of quantum channels to mathematically describe such noise. A quantum channel is a completely positive (CP) and trace-non-increasing (TNI) map that transforms an input density matrix $\rho$ into an output state $\mathcal{E}(\rho)$.

According to Kraus' theorem \cite{kraus1983states}, any completely positive trace-preserving (CPTP) map can be expressed as
\begin{equation}
\mathcal{E}(\rho) = \sum_{k} E_k \rho E_k^\dagger,
\end{equation}
where $\{E_k\}$ are the Kraus operators, satisfying $\sum_k E_k^\dagger E_k = I$ (trace-preserving condition). For trace-non-increasing maps (lossy channels), $\sum_k E_k^\dagger E_k \leq I$.

When multiple noise channels $\mathcal{E}_1, \mathcal{E}_2, \dots, \mathcal{E}_N$ act sequentially, the overall effect is described by the composite channel 
\begin{equation}
\mathcal{E}_{\text{Total}} = \mathcal{E}_N \circ \mathcal{E}_{N-1} \circ \dots \circ \mathcal{E}_1,
\end{equation}
whose Kraus operators are given by chained products:
\begin{equation}
K_j = E_{N,k_N} E_{N-1,k_{N-1}} \cdots E_{1,k_1},
\end{equation}
where $j$ indexes all possible choices of $\{k_1,\dots,k_N\}$.

\subsection*{1. Representative Noise Models}

In our analysis of remote state preparation (RSP), we focus on four dominant sources of noise: photon loss, decoherence, multimode fiber (MMF) induced noise, and detector imperfections.

\subsubsection*{1.1. Amplitude Damping Channel (ADC)}

Photon loss during transmission (e.g., absorption or scattering in optical fibers or free-space channels) can be modeled by the amplitude damping channel.

For a single qubit with loss probability $p$, the Kraus operators are
\begin{equation}
E_0 = 
\begin{pmatrix}
1 & 0 \\
0 & \sqrt{1-p}
\end{pmatrix},
\quad
E_1 =
\begin{pmatrix}
0 & \sqrt{p} \\
0 & 0
\end{pmatrix}.
\end{equation}
Here $E_0$ corresponds to no photon loss, while $E_1$ describes decay from $\ket{1}$ to $\ket{0}$.

If Alice and Bob's photons experience independent losses with probabilities $p_A$ and $p_B$, the overall channel is
\begin{equation}
\mathcal{E}_{\text{Loss}}(\rho_{AB}) = 
\left( \mathcal{E}_{AD}^{(A)}(p_A) \otimes \mathcal{E}_{AD}^{(B)}(p_B) \right)(\rho_{AB}),
\end{equation}
with Kraus operators $E_{ij}^{(\text{Loss})} = E_i^{(A)} \otimes E_j^{(B)}$, $i,j \in \{0,1\}$.

\subsubsection*{1.2. Phase Damping Channel (PDC)}

Phase damping accounts for decoherence without energy exchange, e.g., random refractive index fluctuations or scattering-induced phase noise.

For a single qubit with decoherence probability $q$, the Kraus operators are
\begin{equation}
F_0 =
\begin{pmatrix}
1 & 0 \\
0 & \sqrt{1-q}
\end{pmatrix},
\quad
F_1 =
\begin{pmatrix}
0 & 0 \\
0 & \sqrt{q}
\end{pmatrix}.
\end{equation}

With decoherence probabilities $q_A$ and $q_B$ for Alice and Bob, respectively:
\begin{equation}
\mathcal{E}_{\text{Decoherence}}(\rho_{AB}) = 
\left( \mathcal{E}_{PD}^{(A)}(q_A) \otimes \mathcal{E}_{PD}^{(B)}(q_B) \right)(\rho_{AB}),
\end{equation}
with Kraus operators $F_{kl}^{(\text{Decoherence})} = F_k^{(A)} \otimes F_l^{(B)}$.

\subsubsection*{1.3. MMF Induced Noise}

When a single-mode input qubit is transmitted through a segment of MMF, the different spatial modes supported by the MMF experience varying propagation constants and mode-mixing. This can lead to mode dispersion, differential modal attenuation, and effectively, a loss of coherence or entanglement if the output modes are not perfectly demultiplexed or if the measurement setup cannot distinguish them. For a qubit encoded in polarization or time bins, the interaction with multiple spatial modes can scramble the encoded information, acting as a depolarizing-like channel or a more complex unitary scrambling. A simplified model treats the MMF as inducing random unitary transformations or causing effective dephasing and amplitude damping through mode coupling and differential loss, ultimately leading to a mixed state for the output single mode.

We model the MMF noise on Bob's qubit as a generalized depolarizing channel or a channel that effectively mixes the input state due to coupling to higher-order modes that are subsequently lost or unmeasured. For simplicity, we model the MMF as a depolarizing channel on Bob's qubit.

The Kraus operators for a depolarizing channel on Bob's qubit are:
\begin{align}
H_0 &= \sqrt{1 - \tfrac{3d_{\text{MMF}}}{4}} I^{(B)}, \\
H_1 &= \sqrt{\tfrac{d_{\text{MMF}}}{4}} X^{(B)}, \\
H_2 &= \sqrt{\tfrac{d_{\text{MMF}}}{4}} Y^{(B)}, \\
H_3 &= \sqrt{\tfrac{d_{\text{MMF}}}{4}} Z^{(B)},
\end{align}
where $d_{\text{MMF}}$ is the depolarization probability due to the MMF, and $\{I,X,Y,Z\}$ are Pauli matrices.

Since the MMF noise is specific to Bob's transmission path, the overall channel for the two-qubit system is:
\begin{equation}
\mathcal{E}_{\text{MMF}}(\rho_{AB}) = \left( I^{(A)} \otimes \mathcal{E}_{\text{Depolarizing}}^{(B)}(d_{\text{MMF}}) \right)(\rho_{AB}),
\end{equation}
with Kraus operators $H_m^{(\text{MMF})} = I^{(A)} \otimes H_m^{(B)}$, for $m \in \{0,1,2,3\}$.

\subsubsection*{1.4. Detector Imperfections}
Detector efficiency ($q$) can be absorbed into the loss model, with effective loss $p=1-q$.

Spurious detector clicks (dark counts) introduce random errors in measurement outcomes. We model this effect as an effective depolarizing channel acting on Bob's qubit:
\begin{align}
G_0 &= \sqrt{1 - \tfrac{3d_{\text{det}}}{4}} I, \\
G_1 &= \sqrt{\tfrac{d_{\text{det}}}{4}} X, \\
G_2 &= \sqrt{\tfrac{d_{\text{det}}}{4}} Y, \\
G_3 &= \sqrt{\tfrac{d_{\text{det}}}{4}} Z,
\end{align}
where $d_{\text{det}}$ is the depolarization probability representing the impact of dark counts and other detector-related errors, and $\{I,X,Y,Z\}$ are Pauli matrices.

For the two-qubit system, we assume detector imperfections manifest primarily on Bob's side:
\begin{equation}
\mathcal{E}_{\text{Detector}}(\rho_{AB}) = \left( I^{(A)} \otimes \mathcal{E}_{\text{Depolarizing}}^{(B)}(d_{\text{det}}) \right)(\rho_{AB}),
\end{equation}
with Kraus operators $G_m^{(\text{Detector})} = I^{(A)} \otimes G_m^{(B)}$.

\subsection*{2. Physical Mechanism of Fidelity Degradation in Multi-Mode Fibers and Noise Modeling Analysis}

This section provides a physical interpretation of the fidelity degradation observed when introducing Multi-Mode Fiber (MMF) into the quantum channel and elucidates the role of the noise model employed in this study within this context.

\subsubsection{Physical Principles of MMF-Induced Degradation}

Multi-mode fibers support the simultaneous propagation of multiple spatial modes when the core radius or the refractive index contrast between the core and cladding is sufficiently large. The approximate number of supported modes, $D$, is given by:
\begin{equation}
D \approx \left(\frac{2\pi}{\lambda}\right)^2 \int_{0}^{R} \left[n^2(r) - n_c^2\right]^{1/2} r \, dr,
\end{equation}
where $n(r)$ represents the radial refractive index profile of the fiber core, $n_c$ is the cladding refractive index, $R$ is the core radius, and $\lambda$ is the optical wavelength \cite{gloge1975multimode}. The number of propagating modes is roughly proportional to the refractive index contrast and $(R/\lambda)^2$.

The introduction of MMF into a quantum transmission channel induces several physical effects that degrade the fidelity of the transmitted quantum states, including mode coupling, polarization mode dispersion (PMD), random phase fluctuations, and mode-dependent loss (MDL).

Specifically, unintended mode coupling can arise from manufacturing imperfections (e.g., non-circular core geometry, rough core-cladding interfaces, or refractive index variations) and external mechanical perturbations (e.g., stress, micro-bending, or twisting). In our experiments, controlled twisting was applied to the MMF to induce perturbations, thereby enhancing inter-mode coupling. Such perturbations facilitate energy exchange between different spatial modes and lead to the random evolution of the transmitted optical field. Consequently, the initially prepared pure quantum states undergo decoherence during propagation, resulting in a reduction in measured fidelity.

\subsubsection{Theoretical Modeling Based on Field Coupling}

The inter-mode coupling effects described above can be quantitatively characterized using the coupled-mode theory. In this framework, the total optical field propagating along the longitudinal $z$-direction is expanded as a superposition of orthogonal eigenmodes. The evolution of the modal amplitudes is governed by the following coupled differential equation:
\begin{equation}
\frac{dA_\mu}{dz} = -j\beta_\mu A_\mu + \sum_{\nu \neq \mu} C_{\mu\nu}(z) A_\mu,
\end{equation}
where $A_\mu$ denotes the complex amplitude of the $\mu$-th mode, $\beta_\mu$ is its propagation constant, and $C_{\mu\nu}(z)$ represents the coupling coefficient induced by structural perturbations in the fiber \cite{marcuse2013theory}. The first term describes the independent phase evolution in the absence of coupling, while the second term accounts for energy transfer between modes. These coupling-induced effects ultimately manifest as crosstalk and decoherence in the transmitted quantum states.

\subsubsection{Discussion on Noise Suppression and Model Validity}

Experimentally, in the absence of MMF (i.e., a direct quantum channel), the average fidelity of the prepared states reaches $99.25\% \pm 0.22\%$. As shown in the main text, the introduction of MMF causes a significant drop in fidelity when no noise suppression model is applied. This degradation reflects the raw impact of multi-mode perturbations prior to correction.

The optimized TQSC (Transformer-based Quantum State Characterizer) noise model used in this work is trained on measurement data from the quantum channel incorporating the MMF. Consequently, the TQSC model inherently accounts for the effective noise introduced by the MMF, including contributions from mode coupling, phase perturbations, and mode-dependent loss. Combined with our Transformer-based post-processing reconstruction, the model recovers the experimental density matrices with high fidelity, aligning them closely with the target states.

While the current framework achieves high reconstruction accuracy by treating the channel as a complex noise process, future work could benefit from incorporating more explicit physical constraints. Specifically, independently modeling the distinct noise contributions-such as the statistical characterization of coupling coefficients $C_{\mu\nu}(z)$, phase noise spectra, and polarization effects-and integrating them into the noise prior or loss function could further refine the post-processing inversion. 

\subsection*{3. Composite Noise Model}

In RSP, the shared entangled state first undergoes transmission, which includes loss, decoherence, and MMF noise on Bob's side, followed by Alice's measurement and Bob's conditional unitary. We model the total noise channel as a sequential application of the individual noise channels.

The composite Kraus operator for the total noise will be a product of the individual Kraus operators, applied in the correct sequence. The entangled state is distributed. Bob's qubit then goes through the MMF, and finally, both Alice and Bob perform detection.

The Kraus operators for Alice's total noise channel are $K_A^{(k,i)} = F_k^{(A)} E_i^{(A)}$.
The Kraus operators for Bob's total noise channel are $K_B^{(m,n,l,j)} = G_m^{(B)} H_n^{(B)} F_l^{(B)} E_j^{(B)}$.

Then, the noisy state is given by:
\begin{equation}
\begin{split}
\rho_{AB}^{\text{noisy}} = \sum_{\substack{k,i \\ m,n,l,j}} 
& \left( K_A^{(k,i)} \otimes K_B^{(m,n,l,j)} \right) \rho_{AB}^{\text{in}} \\
& \left( K_A^{(k,i)\dagger} \otimes K_B^{(m,n,l,j)\dagger} \right).
\end{split}
\end{equation}

Here, $\rho_{AB}^{\text{in}} = \ket{\Phi^+}\bra{\Phi^+}$ is the initial Bell state.

\subsection*{4. Remote State Preparation under Noise}

Alice performs a Bell-state measurement (BSM) on her subsystem. For outcome $k'$, with projector $\Pi_{k'}^{(A)}$, the probability is
\begin{equation}
P(k') = \mathrm{Tr}\left[(\Pi_{k'}^{(A)} \otimes I^{(B)}) \rho_{AB}^{\text{noisy}} \right],
\end{equation}
and the post-measurement state of Bob's qubit is
\begin{equation}
\rho_B^{(k')} = \frac{\mathrm{Tr}_A\left[(\Pi_{k'}^{(A)} \otimes I^{(B)}) \rho_{AB}^{\text{noisy}} (\Pi_{k'}^{(A)} \otimes I^{(B)})\right]}{P(k')}.
\end{equation}

Depending on Alice's reported result $k'$, Bob applies a correction unitary $U_{k'} \in \{I,X,Y,Z\}$, leading to the final ensemble state:
\begin{equation}
\rho_B^{\text{final}} = \sum_{k'} P(k')\, U_{k'} \rho_B^{(k')} U_{k'}^\dagger.
\end{equation}

\subsection*{5. Performance Metric: Fidelity}

The performance of RSP is quantified by the fidelity between Bob's final state and the target state $\ket{\psi}$:
\begin{equation}
F = \braket{\psi|\rho_B^{\text{final}}|\psi}.
\end{equation}
This allows us to systematically evaluate the impact of loss, decoherence, MMF induced noise, and detector imperfections on RSP.

\section{Theoretical Framework of the TQSC Model}
\label{app:TQSC}

\subsection{Problem Formulation: Learning an Inverse Quantum-to-Classical Map}

The experimental realization of RSP is inevitably influenced by a multitude of physical imperfections. The entire process can be conceptualized as a complex quantum-to-classical channel, denoted by $\mathcal{C}$, which maps an ideal target quantum state $\rho_{\text{target}}$ to a set of classical measurement outcomes. In our work, each data record of these outcomes is tokenized into a 13-element feature vector $\mathbf{M} \in \mathbb{R}^{13}$. This vector acts as a comprehensive, noise-corrupted classical ``snapshot" of the quantum state:
\begin{equation}
\mathbf{M} = \mathcal{C}(\rho_{\text{target}}).
\end{equation}
To intuitively understand this formulation, $\mathbf{M}$ aggregates the multidimensional profile of the state observed from 4 distinct perspectives. Specifically, it consists of direct physical observables: the coincidence counts between Alice and Bob, the corresponding single-photon counts for each party, all measured across four different measurement bases, alongside the total coincidence measurement time. This specific design ensures that the vector is informationally complete, encapsulating the full spectrum of data required for quantum state tomography. Through this numerical encoding, we map raw physical variables into a structured feature space, enabling our TQSC Transformer model to process quantum data with the same architectural logic used for word embeddings in natural language processing.

The map $\mathcal{C}$ encapsulates both coherent errors, such as misalignments in optical components, and incoherent noise processes, including channel decoherence, detector dark counts, and Poissonian shot noise.

Conventional quantum state tomography (QST) techniques attempt to reconstruct the state by inverting this map from the data $\mathbf{M}$. However, they often falter when faced with complex, non-Gaussian, or correlated noise structures inherent in real-world systems.
To overcome this limitation, we propose a data-driven TQSC model.
The TQSC is a deep neural network, parameterized by a set of weights $\theta$, that learns an effective inverse map $\Phi_{\text{TQSC}}(\cdot; \theta)$. This learned function directly transforms the noisy classical measurement data $\mathbf{M}$ into a high-fidelity estimate of the density matrix, $\rho_{\text{pred}}$:
\begin{equation}
    \rho_{\text{pred}} = \Phi_{\text{TQSC}}(\mathbf{M}; \theta).
    \label{eq:inverse_map}
\end{equation}
The core objective is to train the TQSC such that the learned map $\Phi_{\text{TQSC}}$ not only inverts the deterministic dynamics of the channel but also actively suppresses the stochastic noise components.

\subsection{Model Architecture and Mathematical Formalism}

The architecture of the TQSC is meticulously designed to leverage the powerful sequence-processing capabilities of the Transformer model \cite{vaswani2017attention}, adapting it to the structured nature of our physics-based input vector.

\paragraph{Input Embedding with Encoding}
The input vector $\mathbf{M} \in \mathbb{R}^{13}$ is interpreted as a sequence of $L=13$ tokens, $S = \{m_1, m_2, \dots, m_L\}$. To provide positional information to the Transformer, we apply a Sinusoidal Positional Encoding $\mathbf{P}_{\text{pos}} \in \mathbb{R}^{L \times d_{\text{model}}}$. This encoding is generated by fixed sine and cosine functions of different frequencies, where each position $pos$ in the sequence is mapped to a unique vector. The initial representation $\mathbf{H}^{(0)} \in \mathbb{R}^{L \times d_{\text{model}}}$ is formed as:
\begin{equation}
    \mathbf{H}^{(0)} = \text{Linear}(S) + \mathbf{P}_{\text{pos}}.
\end{equation}
This injects absolute positional information but does not incorporate explicit knowledge of physical feature types.

\paragraph{Multi-Head Self-Attention Core}
The heart of the TQSC is a stack of $N=3$ identical encoder layers. The central mechanism in each layer is Multi-Head Self-Attention (MHSA) with $h=4$ heads, which allows the model to weigh the importance of all input tokens relative to each other. Each head independently learns different subspace features of the input, such as photon counting patterns, noise correlations, and temporal correlations in quantum signals. For an input representation $\mathbf{X} \triangleq \mathbf{H}^{(l-1)} \in \mathbb{R}^{n \times d}$ at layer $l$ (where $n$ is the sequence length, $d$ the feature dimension), the MHSA output is computed as:
\begin{equation}
    \text{MHSA}(\mathbf{X}) = \text{Concat}(\text{head}_1, \dots, \text{head}_h)\mathbf{W}^O,
\end{equation}
where $\mathbf{W}^O \in \mathbb{R}^{hd_v \times d}$ is the output projection matrix, and each head is defined by:
\begin{equation}
    \text{head}_i = \text{Attention}(\mathbf{Q}_i, \mathbf{K}_i, \mathbf{V}_i).
\end{equation}
Here the query, key, and value matrices for head $i$ are computed as:
\begin{align}
    \mathbf{Q}_i &= \mathbf{X}\mathbf{W}_i^Q, \quad \mathbf{W}_i^Q \in \mathbb{R}^{d \times d_k} \\
    \mathbf{K}_i &= \mathbf{X}\mathbf{W}_i^K, \quad \mathbf{W}_i^K \in \mathbb{R}^{d \times d_k} \\
    \mathbf{V}_i &= \mathbf{X}\mathbf{W}_i^V, \quad \mathbf{W}_i^V \in \mathbb{R}^{d \times d_v}
\end{align}
with $d_k = d_v = d/h$. The attention function is then given by:
\begin{equation}
    \text{Attention}(\mathbf{Q}_i, \mathbf{K}_i, \mathbf{V}_i) = \text{softmax}\left(\frac{\mathbf{Q}_i\mathbf{K}_i^\top}{\sqrt{d_k}}\right)\mathbf{V}_i.
\end{equation}

This attention mechanism enables the model to identify and exploit complex, non-linear correlations within the measurement data---for instance, the relationship between single-photon counts and coincidence events, which is a key indicator of channel loss and noise. The output of the attention block is then processed through a position-wise Feed-Forward Network (FFN), with residual connections and layer normalization applied after each sub-layer to ensure stable training.

\paragraph{State Vector Prediction Head}
Following the final encoder layer, we obtain the output representation $\mathbf{H}^{(N)} \in \mathbb{R}^{L \times d_{\text{model}}}$. To aggregate the information from the entire sequence into a single vector, we utilize the fixed positional encoding scheme inherent in the Transformer architecture. Specifically, the sinusoidal positional encoding $\mathbf{P}_{\text{pos}}$ provides absolute position information for each element in the sequence. The aggregated representation $\mathbf{h}_{\text{agg}} \in \mathbb{R}^{d_{\text{model}}}$ is computed as:
\begin{equation}
    \mathbf{h}_{\text{agg}} = \frac{1}{L} \sum_{i=1}^{L} \mathbf{h}_i^{(N)}
\end{equation}
where $L$ is the sequence length. This mean-pooling strategy leverages the position-aware representations created by the positional encoding module.
\subsection{A brief non-technical explanation of Key and Query.}
Keys and queries are fundamental components of the attention mechanism widely used in modern neural networks. Intuitively, a query represents the question or probe used to search for relevant information, while a key is a descriptor associated with each candidate item. The value stores the actual information to be retrieved. In simple terms, the query asks ``what am I looking for?", the keys label the available items, and the values contain the corresponding content. Items whose keys better match the query receive higher weights and thus contribute more strongly to the output representation.

A useful analogy is searching for a book in a library. The query corresponds to what one is searching for, for example, ``an introductory book on machine learning.'' Each book in the library is associated with a key that summarizes its main characteristics, such as subject, difficulty level, or keywords. The value represents the actual content of the book. The attention mechanism compares the query with the keys of all books and assigns higher weights to those whose keys better match the query. As a result, books that are more relevant to the search request contribute more to the final output, while irrelevant books receive negligible weights.

From this perspective, the attention mechanism can be viewed as a soft, learnable retrieval process, rather than a hard selection of a single item.

\subsection{Physically-Constrained Output Layer}

A critical challenge is ensuring that the neural-network output corresponds to a physically valid density matrix. A density matrix $\rho$ must satisfy three
constraints: (i) Hermiticity, $\rho=\rho^\dagger$, (ii) unit trace,
$\operatorname{Tr}(\rho)=1$, and (iii) positive semidefiniteness,
$\rho\geq 0$.

To enforce these constraints by construction, we adopt a Cholesky-style
parametrization, as commonly used in maximum-likelihood quantum-state
tomography. Instead of directly predicting the entries of $\rho$, the network
outputs an unconstrained real vector $\bm{r}\in\mathbb{R}^{d^2}$, which is used
to construct a lower-triangular complex matrix $C$. The output density matrix is then reconstructed as
\begin{equation}
    \rho_{\mathrm{out}}
    =
    \frac{C C^\dagger}{\operatorname{Tr}(C C^\dagger)} .
\end{equation}

For the single-qubit case considered in this work, $d=2$ and the network outputs four real parameters $\bm{r}=(r_0,r_1,r_2,r_3)$. We define
\begin{equation}
    \begin{split}
        C &=
        \begin{pmatrix}
            s_0 & 0 \\
            r_2 + \mathrm{i} r_3 & s_1
        \end{pmatrix}, \\[1.5ex] 
        s_j &= \operatorname{softplus}(r_j)+\epsilon,\quad j=0,1,
    \end{split}
\end{equation}
where $\epsilon>0$ is a small constant used for numerical stability. This choice ensures that $\operatorname{Tr}(C C^\dagger)>0$ for every finite network output.

This parametrization guarantees Hermiticity because
\begin{equation}
    (C C^\dagger)^\dagger = C C^\dagger .
\end{equation}
It also guarantees unit trace by construction:
\begin{equation}
    \operatorname{Tr}(\rho_{\mathrm{out}})
    =
    \frac{\operatorname{Tr}(C C^\dagger)}
    {\operatorname{Tr}(C C^\dagger)}
    =
    1 .
\end{equation}
Finally, for any complex vector $v$, we have
\begin{equation}
    v^\dagger \rho_{\mathrm{out}} v
    =
    \frac{v^\dagger C C^\dagger v}
    {\operatorname{Tr}(C C^\dagger)}
    =
    \frac{\|C^\dagger v\|^2}
    {\operatorname{Tr}(C C^\dagger)}
    \geq 0 .
\end{equation}
Therefore, $\rho_{\mathrm{out}}$ is positive semidefinite for any real-valued
network output $\bm{r}$.

For $d=2$, the reconstructed density matrix can be written explicitly as
\begin{equation}
    \begin{split}
        \rho_{\mathrm{out}} &= \frac{1}{s_0^2+s_1^2+r_2^2+r_3^2} \\
        &\quad \cdot \begin{pmatrix}
            s_0^2 & s_0(r_2 - \mathrm{i} r_3) \\[1.5ex]
            s_0(r_2 + \mathrm{i} r_3) & s_1^2+r_2^2+r_3^2
        \end{pmatrix}.
    \end{split}
\end{equation}
Thus, the physically constrained output layer maps every unconstrained
real-valued output of the Transformer backbone to a bona fide density matrix,
while preserving end-to-end differentiability.

\subsection{Optimization and Theoretical Merit}
The network parameters $\theta$ are optimized by minimizing the Mean Squared Error (MSE) loss function, defined as the squared Frobenius norm between the predicted state $\rho_{\text{pred}}$ and the known ideal target state $\rho_{\text{target}}$. Over a training dataset of $K$ samples $\{(\mathbf{M}_k, \rho_{\text{target},k})\}_{k=1}^K$, the loss is:
\begin{equation}
    \mathcal{L}(\theta) = \frac{1}{K} \sum_{k=1}^K \left\| \Phi_{\text{TQSC}}(\mathbf{M}_k; \theta) - \rho_{\text{target},k} \right\|_{\text{F}}^2.
    \label{eq:mse-loss}
\end{equation}
Optimization is performed using the Adam optimizer, which implements a variant of stochastic gradient descent.

From a theoretical standpoint, the TQSC framework offers a significant advantage. The Transformer architecture is a universal function approximator, guaranteeing its capacity to model the highly complex inverse map $\Phi_{\text{TQSC}} \approx \mathcal{C}^{-1}$. More pointedly, the self-attention mechanism functions as a learned, context-aware adaptive filter. It can dynamically identify and down-weight measurement values that are inconsistent with the context provided by the rest of the data, effectively learning to suppress noise-induced anomalies. Therefore, the minimization of the MSE loss in Eq.~\eqref{eq:mse-loss} does not merely fit a curve; it drives the model to learn the underlying statistical structure of the noise and systematically correct for it. This establishes a robust framework where deep learning directly enhances quantum state fidelity by learning noise-resilient inversion of physical measurements.

The TQSC framework derives its theoretical strength from two fundamental properties of the Transformer architecture. First, as a universal function approximator, the Transformer possesses the inherent capacity to model arbitrarily complex inverse mappings with any desired precision. This mathematical guarantee ensures that TQSC can in principle learn the exact inverse transformation from measurement data to quantum states.
Second, the self-attention mechanism provides a sophisticated context-aware processing capability. By dynamically evaluating the consistency of each measurement value against the full context of the dataset, it learns to identify and suppress noise-induced anomalies while preserving genuine quantum signatures. This adaptive filtering operates as a learned denoising function that transcends traditional signal processing approaches.
Consequently, minimizing the mean squared error loss accomplishes more than simple data fitting. It drives the model to internalize the statistical patterns of measurement noise and implement physically meaningful corrections. This creates a principled framework where deep learning directly enhances the accuracy of quantum state reconstruction through intelligent noise-resilient inversion of experimental data.

\section{Experimental Setup}
\label{app:setup_details}

The core of the system is a Sagnac interferometric loop housing a spontaneous parametric down-conversion (SPDC) source, designed for generating polarization-entangled photon pairs with inherent phase stability. A femtosecond laser operating at a repetition rate of 60 MHz undergoes second-harmonic generation to produce 775 nm pump light. This pump beam is injected into the polarization-entangled Sagnac loop, where SPDC occurs in a periodically poled lithium niobate (PPLN) crystal, generating degenerate photon pairs at 1550 nm wavelength. The half-wave plate (HWP) and quarter-wave plate (QWP) within the Sagnac loop were initially aligned with their fast axes at 0\textdegree\ relative to the horizontal polarization axis to ensure maximal generation of the target Bell state.

Following generation, wavelength-division multiplexing (WDM) filters select photons within the approximately 1550 nm band. These photons are then routed through specific dense wavelength-division multiplexing (DWDM) channels: channel 31 (1552.54 nm) directs photons to Alice, and channel 33 (1550.92 nm) directs photons to Bob.

In Alice's detection module, the incoming photon is first split by a 9:1 beam splitter (BS1). One output arm incorporates a tunable optical delay line for precise temporal synchronization before recombination at the second beam splitter (BS2). The alternate output path from BS1 contains a variable optical attenuator used to adjust the intensity ratio `$\eta=\alpha^{2}/\beta^{2}$' between the two measurement bases. Both arms subsequently pass through fiber-based polarizing beam splitters (FPBS) for projection onto the horizontal (H) or vertical (V) polarization basis. For the critical state analysis step, a hybrid detection scheme is implemented by combining the H-polarized output from the upper FPBS and the V-polarized output from the lower FPBS through BS2. Polarization controllers (PC) in both arms compensate for polarization rotation induced by the optical fibers. Following BS2, projective measurements are performed using a 22.5\textdegree~half-wave plate (HWP) and polarizing beam splitters (PBS) before coupling into fiber-coupled single-mode collection paths.

Bob's receiver module employs a polarization controller (PC) for initial compensation of fiber-induced birefringence. A commercially available OM2 graded-index MMF (core diameter \qty{50}{\micro\metre}, cladding diameter \qty{125}{\micro\metre}, length 1.5 m) is used in the experiment. To simulate the spatial mode scrambling encountered in complex scattering environments, the photon is transmitted through this fiber section. Full quantum state tomography is implemented using a polarization analyzer consisting of a quarter-wave plate (QWP), a half-wave plate (HWP), and a polarizing beam splitter (PBS) in sequence, with the outputs coupled via single-mode fibers to detectors.

Both Alice and Bob employ superconducting nanowire single-photon detectors (SNSPDs) for photon detection, characterized by a quantum efficiency of approximately 80\% at the operating wavelength of 1550 nm. Detection events are recorded and time-tagged using time-correlated single-photon counting (TCSPC) electronics offering a timing resolution of 81 ps. The coincidence window was set to 500 ps, optimized relative to the 16.7 ns repetition period of the femtosecond laser source.

\section{Characterization of the SPDC Source}
\label{sec:spdc_characterization}
To systematically quantify the performance of the Spontaneous Parametric Down-Conversion (SPDC) source, we provide a comprehensive experimental characterization covering its efficiency and other key experimental parameters. The TQSC model, trained on experimental data, implicitly captures SPDC failure events such as vacuum and multi-pair contributions, learning to suppress noise and restore fidelity without requiring an analytic noise model. The characterization results and analysis are organized as follows:

Firstly, the spectral properties and normalized conversion efficiency of the PPLN waveguide are investigated to clarify the phase-matching bandwidth and the generation efficiency of the source. Secondly, the signal-to-noise performance is evaluated by measuring the coincidence-to-accidental ratio (CAR) as a function of pump power. Thirdly, the single-photon purity and suppression of multi-pair emissions are verified using the second-order correlation function $g^{(2)}(0)$. Fourthly, the high quality of the generated states is demonstrated through polarization entanglement characterization. Finally, regarding SPDC failure events, the TQSC model implicitly learns to suppress the resulting noise and restore fidelity, without the need for an explicit analytic model of the imperfections.

Detailed discussions and measurement results for each of these parameters are presented below.
\subsection{Spectral Properties and Normalized Conversion Efficiency}
\begin{figure*}[hbtp!]
    \centering
    \includegraphics[width=0.8\textwidth]{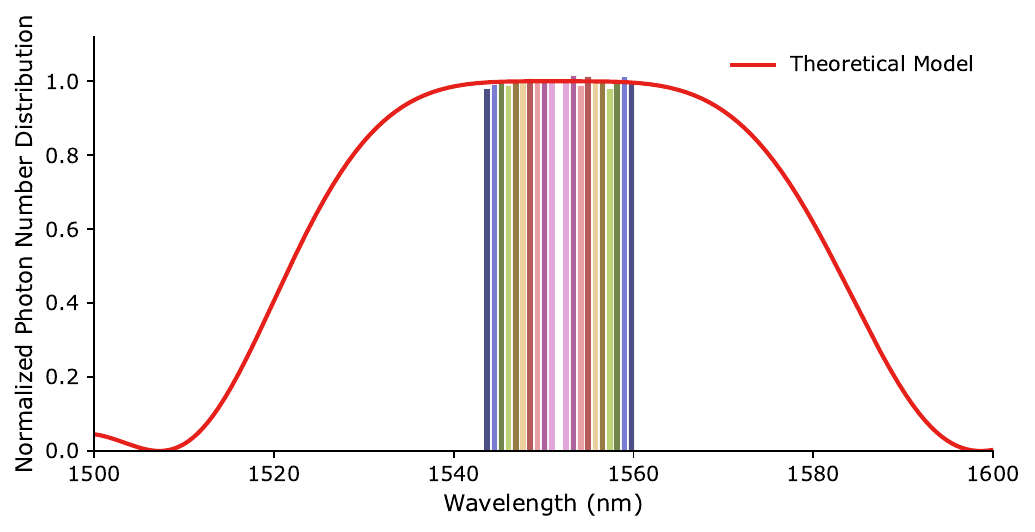} 
    \caption{Characterization of SPDC spectral properties. The red solid line represents the theoretical normalized conversion efficiency (phase-matching curve). The colored bars indicate the measured normalized photon counts for 10 symmetric channel pairs centered at CH32 (from CH22 \& CH42 to CH31 \& CH33). The measured Full-Width at Half-Maximum (FWHM) bandwidth is 60.74 nm.}
    \label{fig:S1}
\end{figure*}

To characterize the phase-matching bandwidth and generation efficiency of the PPLN waveguide, we investigated the spectral distribution of the SPDC photons and compared it with the theoretical model. Fig.~\ref{fig:S1} presents the measured normalized photon counts for 10 symmetric signal-idler channel pairs centered at ITU Channel 32 (CH32), ranging from the outermost pair (CH22 \& CH42) to the innermost pair (CH31 \& CH33). In the experiment, the pump wavelength was tuned to generate degenerate photon pairs centered at CH32.

The experimental data (colored bars) show excellent agreement with the theoretical phase-matching curve (red solid line), confirming the reliability of the waveguide design. Based on the measured spectral distribution, we calculated a Full-Width at Half-Maximum (FWHM) bandwidth of 60.74 nm. This broad phase-matching bandwidth ensures a relatively flat and high normalized conversion efficiency across a wide range of channels. This characteristic verifies the uniformity of photon generation and confirms that the source is well-suited for multi-channel quantum communication protocols requiring high spectral purity and broad wavelength tunability.

\subsection{Source Noise Performance and Coincidence-to-Accidental Ratio (CAR)}
\label{subsec:noise_car}

To evaluate the signal-to-noise ratio performance of the source, we measured the Coincidence-to-Accidental Ratio (CAR) as a function of pump power for the 10 symmetric channel pairs centered at CH32 (spanning from CH22/CH42 to CH31/CH33), as summarized in Fig.~\ref{fig:S2}.

As shown, the CAR values for all channel pairs exhibit a consistent dependence on pump power, strictly following the theoretical inverse relationship $\text{CAR} \propto 1/P_{\text{pump}}$. This behavior indicates that accidental coincidence counts are dominated by multi-pair emissions rather than background noise or dark counts, confirming the low-noise nature of the experimental setup.

Quantitatively, at low pump powers (in the $\mu$W regime), the maximum CAR values for all measured channel pairs exceed 20,000, with the central channel pair (CH30 and CH34) reaching a peak of 30,446. Even at higher pump powers, where the statistical significance of multi-photon events increases, the fitted curves maintain excellent agreement with the experimental data across the entire spectral range. These results demonstrate that the device maintains high entanglement purity and a high signal-to-noise ratio over a broad bandwidth, validating the robust performance of the source for multi-channel quantum communication applications.

\subsection{Single-Photon Purity and Second-Order Correlation Function \texorpdfstring{$g^{(2)}(0)$}{g2(0)}}
\label{subsec:g2_purity}

\begin{figure*}[hbt!]
    \centering
    \includegraphics[width=0.9\textwidth]{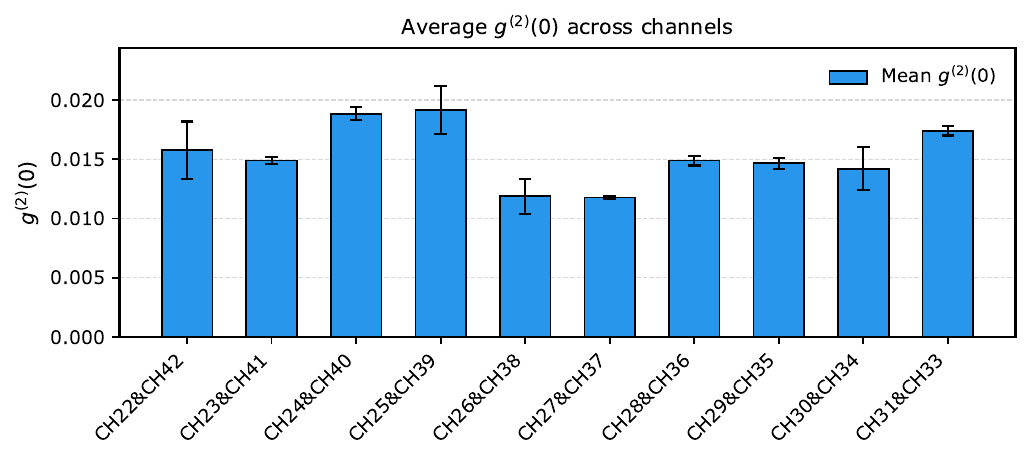} 
    \caption{Measurement of the second-order correlation function $g^{(2)}(0)$ for 10 symmetric channel pairs. The bar chart shows the average $g^{(2)}(0)$ values from the outer pairs (CH22 \& CH42) to the inner pairs (CH31 \& CH33). Error bars represent the standard deviation calculated from three independent measurements.}
    \label{fig:S3}
\end{figure*}

To verify the single-photon purity and the suppression of multi-pair emissions, we measured the zero-delay second-order correlation function, $g^{(2)}(0)$. The results for the 10 symmetric channel pairs (from CH22/CH42 to CH31/CH33) are summarized in Fig.~\ref{fig:S3}. To ensure statistical reliability, each value represents the mean of three independent repeated measurements, with error bars indicating the standard deviation.

As shown in Fig.~\ref{fig:S3}, all channel pairs exhibit consistently low $g^{(2)}(0)$ values, ranging from approximately $0.012$ to $0.019$. These values are well below the classical limit ($g^{(2)}(0) \ge 1$) and significantly lower than the typical benchmark for high-quality single-photon sources ($g^{(2)}(0) < 0.1$), clearly demonstrating strong non-classical statistical properties and effective suppression of multi-channel noise. Furthermore, the small standard deviations and the narrow distribution of means across different channel pairs confirm the uniformity and stability of the multi-channel system, indicating that high-quality quantum properties are achieved simultaneously across the entire spectral range without significant channel-dependent systematic errors.

\subsection{Polarization Entanglement Characterization}
\label{subsec:entanglement}

To demonstrate the high quality of the generated states, we performed polarization entanglement characterization across all 10 symmetric channel pairs. Fig.~\ref{fig:S4} presents the polarization-dependent fringe measurements. To ensure experimental rigor, entanglement fringes were measured in two non-orthogonal polarization bases ($0^\circ$ and $22.5^\circ$). Each data point represents the average of three independent trials.

The entanglement fringes exhibit robust and consistent sinusoidal behavior across all channels, in excellent agreement with theoretical predictions. Quantitatively, we achieved an average visibility of $98.74\% \pm 0.69\%$ across all channels and settings. Specifically, the average visibility was $98.91\% \pm 0.37\%$ in the $0^\circ$ basis and $98.56\% \pm 0.89\%$ in the $22.5^\circ$ basis. The close agreement between these two bases, combined with low statistical uncertainty, confirms that the SPDC source maintains high-fidelity entanglement and uniform performance across all channel pairs with no observable systematic bias.

\subsection{Accounting for SPDC Failure Events in the Noise Model}
\label{subsec:noise_modeling_concept}

In a coincidence-based post-selection scheme, coincidence measurements can filter out the most fundamental photon losses and the vast majority of $|0,0\rangle$ vacuum noise. However, the SPDC source still faces inherent failure issues. Specifically, high-order multi-photon emissions intrinsic to the SPDC Hamiltonian, such as $|2,2\rangle$ states, can masquerade as valid single-photon pairs when coupled with detectors lacking photon-number resolution and partial channel losses. Furthermore, detector dark counts within the coincidence window also lead to accidental coincidences. Mathematically, these phenomena manifest as a finite CAR, and a non-zero $g^{(2)}(0)$, which is consistent with the characterization results discussed earlier.

Instead of relying on idealized theoretical assumptions, our model learns from the actual photon statistics where these imperfections are naturally present. In the resulting coincidence data, these physical failures manifest as specific noise patterns. For instance, reduced total counts from a finite heralding efficiency introduce Poissonian shot noise, while accidental coincidences mix the pure state with classical white noise to create depolarizing noise. The source imperfections and intrinsic noise effects of the SPDC process are naturally reflected in the measurement data. By using these noisy measurements as the input to train the TQSC model, we ensure that the model's predictions faithfully reflect the performance of the physical hardware.

The core advantage of our Transformer model is that it does not require an explicit analytical inversion of these physical failures. Instead, through the self-attention mechanism trained on large datasets generated by this noise model, the Transformer automatically learns the complex mapping from the noise-corrupted projection data back to the high-fidelity, physically valid (positive semi-definite) density matrix $\rho_{\mathrm{ideal}}$. It inherently learns to suppress the statistical noise caused by SPDC low efficiency and correct the fidelity drop caused by accidental coincidences.

\section{Quantum States Prepared and Measurement Statistics}
First, it is clarified that the reported five repetitions do not correspond to five single-shot projective measurements. Instead, they refer to five independent experimental batches. In each batch, coincidence counts and the corresponding single-photon counts are accumulated across all projection bases within a specific time. Given the source brightness, approximately $10^2$ coincidence counts per second are collected, while single-photon counts per user range from $10^2$ to $10^3$ depending on the basis. Conducting multiple independent batches allows for evaluating the long-term stability of the setup and for averaging out random fluctuations or system drifts.

\begin{figure*}[hbt!]
    \centering
    \includegraphics[width=0.9\textwidth]{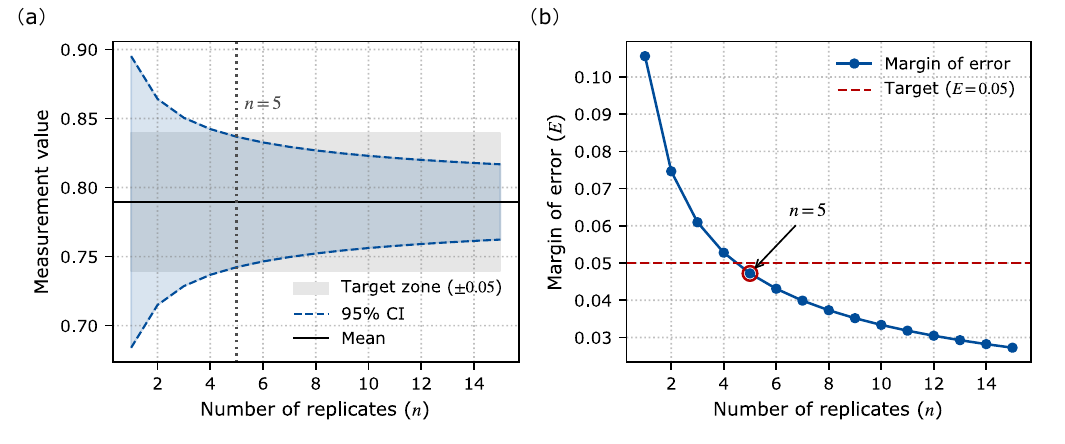}
    \caption{Dependence of measurement precision on replicate number n. (a) Measured values with 95\% confidence intervals. (b) Margin of error.}
    \label{fig:Measurement_precision}
\end{figure*}

Second, to bound the margin of error within $E = 0.05$ at a $95\%$ confidence level ($Z \approx 1.96$), the minimum required sample size is given by 
\begin{equation}
    n = (Z\sigma / E)^2. 
\end{equation}
Taking the mixed state $\eta = 1$ as an example, preliminary tests ($n = 3$) indicated that five independent replicates would be sufficient. Subsequent formal measurements at $n = 5$ yielded $\sigma \approx 0.0539$, confirming this design. As validated in Fig.~\ref{fig:Measurement_precision}(a) and Fig.~\ref{fig:Measurement_precision}(b), both the 95\% confidence interval and the margin of error converge securely within the predefined tolerance at this exact sample size.

\begin{figure}[!tbp]
    \centering
    \includegraphics[width=0.9\linewidth]{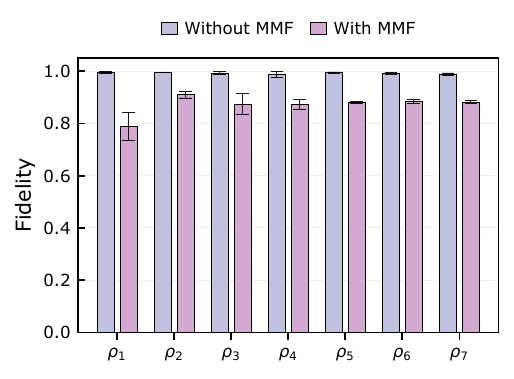}
    \caption{State fidelities without and with the MMF.}
    \label{fig:Fidelity_compare}
\end{figure}

The observed fidelity decrease reflects the physical impact of the MMF, rather than an omission in the noise model. Experimentally, in the absence of the MMF (i.e., when the quantum channel is directly connected without multimode transmission), the average fidelity of the seven prepared quantum states reaches $(99.25 \pm 0.22)\%$. As illustrated in Fig.~\ref{fig:Fidelity_compare}, the baseline fidelity of the calibrated setup without the MMF (blue bars) is slightly below unity due to inherent experimental imperfections, such as Poissonian statistical noise, detector dark counts, and imperfect optical components. Upon introducing the MMF, the uncorrected fidelity (red bars) decreases to an average of $(87.04 \pm 3.77)\%$ due to multimode perturbations.

The optimized TQSC model is trained directly on measurement data from the quantum channel with the MMF already incorporated. Consequently, it inherently accounts for both inherent experimental imperfections and the MMF-induced effective noise (e.g., mode coupling, phase perturbations, and mode-dependent loss). When paired with the Transformer-based post-processing, this framework reliably recovers the experimentally measured density matrices with high fidelity toward the target states.
\subsection{Measurement statistics and BER evaluation in the MNIST demonstration}
\label{sec:supp-mnist-statistics}

In the MNIST transmission demonstration, the two logical bit values are encoded into the mixed states
\begin{equation}
\rho_{\eta}
=
\frac{1}{1+\eta}
\left(
|H\rangle\langle H|
+
\eta |V\rangle\langle V|
\right),
\end{equation}
with $\eta=2$ and $\eta=3$. Using the squared Uhlmann fidelity convention, their fidelity is
\begin{equation}
F(\rho_2,\rho_3)
=
0.9916 .
\end{equation}
Thus, the two ideal encoded states are intrinsically close to each other. For equal prior probabilities, the Helstrom minimum error probability for direct single-copy discrimination is
\begin{equation}
P_{\mathrm{e}}^{\mathrm{Hel}}
=
\frac{1}{2}
\left(
1-\frac{1}{2}\|\rho_2-\rho_3\|_1
\right)
=
45.83\%,
\end{equation}
where $\|\rho_2-\rho_3\|_1=1/6$. This value quantifies the minimum possible error probability for directly distinguishing one ideal copy of $\rho_2$ from one ideal copy of $\rho_3$ using the optimal quantum measurement. It therefore reflects the intrinsic closeness of the two encoded ideal states, but it is not the operational BER bound for the experimental MNIST transmission task, where each sample is represented by a finite-time acquisition record.

We emphasize that the $N=100$ measurements shown in Fig.~4(b) are used only to characterize and visualize the marginal projection-outcome distributions of the two encoded states. These measurements are not concatenated, averaged, or majority-voted for the bit-error-rate evaluation.

In the bit error rate (BER) evaluation, each transmitted bit is represented by one independent experimental measurement record,
\begin{equation}
\mathbf{x}
=
(C_1,\ldots,C_4,S_1,\ldots,S_8,\tau)
\in \mathbb{R}^{13},
\end{equation}
where $C_1,\ldots,C_4$ and $S_1,\ldots,S_8$ denote four coincidence and eight single-photon counts, respectively, and $\tau$ is the timing feature. The TQSC model reconstructs the output state and assigns the corresponding label from this single 13-dimensional experimental record. Therefore, the BER result is not obtained by statistically averaging the $N=100$ repeated measurements shown in Fig.~4(b) of the main text, but by evaluating independent experimental records after MMF transmission.

For comparison, the conventional maximum-likelihood reconstruction baseline uses the four coincidence-count measurements. For the $i$th experimental record, we denote the state reconstructed by method $m\in\{\mathrm{ML},\mathrm{TQSC}\}$ as $\hat{\rho}_{m,i}$, and evaluate its fidelity to the target state $\rho_{\eta}$ as $F(\hat{\rho}_{m,i},\rho_{\eta})$. Averaging over 700 experimental realizations for each target state gives
\begin{align}
    \bar{F}_{\mathrm{ML}}^{(\eta=3)} &= 0.8732 \pm 0.0360,  \\
    \bar{F}_{\mathrm{TQSC}}^{(\eta=3)} &= 0.9625 \pm 0.0020, \\[1ex] 
    \bar{F}_{\mathrm{ML}}^{(\eta=2)} &= 0.9353 \pm 0.0930, 
    \\
    \bar{F}_{\mathrm{TQSC}}^{(\eta=2)} &= 0.9843 \pm 0.0005 .
\end{align}
Here, the values after $\pm$ denote the standard deviations over different experimental realizations.

The decoded label is assigned according to the reconstructed state's fidelity to the two target states,
\begin{equation}
\hat{\eta}
=
\underset{\eta\in\{2,3\}}{\operatorname{arg\,max}}
\,
F(\hat{\rho},\rho_{\eta}) .
\end{equation}
On the 1400 test records, including 700 realizations of $\rho_2$ and 700 realizations of $\rho_3$, the TQSC reconstruction gives a BER of $0\%$, whereas the conventional ML reconstruction gives a BER of $49.57\%$.

The experiment evaluates the robustness of different reconstruction pipelines when applied to noisy experimental records after multimode-fiber transmission and detection. The performance improvement of TQSC originates from its use of the complete 13-dimensional experimental measurement record, rather than from repeated-measurement averaging of the $N=100$ statistics shown in Fig.~4(b).
\section{Detailed TQSC Architecture and Training}
\label{app:tqsc_details}
The TQSC processes a 13-dimensional input vector composed of three key components for each measured mixed state: the coincidence counts of relevant projection bases, the corresponding photon counts recorded at Alice's and Bob's measurement stations, and the measurement coincidence time. The target output for the fidelity optimization task is the density matrix representing the quantum state.

Unlike conventional quantum state tomography approaches that rely on repeated measurements to estimate expectation values, the proposed TQSC model infers the quantum state from a single set of measurement outcomes.
Specifically, for each quantum state, only one measurement shot per measurement setting is performed, and no averaging over multiple measurement rounds is required. The Transformer model learns to reconstruct the underlying quantum state directly from this single-shot measurement data.

Each element in the input vector is treated as a token. The input tensor first undergoes preprocessing where it passes through a linear transformation layer that projects it into a higher-dimensional feature space. This is followed by an unsqueeze operation to adjust the tensor dimensions for compatibility with the subsequent Transformer architecture.

The preprocessed tensor is then fed into a stack of three identical Transformer encoder layers. Each encoder layer begins by incorporating positional encoding to preserve the sequence order information. The core component is a multi-head self-attention mechanism that operates with four parallel attention heads. Each head learns distinct types of correlations within different representation subspaces, significantly enhancing the model's capacity to capture complex dependencies between the input tokens. The outputs from the attention heads are integrated and passed through a residual connection followed by layer normalization (Add \& Norm block), which stabilizes training and facilitates convergence. The normalized output is then processed by a position-wise feedforward neural network with nonlinear activation, enabling deeper extraction of complex features from the attention outputs.

After processing through the three encoder layers, the enriched feature representation enters the post-processing stage. The feature dimension is first adjusted through a final linear transformation. The resulting tensor is then decomposed into two components: one corresponding to the diagonal elements of the density matrix and the other to the off-diagonal elements. The diagonal elements undergo softmax normalization to ensure they sum to unity and represent valid probabilities. The off-diagonal elements are processed through a hyperbolic tangent (tanh) activation function, constraining their values to the range [-1, 1] to reflect the properties of coherence terms. Finally, the processed diagonal and off-diagonal components are recombined appropriately to reconstruct the full Hermitian density matrix, which constitutes the final output of the model.

For model development and evaluation, the experimental dataset was partitioned using stratified random sampling, allocating 80\% of the data to the training set and 20\% to the independent test set. This approach preserves the underlying distribution of quantum states in both subsets. During training, the model was optimized using the Adam optimizer, an adaptive variant of stochastic gradient descent. The MSE loss function was employed to quantify the discrepancy between the predicted density matrix ($\rho_{\text{pred}}$) and the true label density matrix ($\rho_{\text{true}}$):
\begin{equation}
    \mathcal{L}_{\text{MSE}} = \frac{1}{N} \sum_{i=1}^{N} \|\rho_{\text{pred}}^{(i)} - \rho_{\text{true}}^{(i)} \|_F^2,
\end{equation}
where $N$ is the batch size and $\|\cdot\|_F$ denotes the Frobenius norm. The initial learning rate was set to 0.001 and training proceeded with a batch size of 64. Over successive training epochs, the loss value steadily decreased as the predicted density matrices converged toward their ground-truth counterparts.
\section{Baseline comparison}
\label{sec:supp_strengthened_baselines}

To evaluate reconstruction performance under the same experimental protocol,
we compared the Transformer--Cholesky model with representative neural,
tomographic, Bayesian, and maximum-likelihood baselines. All learning-based
models used the same experimental input vectors, normalization protocol,
train/validation/test split, and fidelity metric. The neural baselines were
also constrained to output physical density matrices through the Cholesky
parametrization, enabling a direct comparison between attention-based and
non-attention learning-based reconstructions.

The conventional reconstruction baselines include HVDR linear inversion,
HVDR least-squares tomography, and HVDR maximum-likelihood estimation (MLE),
which are standard density-matrix reconstruction approaches in quantum-state
tomography. For the HVDR-MLE
baseline, the maximum-likelihood procedure was applied to the experimental
remote-state-preparation projective measurements obtained under dynamic
Gaussian noise superimposed with MMF noise. For each target
state, five repeated projective measurements were performed, the density matrix
was reconstructed for each repetition, and the reported fidelity is the average
fidelity between the reconstructed density matrices and the corresponding target states.

The baseline set covers several complementary comparisons: (i) a
parameter-matched non-attention MLP with the same Cholesky physicality
constraint as the Transformer; (ii) a smaller Cholesky-constrained MLP; (iii)
closed-form HVDR linear-inversion and least-squares tomography from the measured
HVDR coincidence counts, followed by positive-semidefinite (PSD) projection;
(iv) a Bayesian known-family estimator over the experimentally used discrete
mixed-state family; and (v) an empirical training-calibrated HVDR-MLE estimator.
Randomized benchmarking and randomized-compiling-based noise tailoring are
gate-sequence characterization and mitigation protocols
\cite{magesan2011scalable,wallman2016noise}; they are therefore outside the
direct scope of the present projection-measurement data set.

The quantitative results are summarized in Table~\ref{tab:baseline}. While the results presented in the main text are obtained using a fixed random seed for consistency, here we evaluate five different random seeds to assess the robustness of the model architectures against variations in random initialization. The Transformer--Cholesky model achieved the highest average test fidelity among the compared methods, reaching $0.999982 \pm 0.000046$. The parameter-matched MLP--Cholesky baseline also achieved high fidelity, reaching $0.999978 \pm 0.000166$ with a comparable number of trainable parameters. This comparison shows that physically constrained learning-based reconstruction is highly effective for the present dynamic Gaussian plus MMF noise condition, and that the Transformer formulation retains a slight average advantage under an intentionally strong non-attention neural baseline. The smaller MLP reached $0.999933 \pm 0.000429$, further confirming the robustness of the Cholesky-constrained neural reconstruction strategy.

The Bayesian known-family estimator reached $0.996200 \pm 0.007437$, confirming
that the discrete state-family information provides a useful prior for this
task. The Transformer--Cholesky and MLP--Cholesky models further improve the
average fidelity by learning directly from the experimental count vectors and
auxiliary channels. The HVDR-LI-PSD, HVDR-LS-PSD, and training-calibrated
HVDR-MLE baselines provide conventional reconstruction references under the
same dynamic-noise data. Overall, the comparison supports the use of
physically constrained neural reconstruction for high-fidelity single-qubit
state estimation in this experiment.

\begin{figure*}[htbp]
    \centering
    \includegraphics[width=\textwidth]{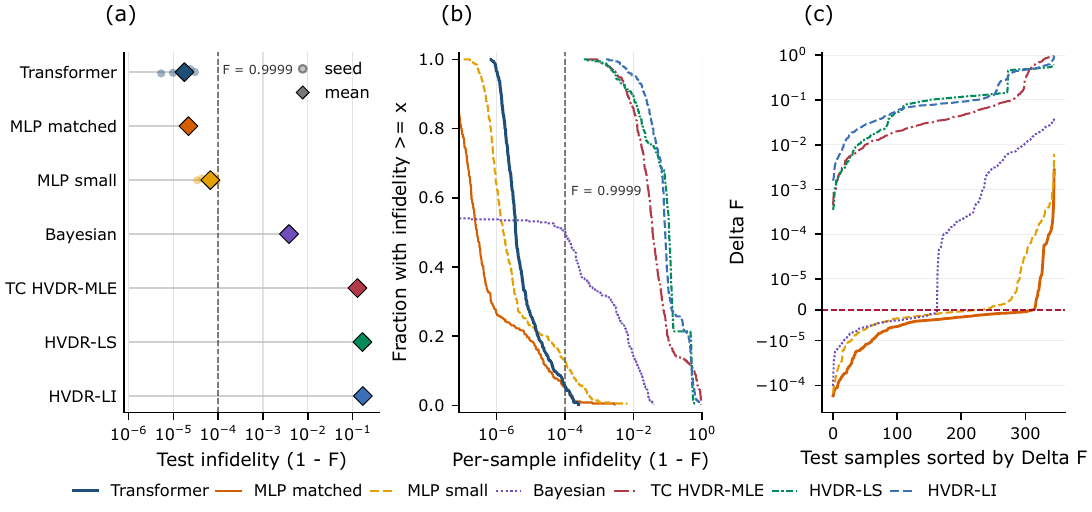} 
    \caption{Comparison with strengthened baselines. (a) Test infidelity $1-F$ on a logarithmic scale; circles denote individual seeds and diamonds denote means. (b) Per-sample infidelity $1-F$ tail distribution. (c) Sorted paired fidelity differences relative to the Transformer; the dashed horizontal line marks parity. Sorted paired fidelity differences are computed by subtracting the Transformer fidelity from each baseline fidelity for the same input state.}
    \label{fig:baseline}
\end{figure*}

Fig.~\ref{fig:baseline} visualizes the baseline comparison
from three complementary perspectives. Fig.~\ref{fig:baseline}(a)
compares the test infidelity, $1-F$, of different reconstruction methods on a
logarithmic scale, showing that the Transformer--Cholesky model attains the
lowest mean infidelity among the evaluated methods.
Fig.~\ref{fig:baseline}(b) presents the per-sample
infidelity-tail distribution, where the Transformer--Cholesky results remain
concentrated in the high-fidelity regime across the test set.
Fig.~\ref{fig:baseline}(c) shows the sorted paired fidelity
differences relative to the Transformer--Cholesky model; positive values
correspond to test samples for which the Transformer--Cholesky reconstruction
gives higher fidelity than the corresponding baseline, and the dashed horizontal
line denotes parity with the Transformer.

\begin{table*}[htbp]
    \centering
    \caption{Quantitative comparison of the strengthened baselines. For neural models, the mean is averaged over five random seeds. The reported standard deviation is the mean per-sample fidelity standard deviation on the test set, not the seed-to-seed standard deviation. 
    The parameter column reports trainable parameters; `--' denotes non-neural baselines without a comparable trainable parameter count.}
    \label{tab:baseline}
    \begin{tabular}{lccc}
        \toprule
        Method & Purpose & Parameters & Test Fidelity (Mean $\pm$ Std) \\
        \midrule
        Transformer-Cholesky & Proposed model & 152,292 & $0.999982 \pm 0.000046$ \\
        MLP-Cholesky-ParamMatched & Non-attention MLP, comparable size & 152,306 & $0.999978 \pm 0.000166$ \\
        MLP-Cholesky-Small & Small MLP & 7,268 & $0.999933 \pm 0.000429$ \\
        Bayesian-Known-Family & Oracle-like known-family posterior & 0 & $0.996200 \pm 0.007437$ \\
        Training-Calibrated HVDR-MLE & Noise-aware reconstruction & -- & $0.871839 \pm 0.235055$ \\
        HVDR-LS-PSD & Least-squares tomography & -- & $0.834470 \pm 0.179945$ \\
        HVDR-LI-PSD & Linear-inversion tomography & -- & $0.830808 \pm 0.193338$ \\
        Legacy HVDR-MLE Ref.~\cite{james2001measurement} & Legacy MLE reference & -- & $0.383100 \pm 0.123000$ \\
        \bottomrule
    \end{tabular}
\end{table*}

The Transformer formulation is well matched to the structure of the present
tomographic input. The HVDR counts and auxiliary channels are basis-specific
observations of the same density matrix under a shared noise environment.
Tokenizing these components allows self-attention to model inter-channel
interactions and provides a natural route to larger measurement sets,
additional monitoring channels, time-dependent records, and potentially
multi-qubit tomography \cite{vaswani2017attention,song2019autoint,gorishniy2021revisiting}.
The attention maps are used as post-hoc diagnostic tools to inspect how the
model weights different measurement channels
\cite{vig2019multiscale,abnar2020quantifying}.

\section{Generalization Capability of the TQSC Model}
\subsection{Experimental generalization within the measured state family}
To further evaluate the generalization capability of the proposed TQSC model beyond the initially selected eight training states, we conducted additional experiments on previously unseen quantum states.

\paragraph{Interpolation Test at an Unseen Intermediate Parameter ($\eta=4.5$).}

A model trained on the original dataset with parameter values 
$\eta \in \{1,2,3,4,5,6,7\}$ was tested on quantum states generated using an intermediate value $\eta=4.5$, which was not included during training. 
The model successfully reconstructed the corresponding quantum states with an average fidelity of $0.993712$ (standard deviation: $4.28\times10^{-6}$), demonstrating its ability to interpolate within the learned parameter manifold.

\paragraph{Leave-One-Out Validation ($\eta=6$).}

To further examine predictive capability, we retrained the model on a reduced dataset that explicitly excluded $\eta=6$, i.e.,
$\eta \in \{1,2,3,4,5,7\}$. 
The retrained model was then evaluated on quantum states corresponding to $\eta=6$. 
The resulting reconstruction achieved an average fidelity of $0.999989$ (standard deviation: $0.32\times10^{-6}$), indicating strong predictive performance on unseen data.

\begin{figure}[thbt!]
    \centering
    \includegraphics[width=0.9\linewidth]{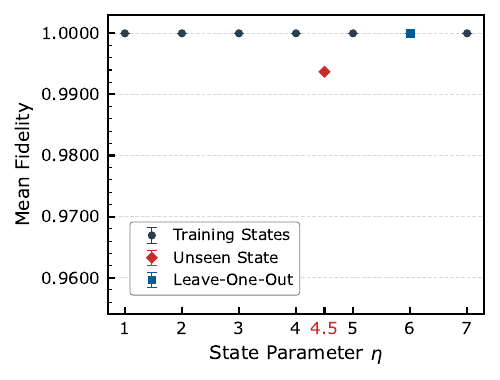}
    \caption{Mean fidelity vs $\eta$. Circles: training. Square: leave-one-out. Diamond: unseen. Error bars: $\pm1\sigma$.}
    \label{fig:Fidelity_vs_eta}
\end{figure}
\paragraph{Discussion.}

As shown in Fig.~\ref{fig:Fidelity_vs_eta}, these results confirm that the TQSC model exhibits strong generalization capability to quantum states not explicitly present in the training set. 
This behavior can be attributed to the regression-based formulation of TQSC, which learns a continuous mapping between measurement data and quantum state parameters rather than memorizing discrete categories.

Nevertheless, the predictive performance is inherently dependent on the statistical proximity between unseen data and the training distribution. 
The model performs best in interpolation regimes (e.g., $\eta=4.5$) or when the test states share underlying physical characteristics with the training data. 
As is typical in regression-based learning tasks, prediction accuracy may degrade for states that deviate significantly from the training manifold~\cite{carrasquilla2017machine,carleo2019machine}. 
Therefore, enlarging and diversifying the training dataset can improve the sampling density of the state space, leading to enhanced predictive accuracy and robustness against distributional shifts~\cite{kaplan2020scaling,caro2022generalization}.

\subsection{Numerical generalization over Bloch-ball states}
\label{sec:supp_bloch_ball_benchmark}

The experimentally acquired photonic data used in the main TQSC demonstration do not constitute an exhaustive sampling of the full single-qubit Bloch ball. In particular, the mixed states in Eq.~(2) of the main text form a one-parameter family that is diagonal in the $\{|H\rangle,|V\rangle\}$ basis, and the experimentally prepared pure-state example in Eq.~(8) represents one selected target state. Therefore, the experimental leave-one-out and interpolation tests should be interpreted as proof-of-principle tests within this restricted measured state family, rather than as a demonstration over arbitrary single-qubit states.

To examine the behavior of the same TQSC pipeline beyond this diagonal family, we performed an additional Qiskit-based numerical benchmark using pure and mixed single-qubit target states distributed on and inside the Bloch ball. The target states are parameterized as
\begin{equation}
\rho(r,\theta,\phi)
=
\frac{1}{2}
\left[
I+r\,\hat{\mathbf n}(\theta,\phi)\cdot\boldsymbol{\sigma}
\right],
\quad
0\leq r\leq 1,
\label{eq:supp_bloch_ball_state}
\end{equation}
where
\begin{equation}
\hat{\mathbf n}(\theta,\phi)
=
(\sin\theta\cos\phi,\sin\theta\sin\phi,\cos\theta).
\end{equation}
Here, $r=1$ corresponds to pure states on the Bloch-sphere surface, while $0\leq r<1$ corresponds to mixed states inside the Bloch ball. This construction includes both diagonal and off-diagonal density matrices in the $\{|H\rangle,|V\rangle\}$ basis.

\begin{figure}[tbp]
\centering
\includegraphics[width=1\linewidth]{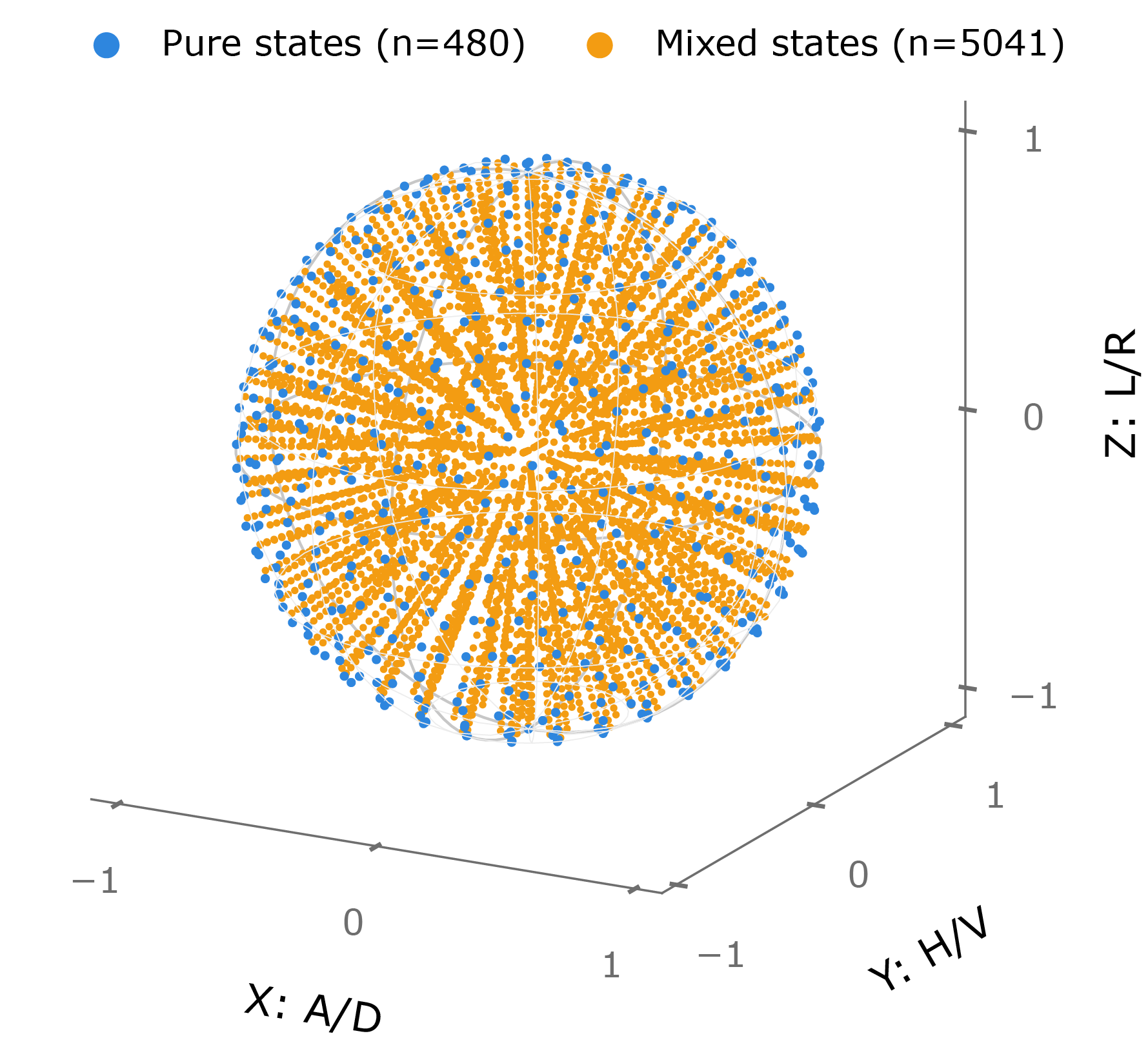}
\caption{
Qiskit-generated Bloch-ball target states used in the supplementary numerical benchmark.
Pure states are sampled on the Bloch-sphere surface, while mixed states are distributed inside the Bloch ball.
The dataset contains 480 pure states and 5041 mixed states, covering both diagonal and off-diagonal single-qubit density matrices.
}
\label{fig:supp_bloch_ball_states}
\end{figure}

Fig.~\ref{fig:supp_bloch_ball_states} shows the state-space coverage used in this supplementary benchmark. We used 480 pure-state directions on the Bloch-sphere surface and 360 angular directions for each of 14 nonzero mixed-state radial shells, together with the maximally mixed state at the origin. Therefore, the total number of target states before repeated noisy measurement sampling is
\begin{equation}
N_{\mathrm{state}}=480+14\times 360+1=5521.
\end{equation}

\begin{figure*}[htbp]
\centering
\includegraphics[width=0.8\linewidth]{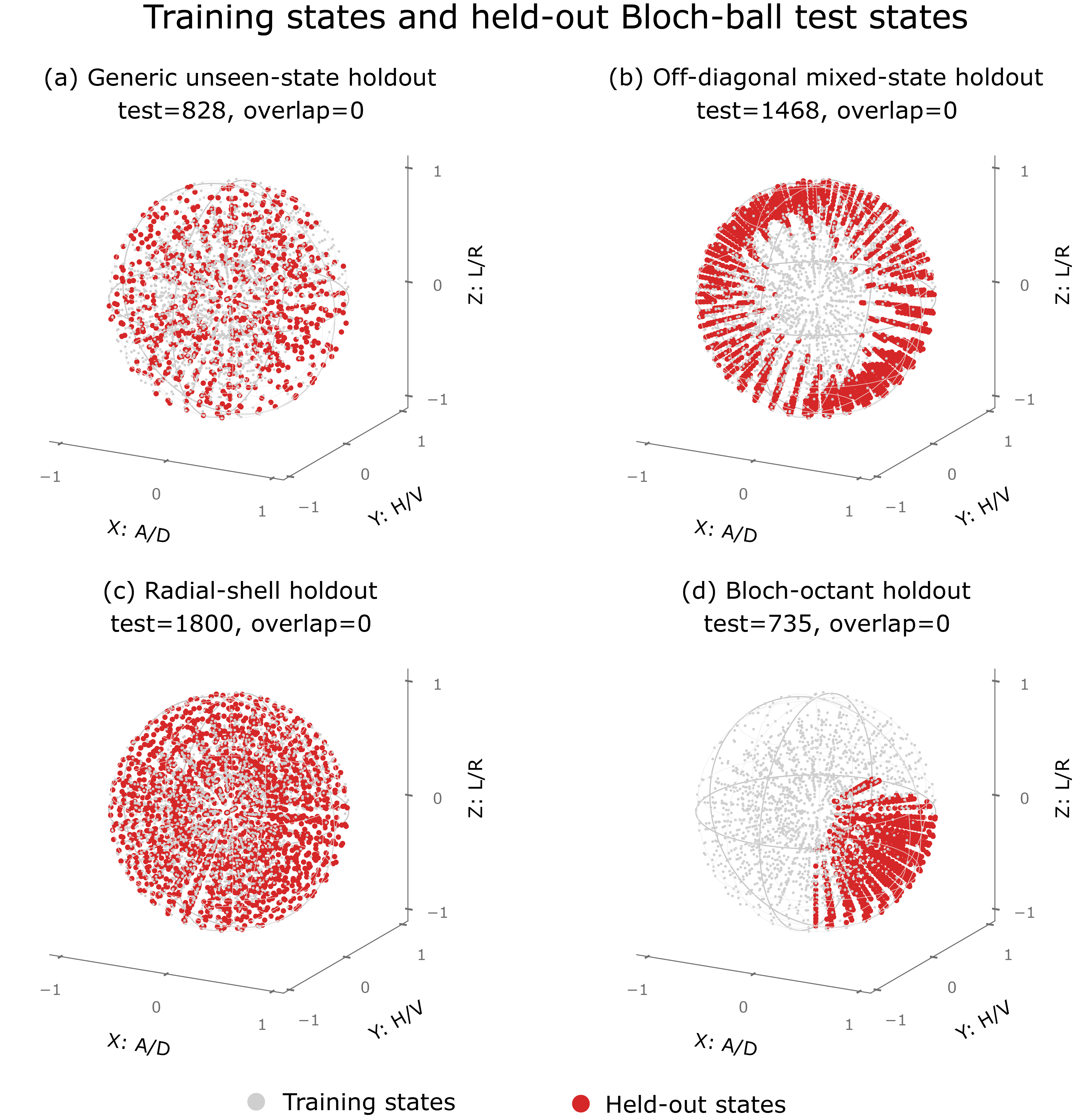}
\caption{
Train and holdout splits in the Bloch ball.
Gray points denote training states and red points denote held-out test states.
The target-state-ID overlap between the training and held-out test sets is zero for all splits.
}
\label{fig:supp_bloch_ball_splits}
\end{figure*}

The noisy measurement samples were generated using an MMF-inspired Bob-channel model. This model is not intended to serve as a device-specific calibration of a particular fiber. Instead, it defines a physically motivated noisy-channel benchmark including transmission loss, polarization-dependent loss, polarization rotation, depolarization, phase damping, amplitude damping, temporal jitter, detector dark counts, and photon-counting shot noise. The parameters used in this simulation are summarized in Table~\ref{tab:supp_mmf_params}.

To avoid overlap between training and testing target states, we evaluated four strict train--test splits: leave-state-out, off-diagonal mixed-state holdout, radial-shell holdout, and Bloch-octant holdout. These splits are illustrated in Fig.~\ref{fig:supp_bloch_ball_splits}, where the training and held-out states are separated by target-state identity.

\begin{figure}[htbp]
\centering
\includegraphics[width=1\linewidth]{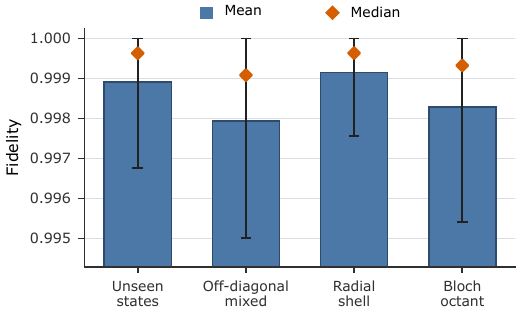}
\caption{Generalization fidelity on held-out Bloch-ball states.
Bars show the mean fidelity, error bars denote one standard deviation, and diamonds indicate the median fidelity.
}
\label{fig:supp_bloch_ball_fidelity}
\end{figure}

The corresponding held-out fidelities are shown in Fig.~\ref{fig:supp_bloch_ball_fidelity}. Across the four strict generalization splits, the mean fidelities are $0.998911$, $0.997943$, $0.999145$, and $0.998286$, respectively. In total, these tests include 4831 held-out target states and 483100 repeated noisy measurement samples.

These results show that, in the supplementary numerical benchmark, the TQSC model can generalize beyond the original one-parameter diagonal mixed-state family. We emphasize, however, that this benchmark is numerical and should not be interpreted as a replacement for experimentally measured Bloch-ball data. The experimental dataset in the main text remains a cost-limited proof-of-principle demonstration under realistic optical measurement noise.

\begin{table*}[htbp]
\centering
\caption{Parameters used in the MMF-inspired noisy Bob-channel simulation.}
\label{tab:supp_mmf_params}
\small
\begin{tabularx}{\textwidth}{l l X}
\toprule 
Parameter & Value & Role in the simulation \\
\midrule 
Pair-generation rate & $2.5\times10^{5}~\mathrm{s}^{-1}$ & Sets the photon-pair flux before detection and post-selection. \\
Acquisition time per basis & $0.25~\mathrm{s}$ & Integration time used to generate finite HVDR coincidence counts. \\
Coincidence window & $5.0~\mathrm{ns}$ & Defines the temporal acceptance for coincidence counting. \\
Alice detection efficiency & $0.55$ & Models finite Alice-side photon detection efficiency. \\
Bob detection efficiency & $0.50$ & Models finite Bob-side photon detection efficiency. \\
RSP success probability & $0.50$ & Accounts for successful remote-state-preparation events. \\
Bob dark-count rate & $80~\mathrm{s}^{-1}$ & Adds detector dark photons and accidental coincidences. \\
Baseline Bob-channel transmission & $0.88$ & Baseline transmission before MMF attenuation and timing acceptance. \\
Depolarizing probability & $0.06$ & Mixes the transmitted state toward $I/2$. \\
Amplitude-damping probability & $\gamma=0.04$ & Models polarization-dependent amplitude relaxation. \\
Phase-damping parameter & $\lambda=0.03$ & Models loss of phase coherence. \\
MMF length & $2.0~\mathrm{m}$ & Propagation length of the multimode-fiber segment. \\
MMF attenuation & $0.015~\mathrm{dB\,m}^{-1}$ & Propagation loss in the MMF channel. \\
Polarization-dependent loss & $0.25~\mathrm{dB}$ & Models unequal attenuation of orthogonal polarization components. \\
MMF rotation angle & $0.18~\mathrm{rad}$ & Birefringence-like polarization rotation angle. \\
MMF rotation azimuth & $0.70~\mathrm{rad}$ & Azimuth of the polarization-rotation axis. \\
MMF modal depolarization & $0.04$ & Additional modal depolarizing contribution. \\
MMF modal dephasing & $0.025$ & Additional modal coherence damping. \\
MMF temporal jitter & $1.0~\mathrm{ns}$ & Temporal broadening that reduces coincidence-window acceptance. \\
\bottomrule 
\end{tabularx}
\end{table*}

\section{Attention-Assisted Post-hoc Attribution and Functional Analysis of L3H4}
\label{sec:sm_l3h4_analysis}
This section provides additional analyses of the attention structure learned by
the Transformer, with particular emphasis on Head~4 in Layer~3 (L3H4).
A self-attention map characterizes how each token representation weights and
integrates information from the other measurement tokens. Whereas earlier
layers operate more directly on input-level measurement statistics, later
layers act on progressively contextualized representations formed through
repeated attention and nonlinear transformations. We therefore examine
deeper-layer attention because it provides a natural setting in which
task-relevant relations among complementary observables may emerge, including
the joint use of $H/V$ population-sensitive statistics, $D/R$
coherence/phase-sensitive statistics, Alice--Bob coincidence information, and
acquisition-time context. This does not imply that deeper layers necessarily
encode ``deeper physical laws''; rather, they allow us to test whether
lower-level measurement statistics have been integrated into more structured
relational representations.

The purpose of these analyses is not to treat an averaged attention map as
direct evidence of model interpretability. Instead, attention is used as a
post-hoc indicator of information routing, whose structure is further examined
through state-dependence analysis, pathway-specific intervention,
whole-model perturbation, Integrated Gradients (IG), and relational analyses.

The input sequence consists of 13 tokens whose ordering is fixed by the
measurement design. These tokens comprise four coincidence-count channels,
Alice- and Bob-side single-count channels in the H/V/D/R bases, and acquisition
time. The individual tokens and composite groups used throughout this analysis
are summarized in Table~\ref{tab:sm_individual_tokens} and
Table~\ref{tab:sm_composite_groups}.
See Sec.~\ref{subsec:sm_token_groups} for further details.
Because the token order is identical for
all samples, the relevant question is not whether attention changes with token
ordering, but whether different physical input states produce systematically
different attention structures under the same attention head.

\subsection{State Dependence of Sample-Specific L3H4 Attention}
\label{subsec:sm_state_dependence}

For each test sample $i$, the sample-specific L3H4 attention matrix is denoted
by
\begin{equation}
    A_i \in \mathbb{R}^{13\times 13}.
\end{equation}
State dependence was analyzed using a matched acquisition-condition subset
containing $N=229$ samples from five ground-truth state groups.

\begin{figure*}[!htbp]
    \centering
    \includegraphics[width=\textwidth]
    {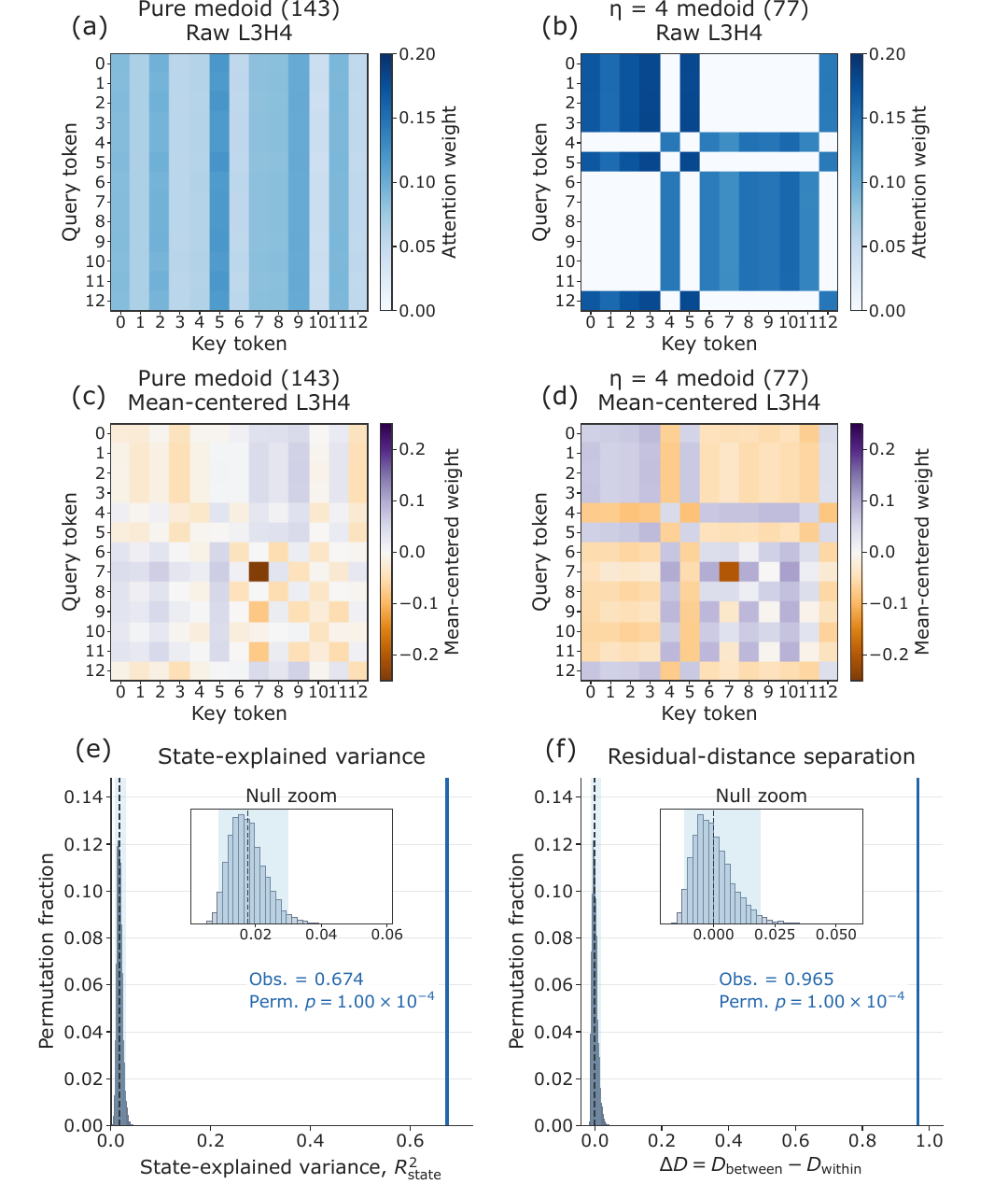}
    \caption{
    State-dependent L3H4 attention.
    (a,b) Raw attention maps for the pure-state medoid (sample~143) and the
    $\eta=4$ medoid (sample~77).
    (c,d) Corresponding mean-centered maps after subtraction of the test-subset
    mean.
    (e) State-explained variance $R_{\mathrm{state}}^{2}$.
    (f) Between-state minus within-state residual distance $\Delta D$.
    Null distributions were obtained from $10^{4}$ state-label permutations;
    the insets show enlarged views of the corresponding null distributions.
    }
    \label{fig:sm_state_dependence}
\end{figure*}

Fig.~\ref{fig:sm_state_dependence}(a,b) shows representative raw L3H4
attention maps for a pure-state sample (sample~143) and an $\eta=4$ sample
(sample~77). These examples were not selected randomly or chosen to maximize
their visual difference. Within each state group, the representative was
defined as the sample having the minimum Frobenius distance from the
corresponding group-mean attention map. The two representatives exhibit
different query--key structures under the same attention head. 
These representative samples illustrate the variation in L3H4 attention
patterns across physical states. The statistical association between
attention structure and state identity is evaluated subsequently using the
full matched subset.

The raw attention maps contain a pronounced vertical-stripe component shared
across the analyzed test subset. To separate this common component from
sample-specific attention variation, the mean-centered attention matrix is
defined as
\begin{equation}
    R_i = A_i - \overline{A},
    \label{eq:sm_centered_attention}
\end{equation}
where $\overline{A}$ denotes the mean L3H4 attention matrix over the analyzed
test subset. The resulting $R_i$ therefore contains deviations from the shared
attention pattern. Positive and negative entries indicate attention weights
above and below the corresponding subset mean, respectively; they do not
represent positive or negative contributions to the final model output.
Representative mean-centered maps are shown in
Fig.~\ref{fig:sm_state_dependence}(c,d).

To quantify the association between the mean-centered attention structure and
the ground-truth physical states, we calculate the state-explained variance
\begin{equation}
    R_{\mathrm{state}}^{2}
    =
    \frac{SS_{\mathrm{between}}}
         {SS_{\mathrm{total}}},
    \label{eq:sm_state_variance}
\end{equation}
where
\begin{equation}
    SS_{\mathrm{between}}
    =
    \sum_g
    n_g
    \left\|
        \overline{R}_g-\overline{R}
    \right\|_{F}^{2},
    \label{eq:sm_ss_between}
\end{equation}
and
\begin{equation}
    SS_{\mathrm{total}}
    =
    \sum_i
    \left\|
        R_i-\overline{R}
    \right\|_{F}^{2}.
    \label{eq:sm_ss_total}
\end{equation}
Here, $n_g$ is the number of samples in state group $g$,
$\overline{R}_g$ is the mean-centered attention centroid of group $g$, and
$\overline{R}$ is the mean of the mean-centered attention matrices over the
analyzed subset. A value of $R_{\mathrm{state}}^{2}=0$ indicates that the
state labels explain none of the residual attention variation, as expected for
a shared fixed attention pattern with state-independent fluctuations. In
contrast, $R_{\mathrm{state}}^{2}=1$ corresponds to the limiting case in which
all residual variation is attributable to differences between the state-group
centroids, with no within-state variation.

For L3H4, the observed value is
\begin{equation}
    R_{\mathrm{state}}^{2}=0.674.
\end{equation}
Thus, 67.4\% of the mean-centered L3H4 attention variance is associated with
differences among the five ground-truth state groups.

Statistical significance was evaluated by randomly permuting the state labels
among the 229 samples while keeping the attention matrices and group sizes
unchanged. A total of $B=10{,}000$ permutations were performed. The resulting
null distribution had a mean of 0.0176, a central 95\% interval of
$[0.0088,\,0.0300]$, and a maximum of 0.0585. None of the $10{,}000$
permutations reached the observed value. Using the finite-permutation
correction
\begin{equation}
    p_{\mathrm{perm}}
    =
    \frac{
        1+
        \displaystyle\sum_b
        I\!\left(S_b\geq S_{\mathrm{obs}}\right)
    }{
        B+1
    },
    \label{eq:sm_perm_p}
\end{equation}
the resulting permutation probability is
\begin{equation}
    p_{\mathrm{perm}}
    =
    9.999\times10^{-5}.
\end{equation}
The observed statistic and permutation distribution are shown in
Fig.~\ref{fig:sm_state_dependence}(e).

As a complementary analysis, we tested whether attention maps from samples
belonging to the same physical state are more similar than maps obtained from
different states. For samples $i$ and $j$, the correlation distance between
their vectorized mean-centered attention maps is defined as
\begin{equation}
    d_{ij}
    =
    1-
    \operatorname{corr}
    \left[
        \operatorname{vec}(R_i),
        \operatorname{vec}(R_j)
    \right].
    \label{eq:sm_corr_distance}
\end{equation}
The distance ranges from 0 to 2:
$d_{ij}=0$ corresponds to perfect positive correlation,
$d_{ij}=1$ to zero Pearson correlation, and
$d_{ij}=2$ to perfect negative correlation.

The average within-state and between-state distances are
\begin{equation}
    D_{\mathrm{within}}
    =
    \operatorname{mean}
    \left(
        d_{ij}\mid g_i=g_j
    \right),
    \label{eq:sm_dwithin}
\end{equation}
and
\begin{equation}
    D_{\mathrm{between}}
    =
    \operatorname{mean}
    \left(
        d_{ij}\mid g_i\neq g_j
    \right),
    \label{eq:sm_dbetween}
\end{equation}
respectively. Their separation is defined as
\begin{equation}
    \Delta D
    =
    D_{\mathrm{between}}
    -
    D_{\mathrm{within}}.
    \label{eq:sm_deltaD}
\end{equation}
If sample-specific attention variation is unrelated to physical state,
$D_{\mathrm{within}}$ and $D_{\mathrm{between}}$ should be comparable and
$\Delta D$ should be close to zero. Conversely, $\Delta D>0$ indicates that
attention maps are more similar within a physical state than between different
states.

For L3H4, the measured distances are
\begin{equation}
    D_{\mathrm{within}}=0.229,\;
    D_{\mathrm{between}}=1.195,\;
    \Delta D=0.965.
    \label{eq:sm_distance_results}
\end{equation}
The corresponding average Pearson correlations are approximately 0.771 within
states and $-0.195$ between states. The average different-state distance is
therefore approximately 5.2 times the average same-state distance.

Sample-label permutation was again used for statistical inference, avoiding
the incorrect treatment of the large number of pairwise distances as
independent observations. The null distribution of $\Delta D$ had a mean of
$-5.95\times10^{-5}$, a central 95\% interval of
$[-0.0121,\,0.0193]$, and a maximum of 0.0565. The observed
$\Delta D=0.965$ exceeded all $10{,}000$ permuted values, yielding
\begin{equation}
    p_{\mathrm{perm}}
    =
    9.999\times10^{-5}.
\end{equation}
The corresponding permutation analysis is shown in
Fig.~\ref{fig:sm_state_dependence}(f).

Together, these results distinguish the shared key-position preference from
sample-specific attention routing. After removal of the shared attention
component, the residual attention structure retains a robust and reproducible
association with the physical input state.

\begin{figure*}[htbp]
    \centering
    \includegraphics[width=0.87\textwidth]
    {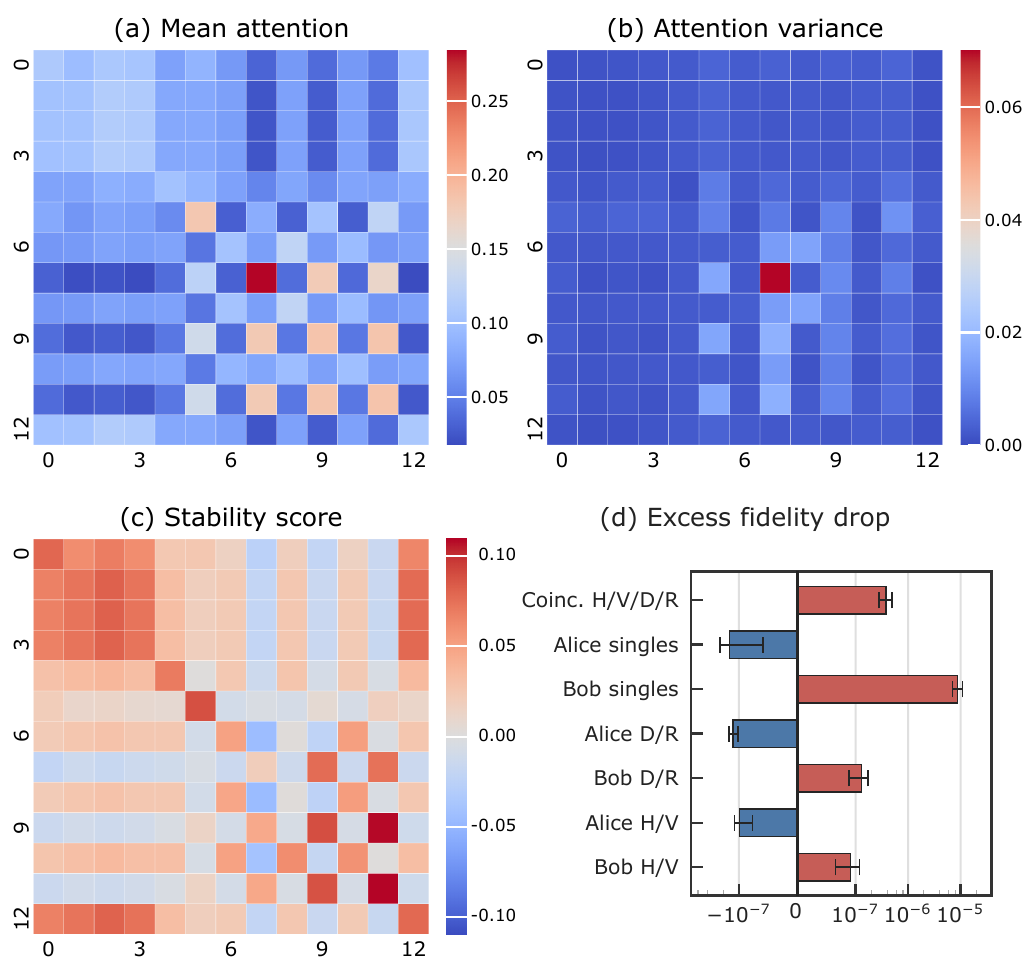}
    \caption{
    Attention characteristics of Layer~3 Head~4 (L3H4) and intervention
    effects beyond random ablation.
    (a) Mean attention weights averaged across all samples.
    (b) Corresponding cross-sample variance, highlighting localized
    input-dependent variability.
    (c) Stability score $S_{ij}$, defined as the mean attention weight minus
    its cross-sample standard deviation.
    (d) Excess fidelity drop for targeted token-group interventions relative
    to size-matched random key ablations.
    }
    \label{fig:sm_l3h4_attention}
\end{figure*}

\subsection{L3H4 Attention Structure and Pathway-Specific Intervention}
\label{subsec:sm_l3h4_intervention}

L3H4 exhibits a clearly structured attention pattern.
Fig.~\ref{fig:sm_l3h4_attention}(a) shows the L3H4 attention weights
averaged over the entire training set. Rather than being uniformly distributed,
the attention weights form distinct block- and stripe-like motifs across
specific query--key token pairs. Because the Alice- and Bob-side single-count
tokens are interleaved in the input sequence, structured organization within
the single-count sector appears as an alternating stripe-like pattern rather
than as a single contiguous block.

A particularly pronounced substructure is visually apparent among the
Bob-side single-count tokens, consistent with coordinated processing of
complementary measurements of the noise-affected Bob output. The coincidence
tokens and the acquisition-time token also show partially overlapping
attention profiles, which motivates examining whether temporal context
participates in the organization of coincidence-related information.
At this stage, these patterns are treated only as descriptive features of the
attention map rather than as evidence of functional or physical causality.

The corresponding cross-sample variance is shown in
Fig.~\ref{fig:sm_l3h4_attention}(b). The variance remains low over most
query--key pairs, with only localized regions showing appreciably larger
fluctuations. The most visible variations occur within the single-count
sector, particularly around several Bob-side relations. Thus, the mean
attention structure is not accompanied by broad sample-to-sample variability,
but instead consists of a largely reproducible backbone together with a
limited number of more sample-dependent relations.

To characterize the consistency of individual query--key relations, the
stability score is defined as
\begin{equation}
    S_{ij}
    =
    \mu_{ij}-\sigma_{ij},
    \label{eq:sm_stability}
\end{equation}
where $\mu_{ij}$ is the mean attention weight of query--key pair $(i,j)$ and
$\sigma_{ij}$ is its standard deviation across samples. A larger positive
stability score identifies a relation that combines relatively strong mean
attention with relatively small cross-sample variability, whereas smaller or
negative values indicate either weak mean attention or larger
sample-dependent variation.

As shown in Fig.~\ref{fig:sm_l3h4_attention}(c), the stability map preserves
several of the structured motifs observed in the mean attention map.
In particular, stable relations are visually apparent within the Bob-side
single-count sector and among coincidence-related entries, whereas several
cross-domain entries are comparatively weaker or more variable. 
These descriptive patterns motivate the quantitative relational analysis
presented in Sec.~\ref{subsec:sm_relational_organization}.

Attention structure alone does not establish functional relevance. To examine
whether selected token groups contribute functionally through the L3H4
pathway, targeted L3H4 interventions were performed. These interventions do
not remove the complete attention head. Instead, only the key columns
corresponding to the selected token group are masked in the L3H4 attention
logits.

The intervened attention is calculated as
\begin{equation}
    A
    =
    \operatorname{softmax}
    \left(
        \frac{QK^{\top}}{\sqrt{d}}
        +
        M
    \right),
    \label{eq:sm_l3h4_mask}
\end{equation}
where the entries of $M$ associated with the ablated keys are assigned a large
negative value. This prevents all queries in L3H4 from attending to the
selected tokens and renormalizes the attention probabilities over the
remaining keys. All other attention heads, residual connections, feed-forward
layers, and subsequent Transformer layers are left unchanged.

The fidelity of the intervened forward pass is then compared with that of the
corresponding unmasked forward pass. The resulting fidelity reduction
therefore measures the functional contribution of the selected token group
specifically through L3H4, rather than its contribution through the complete
network.

To control for the number of tokens included in each predefined group, each
targeted intervention is compared with a size-matched random key ablation.
The excess fidelity drop is defined as
\begin{equation}
    \Delta F_{\mathrm{excess}}
    =
    \Delta F_{\mathrm{target}}
    -
    \Delta F_{\mathrm{random}},
    \label{eq:sm_excess_fidelity}
\end{equation}
where $\Delta F_{\mathrm{target}}$ denotes the fidelity drop induced by the
predefined token group and $\Delta F_{\mathrm{random}}$ denotes the
corresponding size-matched random-ablation baseline. Positive values indicate
that the predefined token group causes a larger fidelity degradation than an
equally sized random group, whereas negative values indicate an effect weaker
than the random baseline.

As shown in Fig.~\ref{fig:sm_l3h4_attention}(d), the intervention effects are
strongly group dependent. Bob-related groups exhibit positive excess fidelity
drops, with the Bob-singles intervention producing the strongest effect among
the analyzed groups. In contrast, the corresponding Alice-related groups
exhibit negative excess values. 
The Coinc.\ H/V/D/R group, where ``Coinc.''
denotes coincidence, also produces a positive excess fidelity drop.

These results cannot be explained solely by differences in the number of
masked tokens. Instead, different token groups make different functional
contributions specifically through the L3H4 pathway. The intervention results
therefore complement the attention-weight analysis by showing that the
structured attention patterns in Fig.~\ref{fig:sm_l3h4_attention}(a--c) are
associated with group-dependent functional effects.

It is important to distinguish this L3H4-specific intervention from the
whole-model input perturbation analyzed below. The L3H4 intervention is local
and pathway specific: selected tokens are blocked only as keys within L3H4,
while the remainder of the network is unchanged. Whole-model input
perturbation instead acts directly on the model input, making the perturbed
information unavailable to all attention heads and all subsequent
computational pathways. These two interventions therefore address
complementary questions: L3H4 intervention evaluates whether selected
information contributes through this particular attention head, whereas
whole-model input perturbation evaluates its overall contribution to the
network.

\subsection{Comparison with Whole-Model Perturbation and Integrated Gradients}
\label{subsec:sm_cross_method}
\begin{figure*}[htbp]
    \centering
    \includegraphics[width=0.8\textwidth]{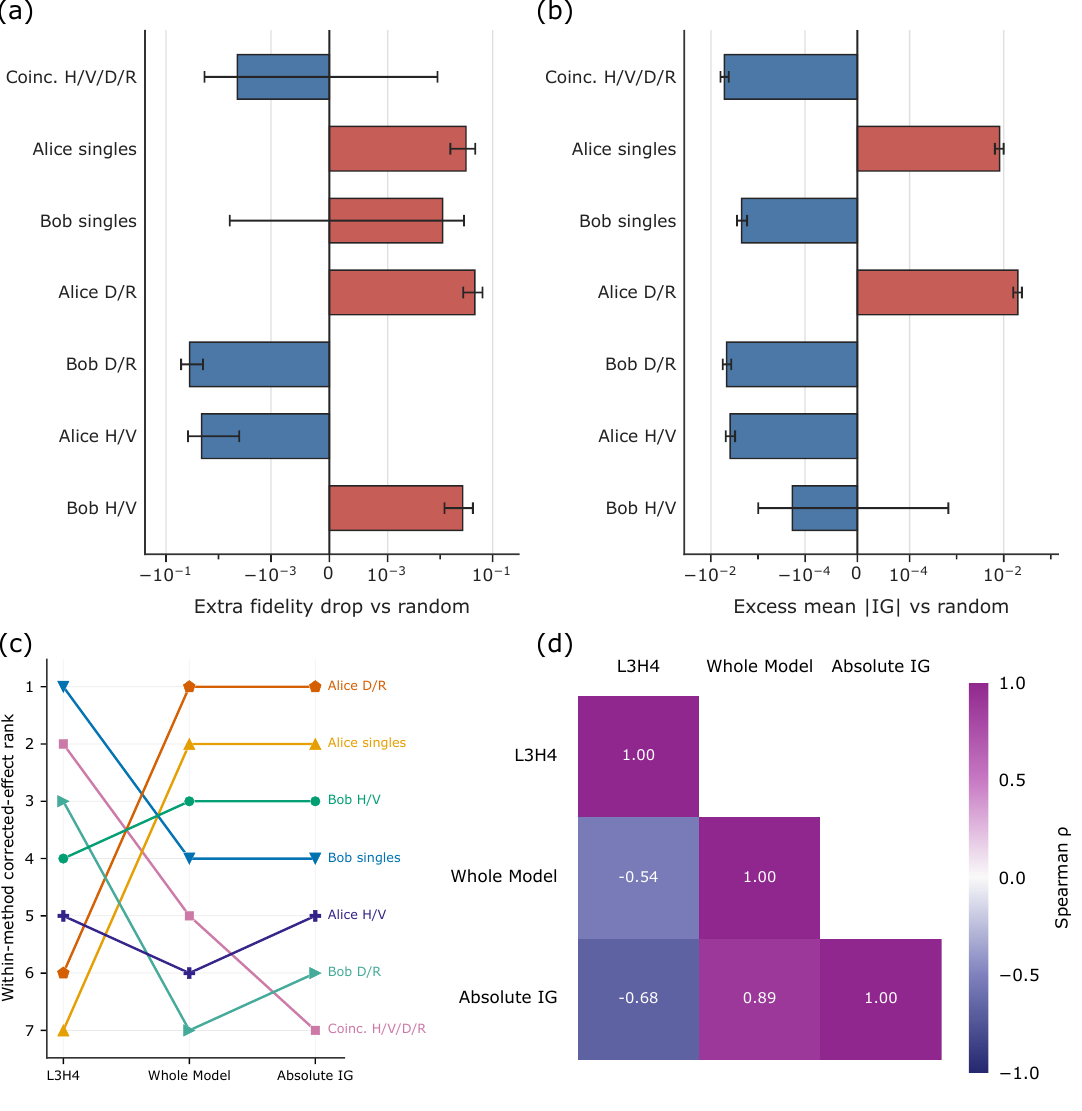}
    \caption{
    Cross-method comparison of token-group importance.
    (a) Size-matched-random-corrected whole-model input permutation effects.
    (b) Corrected absolute Integrated Gradients (IG) attribution.
    (c) Corresponding within-method rankings across L3H4 intervention,
    whole-model perturbation, and IG.
    (d) Spearman rank correlations between methods.
    Whole-model perturbation and IG yield similar end-to-end rankings,
    whereas L3H4 exhibits a distinct head-specific pattern.
    }
    \label{fig:sm_cross_method}
\end{figure*}

To clarify the relationship between the pathway-specific L3H4 intervention and
global input-importance measures, three analyses were compared using the same
346 held-out test samples:
(i) L3H4-specific intervention,
(ii) whole-model input permutation/ablation, and
(iii) Integrated Gradients (IG).

For an input $x$ and a reference baseline $x'$, standard IG attributes the
final model output to each input feature by integrating the gradient along the
straight-line path between $x'$ and $x$ \cite{sundararajan2017axiomatic}. For input
feature $i$, the attribution is
\begin{equation}
    IG_i(x)
    =
    \left(x_i-x_i'\right)
    \int_0^1
    \frac{
        \partial
        F\!\left(
            x'
            +
            \alpha(x-x')
        \right)
    }{
        \partial x_i
    }
    \,d\alpha,
    \label{eq:sm_ig}
\end{equation}
where $F$ denotes the final model output.

Because the gradient is taken with respect to the final output, IG aggregates
the influence of an input token through all computational paths connecting
that token to the prediction, including multiple attention heads, residual
connections, feed-forward blocks, and subsequent Transformer layers. IG
therefore provides an end-to-end attribution of the model output to the input
rather than an attribution restricted to a particular internal attention
pathway.

Whole-model input perturbation is conceptually closer to IG than to the
L3H4-specific intervention because it also evaluates end-to-end dependence on
the original input features. Whole-model perturbation changes a token group at
the model input and measures the resulting change in final prediction or
fidelity after the information has been disrupted throughout the network.
L3H4 intervention, by contrast, masks only selected key columns within one
attention head while leaving other heads and computational pathways unchanged.

Nevertheless, IG and whole-model perturbation are not mathematically
equivalent. IG is a path-integrated gradient attribution, whereas explicit
ablation or permutation measures a finite output change caused by an input
perturbation \cite{ancona2018towards}. Perturbation-based analyses consequently
provide a complementary means of assessing the functional relevance of
features identified by attribution methods \cite{hooker2019benchmark}. Nonlinearity,
feature interactions, and redundancy can lead to quantitative differences
between gradient-based attribution and explicit perturbation.

For both the whole-model and IG analyses, token-group size was controlled by
subtracting the mean effect obtained from 20 size-matched random groups.
Consequently, the corrected quantities in
Fig.~\ref{fig:sm_cross_method}(a,b) indicate whether a predefined token group
produces a stronger effect than would be expected from an equally sized random
group. Uncertainty was estimated using $10{,}000$ percentile-bootstrap
resamples of the 346 held-out samples, and 95\% confidence intervals are
reported in the figure.

As shown in Fig.~\ref{fig:sm_cross_method}(a), the whole-model analysis
assigns the strongest corrected fidelity effect to Alice D/R, followed by
Alice singles and Bob H/V. Bob D/R and Alice H/V fall below their respective
size-matched random baselines, whereas the confidence intervals for
Coinc.\ H/V/D/R and Bob singles include zero. These values characterize the
end-to-end sensitivity of the network to disruption of the corresponding
input information.

The absolute IG analysis in Fig.~\ref{fig:sm_cross_method}(b) yields a broadly
similar global pattern. Alice D/R and Alice singles again receive the largest
corrected importance values. Several groups exhibit corrected IG values below
zero. Because absolute IG magnitude is used here, a negative corrected value
does not indicate a negative attribution direction. It indicates only that the
absolute attribution magnitude of the predefined group is smaller than the
mean absolute attribution magnitude of size-matched random groups.

The ranking comparison in Fig.~\ref{fig:sm_cross_method}(c) further
distinguishes global and head-specific importance. Whole-model perturbation and
absolute IG both rank Alice D/R and Alice singles first and second,
respectively. L3H4 instead assigns relatively greater importance to Bob
singles, Coinc.\ H/V/D/R, and Bob D/R. The L3H4 ranking is therefore not simply
a noisier version of the global ranking; rather, it reflects a different
head-specific pattern of information dependence.

The relationships among the three rankings are quantified by Spearman
correlation in Fig.~\ref{fig:sm_cross_method}(d). Whole-model perturbation and
absolute IG are strongly positively correlated,
\begin{equation}
    \begin{gathered}
        \rho_{\mathrm{Whole,IG}} = 0.893, \\
        95\%~\mathrm{CI} = [0.750,\,0.964].
    \end{gathered}
\end{equation}
By contrast, the L3H4 ranking is negatively correlated with the whole-model
analysis,
\begin{equation}
    \begin{gathered}
        \rho_{\mathrm{L3H4,Whole}} = -0.536, \\
        95\%~\mathrm{CI} = [-0.643,\,-0.214],
    \end{gathered}
\end{equation}
and with absolute IG,
\begin{equation}
    \begin{gathered}
        \rho_{\mathrm{L3H4,IG}} = -0.679, \\
        95\%~\mathrm{CI} = [-0.714,\,-0.571].
    \end{gathered}
\end{equation}
The confidence intervals were obtained by paired bootstrap over the test
samples. These correlations are treated as descriptive cross-method
comparisons rather than conventional significance tests across groups,
because the seven predefined token groups partially overlap and are therefore
not statistically independent.

The strong agreement between whole-model perturbation and absolute IG is
consistent with their shared end-to-end scope, despite their different
mathematical mechanisms. The weaker and opposite relationship between L3H4 and
the two global measures is consistent with L3H4 probing a more localized
internal information-routing pathway. More generally, attention-level
importance and prediction-level or gradient-based feature importance need not
coincide \cite{jain2019attention}.

Accordingly, the three analyses should be regarded as complementary rather than
interchangeable. Whole-model perturbation provides an explicit end-to-end
perturbation test, absolute IG provides gradient-based end-to-end attribution,
and L3H4 intervention isolates the contribution of selected tokens through a
specific internal attention pathway. The agreement between whole-model
perturbation and IG, together with the distinct L3H4 ranking, supports the
interpretation that L3H4 exhibits a specialized information-routing pattern
rather than simply reproducing global token importance.

\subsection{Selective Relational Organization in L3H4}
\label{subsec:sm_relational_organization}

To further characterize the role of L3H4, we analyzed pairwise interactions,
attention-scope localization, group-level attention reorganization, and
standalone intervention importance.

The 13 input tokens encode physically distinct information. Alice-side single-count channels provide preparation-side reference information, whereas Bob-side single-count channels describe measurements after dynamic Gaussian phase modulation and MMF noise. Coincidence-count channels retain conditional correlation information, while acquisition time specifies the measurement duration and provides information needed to interpret the associated counting statistics and their statistical reliability. In addition, H/V and D/R measurements provide complementary population-sensitive and coherence/phase-sensitive information.

The relational analysis indicates that L3H4 exhibits selective rather than
uniform information integration. Here, $A\leftrightarrow B$ denotes a
bidirectional relational analysis between predefined token groups $A$ and $B$.
The interaction excess is defined relative to a size-matched random-token
baseline, such that positive values indicate enrichment relative to the random
baseline and negative values indicate depletion.

\begin{figure*}[htbp]
    \centering
    \includegraphics[width=0.8\textwidth]
    {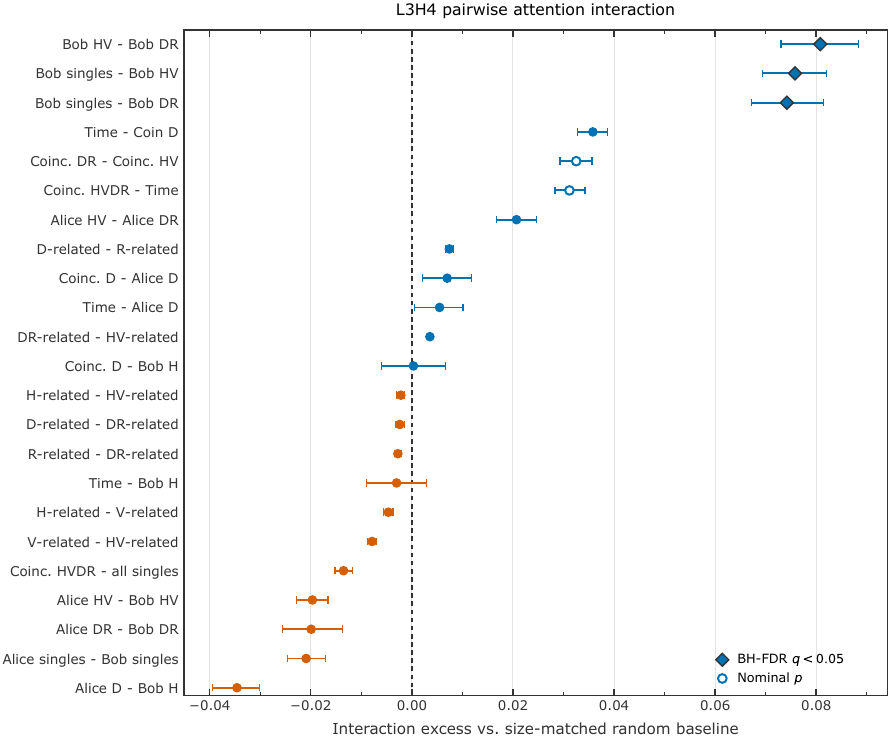}
    \caption{
    Pairwise attention interactions at L3H4.
   Interaction excess denotes the observed bidirectional interaction relative to a size-matched random-token baseline. Error bars show 95\% confidence intervals, and the dashed line marks zero excess. Filled diamonds indicate interactions that remain significant after Benjamini--Hochberg false-discovery-rate correction ($q<0.05$);
open circles indicate nominal significance only ($p<0.05$, $q\geq0.05$);
filled circles indicate all remaining interactions.
    }
    \label{fig:sm_pairwise_interaction}
\end{figure*}

Fig.~\ref{fig:sm_pairwise_interaction} summarizes the pairwise L3H4
interaction results. The most pronounced enriched interaction is between
Bob H/V and Bob D/R. The observed bidirectional interaction at L3H4 is 0.1527,
compared with a size-matched random baseline of 0.0719, giving an interaction
excess of
\begin{equation}
    I_{\mathrm{excess}}
    =
    0.1527-0.0719
    =
    0.0808.
\end{equation}
This corresponds to an approximately 112.4\% enhancement relative to the
random expectation. The interaction remains significant after
Benjamini--Hochberg correction across the 23 unique pairwise tests,
\begin{equation}
    p=0.00498,
    \qquad
    q\simeq0.038.
\end{equation}

The strong Bob H/V $\leftrightarrow$ Bob D/R relation
($\leftrightarrow$ denotes bidirectional relational analysis) is consistent
with localized joint use of complementary output-state information.
H/V measurements primarily provide population-sensitive statistics, while D/R
measurements provide complementary coherence- or phase-sensitive information.
The observed coupling is therefore consistent with an L3H4 representation that
jointly organizes these complementary measurements. 

Coincidence-related relations also exhibit selective enrichment. In particular, the Coinc.\ D/R $\leftrightarrow$ Coinc.\ H/V interaction and the Coinc.\ H/V/D/R $\leftrightarrow$ Time interaction are enhanced when the analysis is localized to L3H4. These patterns are consistent with a localized organization of complementary-basis coincidence information and with the joint interpretation of coincidence statistics and their acquisition duration.

In contrast, several interactions between distinct information domains are
depleted relative to their corresponding size-matched random baselines.
Alice singles $\leftrightarrow$ Bob singles,
Alice H/V $\leftrightarrow$ Bob H/V, and
Alice D/R $\leftrightarrow$ Bob D/R show reductions of approximately
29.3\%, 27.5\%, and 27.6\%, respectively. The
Coinc.\ H/V/D/R $\leftrightarrow$ all-singles relation is reduced by
approximately 18.9\%. The coexistence of enriched and depleted interactions
argues against interpreting L3H4 as a generic global mixing head.

\subsubsection{Localization of L3H4 Relational Structure}
\label{subsubsec:sm_scope_localization}

To determine whether the interaction structure described above reflects
model-wide relations or is preferentially localized to L3H4, the analysis was
evaluated at three nested scopes: all layers and heads, all heads in Layer~3,
and L3H4 alone.

As shown in Fig.~\ref{fig:sm_scope_localization}(a), the Bob H/V
$\leftrightarrow$ Bob D/R relation becomes progressively enriched as the
analysis is localized. Its interaction excess increases from $+0.0397$ across
all layers and heads to $+0.0492$ across Layer~3 and finally to $+0.0808$ at
L3H4. The prominent L3H4 interaction therefore cannot be attributed solely to
a uniform model-wide background relation; instead, it becomes increasingly
concentrated toward Layer~3 and most prominently toward L3H4.

Selected coincidence-related relations show an even more localized pattern.
The Coinc.\ D/R $\leftrightarrow$ Coinc.\ H/V interaction excess changes from
$-0.0247$ across all layers and $-0.0092$ in Layer~3 to $+0.0325$ at L3H4.
The L3H4 value corresponds to an approximately 45.7\% enhancement relative to
its size-matched random baseline.

Similarly, the Coinc.\ H/V/D/R $\leftrightarrow$ Time interaction changes from
$-0.0074$ across all layers and $-0.0095$ in Layer~3 to $+0.0312$ at L3H4,
corresponding to an approximately 42.9\% enhancement relative to random
expectation. The sign reversals show that these relations are not uniformly
elevated throughout the Transformer, but emerge specifically as the analysis
is restricted to L3H4.

The group-level statistics in
Fig.~\ref{fig:sm_scope_localization}(c,d) provide complementary evidence for
this redistribution. Relative to the all-layer baseline, the attention key
mass of Coinc.\ H/V, Coinc.\ D/R, and Coinc.\ H/V/D/R increases by
approximately 42.4\%, 61.3\%, and 51.5\%, respectively. Their corresponding
within-group attention densities increase by approximately 78.7\%, 106.4\%,
and 94.6\%. The L3H4 coincidence-related pattern therefore involves both
increased recruitment of coincidence information and stronger internal
organization of that information.

The opposite localization trend is observed for several cross-domain
relations, as shown in Fig.~\ref{fig:sm_scope_localization}(b). Alice singles
$\leftrightarrow$ Bob singles decreases from $+0.0115$ across all layers to
$+0.0021$ within Layer~3 and $-0.0209$ at L3H4. Alice H/V
$\leftrightarrow$ Bob H/V changes from $+0.0214$ to $+0.0127$ and then to
$-0.0197$, while Alice D/R $\leftrightarrow$ Bob D/R changes from $+0.0033$
to $-0.0091$ and finally to $-0.0199$.

Localization to L3H4 therefore produces opposite effects for different
classes of relations: selected within-domain interactions become enriched,
whereas several cross-domain interactions become depleted. This pattern is
more consistent with selective within-domain integration accompanied by
cross-domain segregation than with indiscriminate aggregation of all input
features.

\subsubsection{Relational Coupling and Standalone Intervention Importance}
\label{subsubsec:sm_relational_vs_standalone}

\begin{figure*}[htbp]
    \centering
    \includegraphics[width=0.9\textwidth]
    {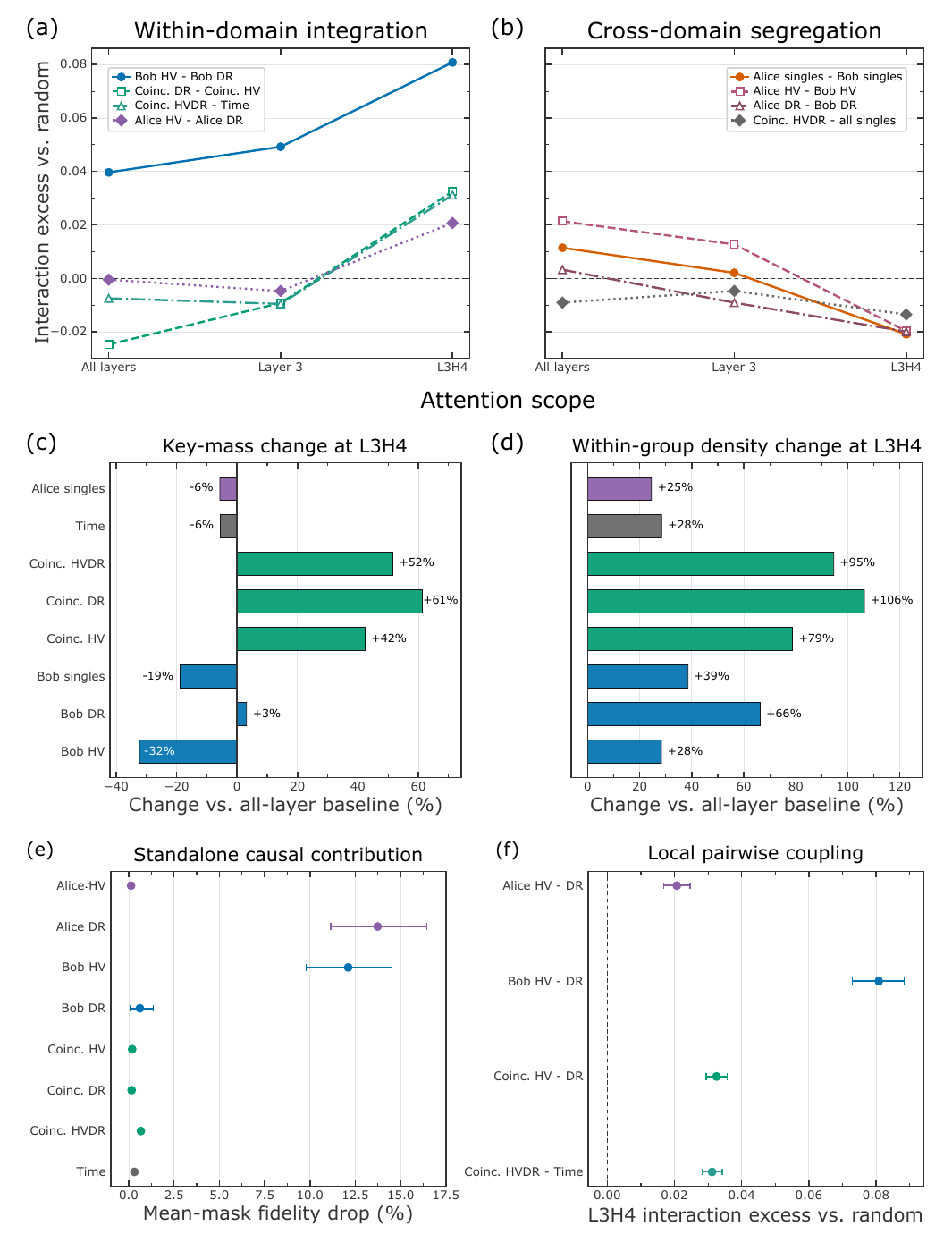}
    \caption{
    Scope localization and functional organization of L3H4 attention
    interactions.
    (a,b) Interaction excess relative to size-matched random-token baselines
    across all layers and heads, all heads in Layer~3, and L3H4 for selected
    (a) within-domain and (b) cross-domain token-group pairs.
    (c,d) Percentage change at L3H4 relative to the all-layer baseline in
    (c) attention key mass and (d) within-group attention density.
    (e) Mean-mask fidelity drop for selected token groups.
    (f) L3H4 interaction excess for selected within-domain and contextual
    pairs.
    Dashed lines denote zero excess where applicable, and horizontal bars in
    panels (e) and (f) indicate 95\% confidence intervals.
    }
    \label{fig:sm_scope_localization}
\end{figure*}

The L3H4 interaction structure is also distinct from the standalone
intervention importance of individual token groups, as illustrated by
Fig.~\ref{fig:sm_scope_localization}(e,f).

Bob H/V produces a mean-mask fidelity drop of approximately 12.11\%, whereas
Bob D/R produces a substantially smaller standalone effect of approximately
0.62\%. Nevertheless, Bob H/V $\leftrightarrow$ Bob D/R forms the strongest
pairwise L3H4 interaction, with an interaction excess of $+0.0808$.

A similar dissociation is observed for coincidence information.
Coinc.\ H/V and Coinc.\ D/R produce standalone fidelity drops of only
approximately 0.19\% and 0.16\%, respectively, whereas their L3H4 pairwise
interaction excess reaches $+0.0325$. Likewise, Coinc.\ H/V/D/R and Time have
standalone fidelity drops of approximately 0.67\% and 0.32\%, while their
interaction excess is $+0.0312$.

The group-level attention redistribution further indicates that these
pairwise effects are not explained by a nonspecific increase in total
attention toward the participating groups. Relative to the all-layer
baseline, the L3H4 key mass of Bob H/V and Bob singles decreases by
approximately 32.3\% and 18.9\%, respectively, while the Bob D/R key mass
changes only modestly by approximately $+3.1\%$. Nevertheless, their
within-group attention densities increase by approximately 28.3\%, 38.6\%,
and 66.3\%, respectively.

The strong Bob H/V $\leftrightarrow$ Bob D/R interaction therefore does not
arise from a nonspecific increase in overall attention toward Bob-related
tokens. Instead, the observations are consistent with redistribution of
Bob-related attention into a more concentrated internal interaction structure
within L3H4.

More generally, these results demonstrate that standalone intervention
importance and localized relational coupling quantify different properties of
the learned computation. A token group can exhibit a comparatively weak
standalone effect while nevertheless participating strongly in a localized
relational structure.

\subsection{Interpretation and Scope of the Attention-Assisted Analysis}
\label{subsec:sm_attention_scope}

The analyses above provide complementary information about the organization
and functional relevance of L3H4.

First, after the attention component shared across samples is removed,
sample-specific L3H4 attention variation remains strongly associated with the
ground-truth physical state. This state dependence is supported both by the
state-explained variance and by the separation between within-state and
between-state residual attention distances.

Second, targeted key interventions demonstrate group-dependent functional
effects specifically through the L3H4 pathway. The use of size-matched random
ablations indicates that these differences cannot be explained simply by the
number of tokens included in each group.

Third, comparison with whole-model input perturbation and IG establishes a
clear distinction between pathway-specific and end-to-end feature importance.
Whole-model perturbation and absolute IG show strongly concordant rankings,
whereas L3H4 exhibits a distinct ranking. This observation is consistent with
L3H4 probing a specialized internal pathway rather than reproducing the global
importance of the original input features.

Finally, the relational analysis shows that L3H4 contains a selective and
domain-structured organization of token-group interactions. The most prominent
features include strong Bob H/V $\leftrightarrow$ Bob D/R coupling, localized
Coinc.\ D/R $\leftrightarrow$ Coinc.\ H/V organization, localized
Coinc.\ H/V/D/R $\leftrightarrow$ Time organization, and reduced interaction
among several Alice--Bob and coincidence--single-count relations.

Within the present experimental setting, these observations are consistent with a role for L3H4 in selectively organizing population-sensitive, coherence/phase-sensitive, coincidence, and acquisition-duration information during reconstruction of the remotely prepared state under dynamic noise. However, these analyses should not be interpreted as establishing attention weights themselves as direct explanations of the final model prediction. Rather, they provide an attention-assisted, post-hoc attribution analysis supported by complementary statistical and intervention-based validation.

\subsection{Token and Group Nomenclature}
\label{subsec:sm_token_groups}

The individual input tokens and composite token groups used in the analyses are
summarized below. The abbreviation ``Coinc.'' is used for coincidence in the
text and figures. H/V and D/R denote paired measurement-basis subsets.
Composite groups are analytical groupings constructed from the listed input
tokens and do not constitute additional model-input tokens.
The notation $A\leftrightarrow B$ denotes a bidirectional relational analysis
between two predefined token groups.

\begin{table*}[htbp]
    \centering
    \caption{
    Individual input-token nomenclature used in the attention and intervention
    analyses.
    }
    \label{tab:sm_individual_tokens}
    \small
    \renewcommand{\arraystretch}{1.08}
    \begin{tabularx}{\textwidth}{
        >{\raggedright\arraybackslash}p{0.20\textwidth}
        >{\raggedright\arraybackslash}X
        >{\centering\arraybackslash}p{0.12\textwidth}}
        \toprule
        \textbf{Recommended name}
        &
        \textbf{Physical meaning}
        &
        \textbf{Token}
        \\
        \midrule
        Coincidence H & Coincidence count in the H basis. & 0 \\
        Coincidence V & Coincidence count in the V basis. & 1 \\
        Coincidence D & Coincidence count in the D basis. & 2 \\
        Coincidence R & Coincidence count in the R basis. & 3 \\
        Alice H       & Alice-side single count in the H basis. & 4 \\
        Bob H         & Bob-side single count in the H basis. & 5 \\
        Alice V       & Alice-side single count in the V basis. & 6 \\
        Bob V         & Bob-side single count in the V basis. & 7 \\
        Alice D       & Alice-side single count in the D basis. & 8 \\
        Bob D         & Bob-side single count in the D basis. & 9 \\
        Alice R       & Alice-side single count in the R basis. & 10 \\
        Bob R         & Bob-side single count in the R basis. & 11 \\
        Time          & Acquisition time. & 12 \\
        \bottomrule
    \end{tabularx}
\end{table*}

\begin{table*}[htbp]
    \centering
    \caption{
    Composite token groups used in the attention and intervention analyses.
    The listed groups are constructed analytically from the original 13 input
    tokens and are not additional model-input tokens.
    }
    \label{tab:sm_composite_groups}
    \small
    \renewcommand{\arraystretch}{1.08}
    \begin{tabularx}{\textwidth}{
        >{\raggedright\arraybackslash}p{0.20\textwidth}
        >{\raggedright\arraybackslash}X
        >{\centering\arraybackslash}p{0.22\textwidth}}
        \toprule
        \textbf{Recommended name}
        &
        \textbf{Physical meaning}
        &
        \textbf{Token(s)}
        \\
        \midrule

        Alice singles
        &
        All Alice-side single-count channels.
        &
        [4, 6, 8, 10]
        \\

        Bob singles
        &
        All Bob-side single-count channels.
        &
        [5, 7, 9, 11]
        \\

        Alice H/V
        &
        Alice-side H/V single-count channels.
        &
        [4, 6]
        \\

        Alice D/R
        &
        Alice-side D/R single-count channels.
        &
        [8, 10]
        \\

        Bob H/V
        &
        Bob-side H/V single-count channels.
        &
        [5, 7]
        \\

        Bob D/R
        &
        Bob-side D/R single-count channels.
        &
        [9, 11]
        \\

        Coinc.\ H/V
        &
        H/V coincidence-count channels.
        &
        [0, 1]
        \\

        Coinc.\ D/R
        &
        D/R coincidence-count channels.
        &
        [2, 3]
        \\

        Coinc.\ H/V/D/R
        &
        All coincidence-count channels.
        &
        [0, 1, 2, 3]
        \\

        All singles
        &
        All Alice- and Bob-side single-count channels.
        &
        [4, 5, 6, 7, 8, 9, 10, 11]
        \\

        H/V-related
        &
        All H/V-associated coincidence and single-count channels.
        &
        [0, 1, 4, 5, 6, 7]
        \\

        D/R-related
        &
        All D/R-associated coincidence and single-count channels.
        &
        [2, 3, 8, 9, 10, 11]
        \\

        H-related
        &
        H-basis coincidence and Alice/Bob single-count channels.
        &
        [0, 4, 5]
        \\

        V-related
        &
        V-basis coincidence and Alice/Bob single-count channels.
        &
        [1, 6, 7]
        \\

        D-related
        &
        D-basis coincidence and Alice/Bob single-count channels.
        &
        [2, 8, 9]
        \\

        R-related
        &
        R-basis coincidence and Alice/Bob single-count channels.
        &
        [3, 10, 11]
        \\

        \bottomrule
    \end{tabularx}
\end{table*}

\subsection{Interpretability Analysis}

High fidelity alone does not guarantee that a machine-learning model has learned physically relevant features. A model may instead exploit spurious correlations or dataset-specific artifacts that are unrelated to the underlying physics. Such shortcut learning has been identified in a variety of contexts, including image classification based on background snow rather than the animal itself~\cite{ribeiro2016should}, source-specific watermarks rather than object features~\cite{lapuschkin2019unmasking}, and textual or acquisition-related markers rather than pathological signatures in radiographic COVID-19 detection~\cite{degrave2021ai}.

This issue is particularly relevant in quantum experiments, where technical noise, experimental imperfections, and correlations introduced during data acquisition may provide unintended predictive cues. Interpretability therefore offers an important complement to fidelity-based benchmarks by testing whether the model response is associated with physically meaningful variables.

Here, we use interpretability analysis to examine the dependence of the denoising model on experimentally relevant variables and its response to noise. The above results suggest that the improvement in fidelity does not arise solely from dataset-specific shortcuts. While interpretability does not by itself prove that the model has learned the complete underlying physics, it provides an additional physical consistency check and strengthens the reliability of the denoising results.

\section{Comparison with Attention-Based Quantum Tomography Methods}
\label{sec:comparison_attention_qst}

The present work is closely related to recent attention-based
machine-learning approaches to quantum state tomography (QST) in
Refs.~\cite{cha2022attention,palmieri2024enhancing,ma2025tomography}.
Since density-matrix reconstruction and denoising are also integral
components of the present framework, the distinction should not be
understood simply as one between QST and RSP. Rather, the relevant differences concern the physical task and
experimental platform, the data and noise regime, the representation
supplied to the learning model, and the level at which the learned
model is validated. A quantitative comparison is summarized in
Table~\ref{tab:technical_comparison}. Because the studies involve
different state families, Hilbert-space dimensions, physical
platforms, and noise conditions, the reported fidelities and resource
requirements should not be interpreted as direct head-to-head
benchmarks.

\begin{table*}[t]
\centering
\caption{Technical comparison between the present work and
Refs.~\cite{cha2022attention,palmieri2024enhancing,ma2025tomography}
The numbers of trainable parameters for
Refs.~\cite{cha2022attention,palmieri2024enhancing} are calculated
from the network architectures specified in the corresponding works.}
\label{tab:technical_comparison}

\renewcommand{\arraystretch}{1.25}
\setlength{\tabcolsep}{4pt}
\footnotesize

\begin{tabularx}{\textwidth}{
    >{\raggedright\arraybackslash}p{0.12\textwidth}
    >{\raggedright\arraybackslash}X
    >{\raggedright\arraybackslash}X
    >{\raggedright\arraybackslash}X
    >{\raggedright\arraybackslash}X
}
\toprule

\textbf{Work}
&
\textbf{Task / platform}
&
\textbf{Model, data scale, and NN input}
&
\textbf{States and noise models}
&
\textbf{Reported performance}
\\

\midrule

\textbf{[51]} Cha \textit{et al.}, MLST 2022
&
Attention-based generative QST; reconstruction of density matrices
from tomographic measurements, including experimental data acquired
on an IBM quantum processor.
&
101,513 trainable parameters; two transformer layers with a
64-dimensional embedding; 2,700 measurements for the 3-qubit IBMQ
benchmark and 20,000 classically generated measurements for the
6-qubit demonstration.
&
GHZ pure-state reconstruction; simulated mixed GHZ states with
single-qubit or distributed bit-flip errors; IBMQ hardware noise.
&
For the 3-qubit IBMQ GHZ benchmark, AQT reports
$\mathcal{F}=0.917$, compared with $\mathcal{F}=0.897$ for MLE.
\\

\midrule

\textbf{[52]} Palmieri \textit{et al.}, PRR 2024
&
NN-enhanced QST; LI/MLE density-matrix estimates are denoised
through matrix-to-matrix neural post-processing using a Cholesky
representation.
&
549,503 trainable parameters; 10,000 training and 1,500 validation
samples; vectorized Cholesky-matrix input of length $d^{2}$,
corresponding to 256 real components for the $d=16$ four-qubit
OAT benchmark.
&
Simulated finite-shot noise, depolarizing noise, and
measurement/calibration bias; Haar/OAT and related tomography
benchmarks.
&
For four-qubit OAT states, the NN-enhanced reconstruction reaches
$(99.3\pm0.2)\%$ at $10^{6}$ trials and
$(87.6\pm4.1)\%$ at $10^{3}$ trials. OOD calibration-noise tests
report $(88.7\pm2.3)\%$ and $(91.0\pm1.9)\%$.
\\

\midrule

\textbf{[53]} Ma \textit{et al.}, IEEE Trans.\ Cybernetics 2025
&
Quantum-aware transformer for QST from structured measurements;
simulations and experiments on IBM quantum computers.
&
Approximately 144K and 152K trainable parameters for the 2- and
4-qubit models, respectively; 95,000 training states and 5,000
evaluation states; measured frequencies and measurement-operator
information are incorporated into the transformer.
&
Pure and mixed 2-, 3-, and 4-qubit states; finite-copy/statistical
noise; IBM hardware noise.
&
On \texttt{ibmq\_manila}, the mean QAT fidelities are
$0.975340$ and $0.994796$ for 100 and 1000 shots, respectively;
on \texttt{ibmq\_belem}, the corresponding values are
$0.976820$ and $0.993845$.
\\

\midrule

\textbf{This work}
&
Interpretable machine-learning-assisted state reconstruction and
denoising in an optical RSP experiment through a MMF channel.
&
152,292 trainable parameters; 1,727 real experimental training
samples; 13-dimensional experimental input consisting of four
coincidence counts, eight Alice/Bob single counts in the H/V/D/R
bases, and acquisition time; density-matrix output.
&
Real optical experimental data; mixed states, static pure states,
and pure states under a time-dependent Gaussian perturbation
physically introduced through an AWG-driven phase modulator;
polarization perturbations and preparation or measurement imperfections.
&
Static-state MLE fidelity: $0.8684\pm0.0093$. Under the dynamic
Gaussian-noise benchmark, the MLE fidelity decreases to
$(38.31\pm12.30)\%$, whereas the proposed model reconstructs
the RSP output with an average fidelity above $99.999\%$.
\\

\bottomrule
\end{tabularx}
\end{table*}

Ref.~\cite{cha2022attention} introduced attention-based quantum
tomography (AQT), in which a transformer-based generative model learns
tomographic measurement statistics and is subsequently used for
density-matrix reconstruction. The method was demonstrated
numerically and using experimental data acquired from an IBM
superconducting quantum processor. The use of self-attention was
motivated by its ability to capture correlations among different
subsystems. Ref.~\cite{ma2025tomography} further incorporated quantum
structure into the learning architecture through a quantum-aware
transformer, in which measured frequencies and measurement-operator
information are jointly encoded and the Bures distance is incorporated
into the learning objective. These works therefore establish important
attention-based and physics-aware strategies for QST reconstruction.

Among Refs.~\cite{cha2022attention,palmieri2024enhancing,
ma2025tomography}, Ref.~\cite{palmieri2024enhancing} is most closely
related to the present work because both approaches use neural networks
to improve density-matrix reconstruction in the presence of noise.
In Ref.~\cite{palmieri2024enhancing}, measurement frequencies are
first processed using a conventional QST estimator, such as linear
inversion (LI) or MLE. The resulting
density-matrix estimate is converted into a vectorized Cholesky
representation and supplied to an attention-based neural network,
which performs matrix-to-matrix post-processing to obtain an improved
density-matrix estimate. The training and evaluation are based on
numerically generated tomography data, including finite-statistics
noise, depolarization, and measurement/calibration errors.

A first major distinction concerns the experimental task and physical
platform. Refs.~\cite{cha2022attention,ma2025tomography} combine
numerical QST studies with experimental demonstrations on IBM
superconducting quantum processors. In particular,
Ref.~\cite{cha2022attention} applies AQT to measurement data acquired
from the IBMQ\_OURENSE device and benchmarks the reconstruction
against MLE tomography, whereas Ref.~\cite{ma2025tomography}
evaluates the quantum-aware transformer using experimental
measurements from the IBM \texttt{ibmq\_manila} and
\texttt{ibmq\_belem} processors. By contrast,
Ref.~\cite{palmieri2024enhancing} presents a simulation-based
QST-denoising study in which experimentally relevant noise mechanisms
are introduced numerically rather than through an operating physical
quantum experiment.

The present work is implemented directly in an optical
remote-state-preparation experiment, in which the remotely prepared
state propagates through a MMF channel and both
training and test data are acquired from the physical optical system.
The measured observables therefore contain the combined effects of
state-preparation and measurement imperfections, MMF-induced modal
mixing and scattering, and associated channel perturbations. In
addition, a controlled time-dependent disturbance is physically
introduced during acquisition by driving a phase modulator with an
arbitrary waveform generator (AWG) generated voltage waveform. The model is consequently required to
reconstruct the output density matrix from experimental observables
while the optical system is simultaneously affected by intrinsic
channel imperfections and externally imposed dynamic perturbations.
The learning task is therefore embedded in an actual
state-preparation, transmission, and measurement process, providing a
direct test of machine-learning-assisted reconstruction and denoising
in a dynamically varying optical RSP environment.

A second major distinction lies in the level at which the learned
model is validated. The previous works already incorporate physical
knowledge or model-level validation in different forms:
Ref.~\cite{cha2022attention} motivates self-attention through its
ability to represent quantum correlations,
Ref.~\cite{palmieri2024enhancing} provides a theoretical
interpretation of the network as a conditional ``debiaser'' and
examines the role of the transformer architecture, and
Ref.~\cite{ma2025tomography} explicitly incorporates
quantum-measurement information into the network design and learning
objective. These analyses, however, address physical motivation,
architectural design, or the performance contribution of model
components rather than the functional role of a specific learned
internal pathway.

As discussed in the preceding interpretability analysis, the present
work additionally probes a specific internal attention pathway using
complementary statistical, intervention-based, and end-to-end
attribution analyses. The results provide mutually consistent evidence
that the identified internal representation contains state-dependent
and experimentally structured information and contributes
functionally to the reconstruction process. Importantly, this
interpretation is not inferred from attention weights alone, but is
supported by independent perturbation and attribution tests. To the
best of our knowledge, Refs.~\cite{cha2022attention,
palmieri2024enhancing,ma2025tomography} do not report a comparable
pathway-level post-hoc analysis combining these complementary forms of
validation. The interpretability analysis therefore provides an
additional physical-consistency check on the learned reconstruction
beyond fidelity-based evaluation alone.

A further, although more implementation-specific, difference concerns
the input representation and data regime. As summarized in
Table~\ref{tab:technical_comparison}, the present model uses 1,727
experimentally acquired training samples and a 13-dimensional input
consisting of four coincidence counts, eight Alice/Bob single counts
in the H/V/D/R bases, and acquisition time, from which the density
matrix is reconstructed. In comparison,
Ref.~\cite{palmieri2024enhancing} uses 10,000 training and 1,500
validation states and applies the neural network to a vectorized
Cholesky representation of an already reconstructed density matrix;
for the $d=16$ four-qubit OAT benchmark, this representation contains
256 real components. The corresponding models contain 152,292 and
549,503 trainable parameters, respectively. These differences
illustrate the compact measurement-level representation and relatively
small experimental training set used in the present implementation.
They should not, however, be interpreted as establishing universal
computational superiority, since the Hilbert-space dimensions,
training distributions, and reconstruction tasks are different.

The distinction in representation is particularly relevant to
Ref.~\cite{palmieri2024enhancing}. Whereas its neural network performs
post-processing on an LI/MLE-reconstructed density matrix, the present
model maps experimentally identifiable observables directly to the
density matrix. Moreover, retaining acquisition time as an explicit
input provides temporal context associated with the dynamically driven
optical system. This measurement-level representation also provides a
natural interface for the interpretability analysis, since the learned
internal structure can be related directly to physically identifiable
quantities such as polarization-resolved coincidence counts,
single-count channels, and temporal information.

The performance values in Table~\ref{tab:technical_comparison} should
be interpreted within these distinct experimental settings. For
example, under the dynamic-noise benchmark considered here, the MLE
fidelity decreases to $(38.31 \pm 12.30)\%$, whereas the proposed
model reconstructs the remotely prepared states with an average fidelity above $99.999\%$. This result does not imply numerical superiority over
Refs.~\cite{cha2022attention,palmieri2024enhancing,ma2025tomography},
whose reported fidelities correspond to different states, noise
models, and platforms. Rather, it demonstrates that the learned
reconstruction remains effective when the measured data are acquired
from an operating optical RSP system subject to continuously varying
physical perturbations.

Taken together, the comparison in
Table~\ref{tab:technical_comparison} shows that the contribution of
the present work does not lie in the use of attention or machine
learning for QST per se. The main distinction is instead the
integration of machine-learning-assisted density-matrix reconstruction
and denoising into a real optical RSP experiment with an MMF channel
and physically imposed dynamic perturbations, together with a
mechanism-level analysis of the learned internal representation.
Relative in particular to Ref.~\cite{palmieri2024enhancing}, the
present framework therefore extends neural QST denoising from
density-matrix post-processing under numerically generated noise
toward measurement-level reconstruction in a dynamically perturbed
physical optical system. The additional interpretability analysis
further tests whether this learned correction is organized in a manner
consistent with physically meaningful experimental information.
\section{Scalability Analysis and MBQC Perspectives}
To better contextualize the practical utility of our proposed Transformer-based approach, it is valuable to assess the TQSC model's scalability and explore its broader implications. In the following subsections, we first examine the classical processing latency and platform-specific applicability of our model. We then analyze the intrinsic scalability of the Transformer architecture itself. Finally, we discuss the potential perspectives of our data-driven noise mitigation strategy for measurement-based quantum computing (MBQC) architectures.
\subsection{Evaluation of TQSC Classical Processing Latency and Platform-Specific Scalability}
The TQSC model requires no additional quantum hardware resources; all TQSC model processing is performed entirely in classical post-processing. Here, the primary scope of this work is quantum state preparation for quantum communication tasks.

Nevertheless, classical latency is a critical factor in real-time resource state generation and quantum error correction (QEC). We therefore evaluated the training and inference times of the TQSC model, as well as its feasibility across different quantum platforms. The analysis is organized as follows:

1. Training and Processing Time on Current Hardware.

The TQSC inference time is benchmarked on a standard GPU (NVIDIA RTX 3060), yielding about $5.15\,\mu\text{s}$ per state. Training is performed offline and incurs no runtime overhead. While this latency exceeds the $\sim\!1\,\mu\text{s}$ throughput target for superconducting qubits, recent Transformer-based surface-code decoders have outperformed conventional algorithmic decoders on physical hardware, confirming the potential of Transformers for practical QEC \cite{bausch2024learning}. Moreover, established hardware-software co-design techniques---lower-precision arithmetic, weight pruning, and deployment on FPGAs or ASICs exploiting fixed-point and sparsity---are expected to bring Transformer decoding into the sub-microsecond regime. For instance, a recent FPGA-based QEC decoder has demonstrated decoding latencies of $65$-$90\,\text{ns}$ and a total feedback latency of $446\,\text{ns}$ \cite{liu2026scalable}.

2. Real-time feasibility across platforms.

- Photonic: Resource state generation operates at a cycle time of approximately $1\,\mu\text{s}$ \cite{aghaee2025scaling}. A $5\,\mu\text{s}$ delay requires roughly $1\,\text{km}$ of coiled fiber, a common approach in photonics that introduces low loss ($<5\%$) \cite{bartolucci2023fusion,li2020advances}.

- Superconducting: The syndrome cycle time is approximately $1.1\,\mu\text{s}$, and the $5.15\,\mu\text{s}$ absolute latency is competitive with recent $63\,\mu\text{s}$ decoding demonstrations \cite{google2025quantum}. Sustained throughput at the $1.1\,\mu\text{s}$ cycle rate will require future FPGA deployment to eliminate GPU scheduling overhead \cite{liu2026scalable}.

- Trapped ions: QEC cycles in trapped-ion platforms extend from tens to hundreds of milliseconds \cite{ryan2021realization}. The $5.15\,\mu\text{s}$ TQSC latency is orders of magnitude below these timescales and is therefore negligible.

\begin{figure}[!tbp]
    \centering
    \includegraphics[width=0.9\linewidth]{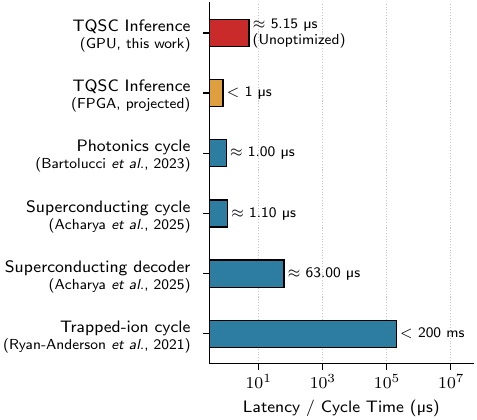}
    \caption{Classical processing latency of the TQSC model compared to characteristic timescales of quantum computing platforms.}
    \label{fig:latency}
\end{figure}

These comparisons are summarized in Fig.~\ref{fig:latency}, which plots the TQSC inference latency against characteristic timescales of the relevant quantum computing platforms.

3. Extension to QEC Protocols.

The TQSC model currently acts as a data-driven state reconstruction and error mitigation tool at the physical level; it is not a logical QEC decoder. However, the underlying mechanism---learning complex noise distributions to map corrupted measurements back to ideal states---shares a fundamental mathematical isomorphism with QEC decoding. Extension of this Transformer architecture to real-time QEC is feasible and has been validated by recent work such as Google DeepMind's AlphaQubit, which successfully decodes surface codes using Transformers \cite{bausch2024learning}.

To adapt the TQSC architecture into a real-time QEC scheme, particularly for photonic quantum computing, several structural modifications would be considered:

- Syndrome decoding adaptation: Rather than reconstructing the full density matrix from projection measurements, the model's input could be modified to process streams of noisy syndrome measurements. The output layer would be restructured to perform a classification task, predicting probable logical errors or required correction operators. The inherent sequence-modeling capabilities of Transformers suggest an advantage in processing the history of stabilizer measurements to extract encoded logical information.

- Mitigation of complex, realistic noise: In physical systems, syndrome information extracted from redundancy checks often suffers from non-local noise, including correlated crosstalk and measurement errors. While traditional algorithms, such as Minimum-Weight Perfect Matching, can face challenges with these asymmetric noise profiles, the Transformer architecture might offer an alternative. By directly processing raw, soft measurement data---such as unthresholded coincidence counts---the model has the potential to adaptively learn complex underlying error distributions, helping to relax the reliance on simplified theoretical noise assumptions.

- Continuous decoding for resource states: For protocols that require continuous resource state generation, common in measurement-based quantum computing, the decoder must handle an ongoing stream of data. The Transformer architecture could be augmented with sliding-window attention or recurrent mechanisms. These additions allow the model to maintain an internal decoder state, process incoming continuous syndrome data, and potentially generalize to deeper circuits without incurring memory overflow.
\subsection{Scalability of the Transformer Architecture}
A potential concern regarding scalability arises from the fact that the standard Transformer's self-attention scales as $O(N^2)$ with the system size $N$ \cite{vaswani2017attention,tay2022efficient}. Nevertheless, our approach remains practical and promising for the following reasons.

\textbf{Asymmetry of Training and Inference Costs.}
As previously noted, the training and inference costs are highly asymmetric: pre-training is a one-time offline cost, whereas inference---generating the remote state preparation protocol---is highly efficient. For our application, inference requires only a single forward pass (approximately $5.15\,\mu\text{s}$), which is orders of magnitude faster than training \cite{tay2022efficient}.

\textbf{Focus on the NISQ Era.}
Our current framework is targeted at the Noisy Intermediate-Scale Quantum (NISQ) regime \cite{preskill2018quantum}. In this regime, as defined by Preskill, the number of available qubits is not yet sufficient for fault tolerance (typically ranging from tens to a few hundreds), and the primary bottleneck is not asymptotic scalability, but rather maximizing the fidelity of operations under constrained circuit depths. The Transformer's ability to capture complex, non-local correlations enables the discovery of highly optimized, hardware-specific state preparation protocols that are difficult to obtain with analytical methods. Although the FLOPs of self-attention scale quadratically as $O(N^2)$, the wall-clock latency on modern GPUs and TPUs does not follow this trend for NISQ-relevant sequence lengths of up to a few hundred. Matrix multiplications in self-attention are highly parallelized and, for $N$ below 1000, are memory-bandwidth bound, leaving many cores under-utilized \cite{kwon2023efficient,pope2023efficiently}. Consequently, increasing $N$ within this regime typically adds little latency, as the extra operations can be largely absorbed by the parallel hardware.

\textbf{Applicability in the Future FTQC Era.}
Looking forward to the Fault-Tolerant Quantum Computing (FTQC) era, where systems will scale to millions of physical qubits, the classical processing bottleneck shifts from physical-level calibration to syndrome decoding and the orchestration of fault-tolerant logical gates \cite{campbell2017roads}. Consequently, the Transformer model would target the logical qubit layer rather than the physical one. Since logical qubits are encoded using thousands of physical qubits, the effective dimensionality of the problem is reduced by orders of magnitude, ensuring that the computational overhead remains tractable.

\textbf{Architectural Pathways for Scalable Inference.}
To further reduce inference time as $N$ grows, several algorithmic enhancements independent of physical hardware accelerations (such as FPGAs) can be employed.
\begin{itemize}
    \item \textbf{Linear Attention and Efficient Transformers.} Integrating mechanisms such as Linformer or Performer can provably reduce the theoretical inference and memory complexity from $O(N^2)$ to $O(N\log N)$ or strictly $O(N)$ \cite{tay2022efficient}.
    \item \textbf{State Space Models.} State-space models such as Mamba substitute attention with linear-time sequence models, achieving $O(N)$ inference with constant memory while matching Transformer performance \cite{gu2023mamba}.
    \item \textbf{Knowledge Distillation and Quantization.} The heavy pre-trained Transformer can be distilled into a lightweight student model, or post-training quantization (e.g., INT8) can be applied \cite{gholami2022survey}. This drastically reduces the number of active parameters during the forward pass, accelerating inference speed without sacrificing the fidelity of the generated quantum protocols.
    \item \textbf{Transfer Learning.} As previously noted \cite{mari2020transfer}, models can be pre-trained on smaller quantum subsystems and efficiently fine-tuned or generalized to larger ones, effectively managing the scaling of classical computation.
\end{itemize}
\subsection{Potential Perspectives for MBQC Architectures}

Standard MBQC relies heavily on the deterministic generation of pristine, large-scale cluster states~\cite{raussendorf2001one}. However, maintaining such states is experimentally demanding: hardware noise and decoherence inevitably degrade state fidelity, and errors in the initial cluster state can propagate detrimentally during computation~\cite{nielsen2006cluster,dawson2006noise}. Our machine-learning approach is inherently data-driven and has the potential to address this challenge directly at the data-processing level. Similar to the hardware-aware noise resilience observed in variational quantum algorithms~\cite{cerezo2021variational}, our model can incorporate specific hardware noise profiles into its training process. By treating noisy projective measurement outcomes as sequence data, the Transformer shows the capability to adaptively learn and mitigate the underlying hardware noise. This demonstrates the potential for efficient reconstruction of high-fidelity density matrices, offering a characterization strategy with inherent robustness to specific decoherence and gate errors, without the need for physical error-correction overhead.

\bibliography{Reference.bib}

\begin{figure*}[hbt!]
    \centering
    \begin{subfigure}[b]{0.48\textwidth}
        \centering
        \includegraphics[width=\textwidth]{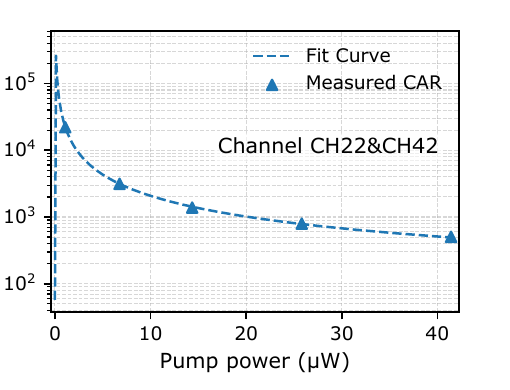}
        \caption{}
        \label{fig:CH22_42}
    \end{subfigure}
    \hfill
    \begin{subfigure}[b]{0.48\textwidth}
        \centering
        \includegraphics[width=\textwidth]{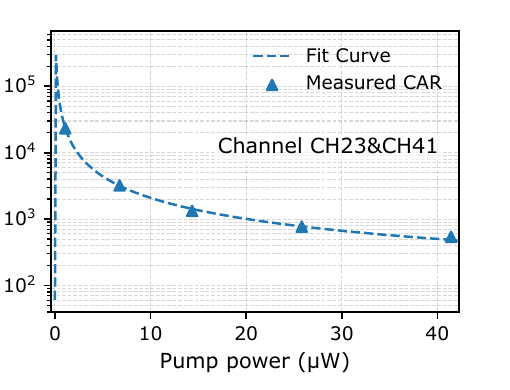}
        \caption{}
        \label{fig:CH23_41}
    \end{subfigure}
    
    \vspace{0cm} 
    \begin{subfigure}[b]{0.48\textwidth}
        \centering
        \includegraphics[width=\textwidth]{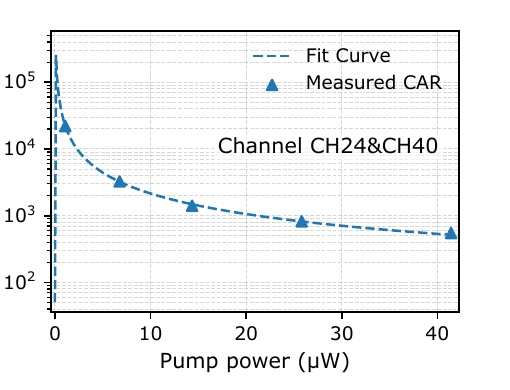}
        \caption{}
        \label{fig:CH24_40}
    \end{subfigure}
    \hfill
    \begin{subfigure}[b]{0.48\textwidth}
        \centering
        \includegraphics[width=\textwidth]{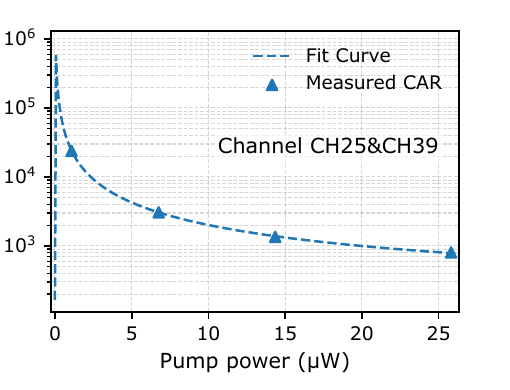}
        \caption{}
        \label{fig:CH25_39}
    \end{subfigure}
    
    \vspace{0cm} 
    \begin{subfigure}[b]{0.48\textwidth}
        \centering
        \includegraphics[width=\textwidth]{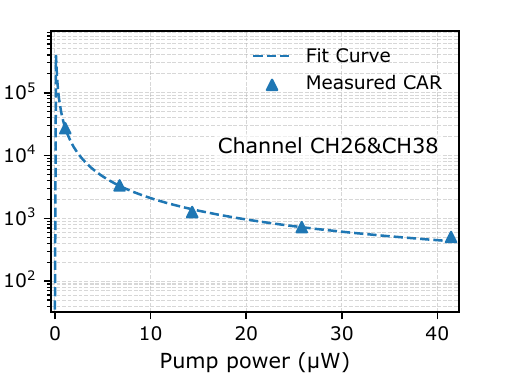}
        \caption{}
        \label{fig:CH26_38}
    \end{subfigure}
    \hfill
    \begin{subfigure}[b]{0.48\textwidth}
        \centering
        \includegraphics[width=\textwidth]{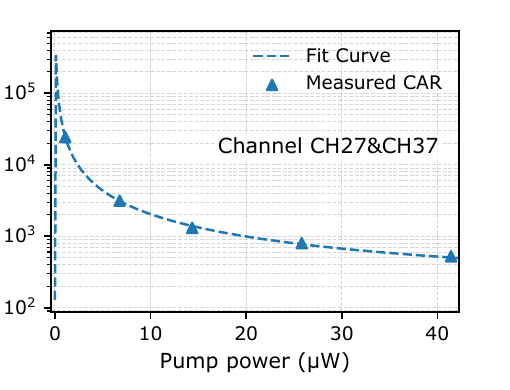}
        \caption{}
        \label{fig:CH27_37}
    \end{subfigure}
    \caption{Characterization of full-spectrum signal-to-noise ratio performance. The Coincidence-to-Accidental Ratio (CAR) as a function of pump power is shown for 10 symmetric channel pairs. (a)-(f) show the outer pairs from CH22 \& CH42 to CH27 \& CH37.}
    \label{fig:S2_part1}
\end{figure*}

\begin{figure*}[hbt!]
    \ContinuedFloat 
    \centering
    \begin{subfigure}[b]{0.48\textwidth}
        \centering
        \includegraphics[width=\textwidth]{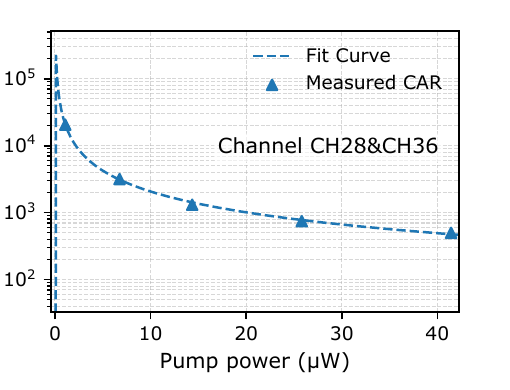}
        \caption{}
        \label{fig:CH28_36}
    \end{subfigure}
    \hfill
    \begin{subfigure}[b]{0.48\textwidth}
        \centering
        \includegraphics[width=\textwidth]{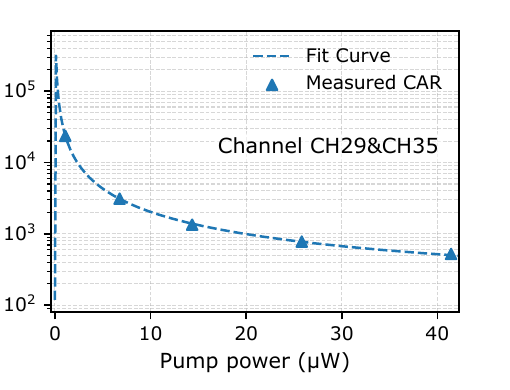}
        \caption{}
        \label{fig:CH29_35}
    \end{subfigure}
    
    \vspace{0cm} 
    \begin{subfigure}[b]{0.48\textwidth}
        \centering
        \includegraphics[width=\textwidth]{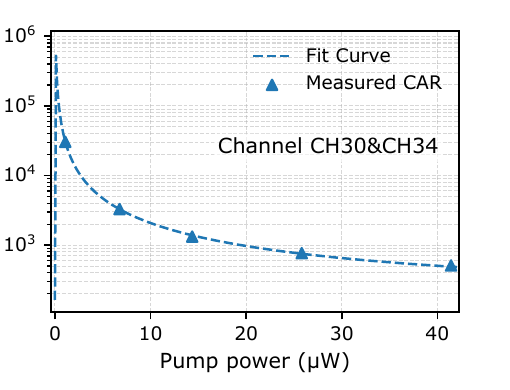}
        \caption{}
        \label{fig:CH30_34}
    \end{subfigure}
    \hfill
    \begin{subfigure}[b]{0.48\textwidth}
        \centering
        \includegraphics[width=\textwidth]{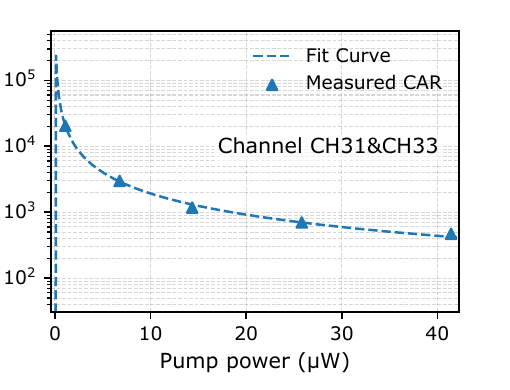}
        \caption{}
        \label{fig:CH31_33}
    \end{subfigure}
    \caption{(Continued) Characterization of full-spectrum signal-to-noise ratio performance. (g)-(j) show the inner pairs from CH28 \& CH36 to CH31 \& CH33. Blue triangles represent experimental data, and dashed lines indicate theoretical fits ($\text{CAR} \propto 1/P_{\text{pump}}$). All channels exhibit CAR values exceeding 200,000 at low pump powers, indicating consistently high signal quality across the entire generated bandwidth.}
    \label{fig:S2}
\end{figure*}

\begin{figure*}[hbt!]
    \centering
    \begin{subfigure}[b]{0.48\textwidth}
        \centering
        \includegraphics[width=\textwidth]{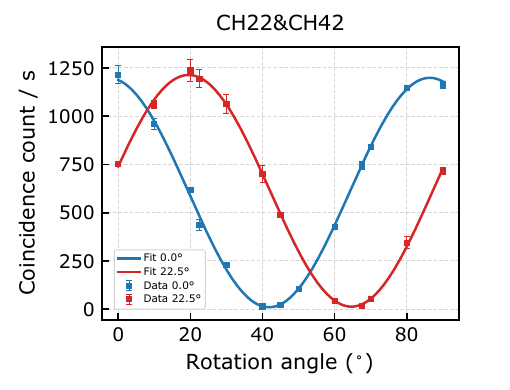}
        \caption{}
        \label{fig:CH22_42}
    \end{subfigure}
    \hfill
    \begin{subfigure}[b]{0.48\textwidth}
        \centering
        \includegraphics[width=\textwidth]{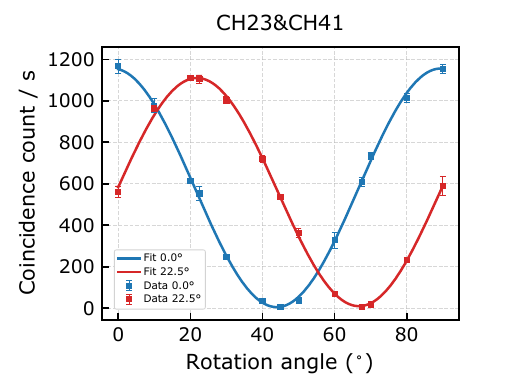}
        \caption{}
        \label{fig:CH23_41}
    \end{subfigure}
    \begin{subfigure}[b]{0.48\textwidth}
        \centering
        \includegraphics[width=\textwidth]{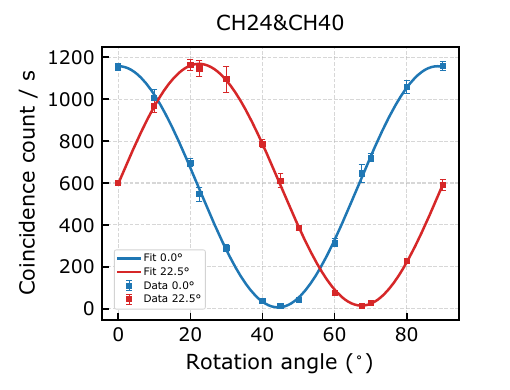}
        \caption{}
        \label{fig:CH24_40}
    \end{subfigure}
    \hfill
    \begin{subfigure}[b]{0.48\textwidth}
        \centering
        \includegraphics[width=\textwidth]{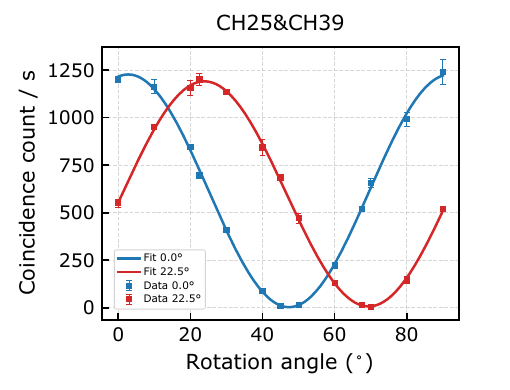}
        \caption{}
        \label{fig:CH25_39}
    \end{subfigure}
    \begin{subfigure}[b]{0.48\textwidth}
        \centering
        \includegraphics[width=\textwidth]{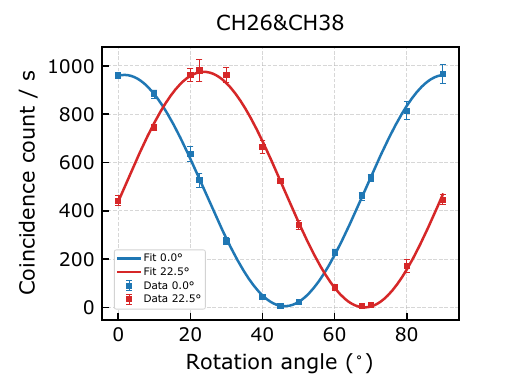}
        \caption{}
        \label{fig:CH26_38}
    \end{subfigure}
    \hfill
    \begin{subfigure}[b]{0.48\textwidth}
        \centering
        \includegraphics[width=\textwidth]{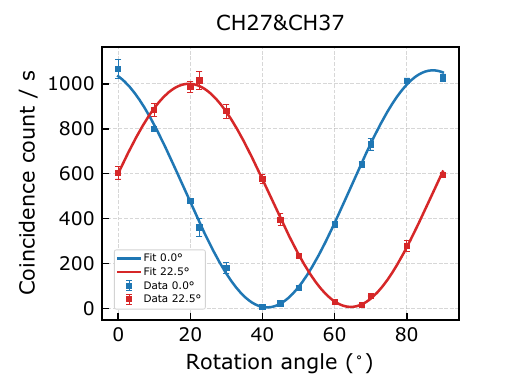}
        \caption{}
        \label{fig:CH27_37}
    \end{subfigure}
    \caption{Polarization entanglement correlation fringes for 10 symmetric channel pairs. (a)-(f) show the outer pairs from CH22 \& CH42 to CH27 \& CH37.}
    \label{fig:S4_part1}
\end{figure*}

\begin{figure*}[hbt!]
    \ContinuedFloat
    \centering
    \begin{subfigure}[b]{0.48\textwidth}
        \centering
        \includegraphics[width=\textwidth]{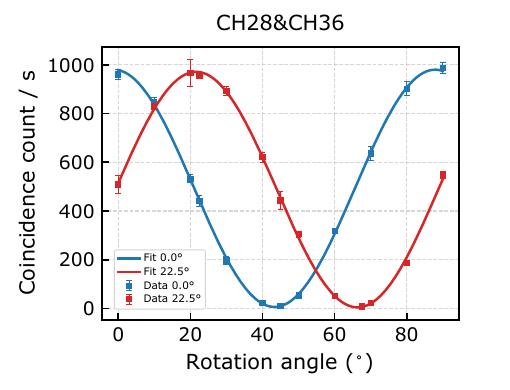}
        \caption{}
        \label{fig:CH28_36}
    \end{subfigure}
    \hfill
    \begin{subfigure}[b]{0.48\textwidth}
        \centering
        \includegraphics[width=\textwidth]{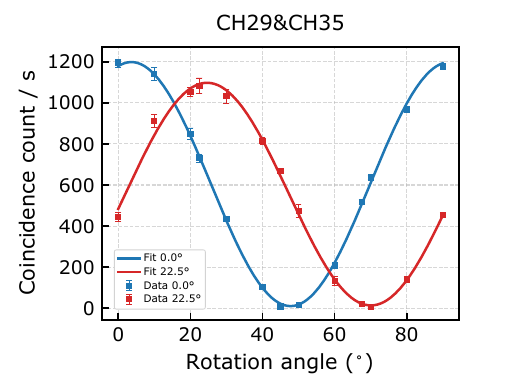}
        \caption{}
        \label{fig:CH29_35}
    \end{subfigure}
    \begin{subfigure}[b]{0.48\textwidth}
        \centering
        \includegraphics[width=\textwidth]{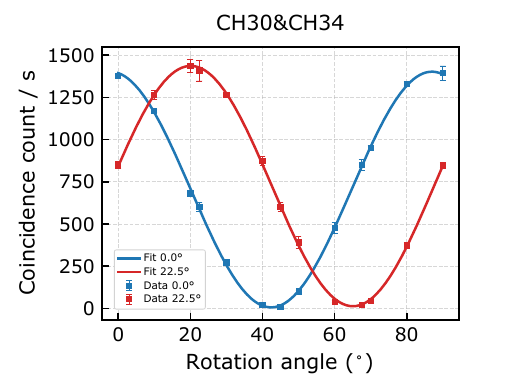}
        \caption{}
        \label{fig:CH30_34}
    \end{subfigure}
    \hfill
    \begin{subfigure}[b]{0.48\textwidth}
        \centering
        \includegraphics[width=\textwidth]{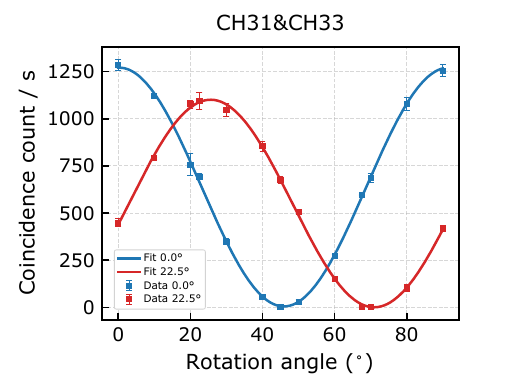}
        \caption{}
        \label{fig:CH31_33}
    \end{subfigure}
    \caption{(Continued) Polarization entanglement correlation fringes for 10 symmetric channel pairs. (g)-(j) show the inner pairs from CH28 \& CH36 to CH31 \& CH33. The plots show coincidence counts as a function of the polarization analyzer angle. Blue and red curves correspond to measurements in the $0^\circ$ and $22.5^\circ$ bases, respectively (points represent experimental data; solid lines represent sinusoidal fits). High interference visibility (average $>98\%$) is observed in both bases for all channels.}
    \label{fig:S4}
\end{figure*}